\documentclass[%
 reprint,
nofootinbib,
 amsmath,amssymb,
 aps,
]{revtex4-2}

\usepackage{tikz}
\usepackage[compat=1.1.0]{tikz-feynman}

\usepackage{xcolor}

\usepackage{slashed}
\usepackage{graphicx}
\usepackage{dcolumn}
\usepackage{bm}
\usepackage{hyperref}

\begin{document}


\title{Can weak-gravity, causality-violation arguments constrain modified gravity?}

\author{Stephon Alexander}
 \email{stephon\_alexander@brown.edu}
 
\author{Heliudson Bernardo}%
 \email{heliudson\_bernardo@brown.edu}
\affiliation{Brown Theoretical Physics Center \& Department of Physics, Brown University,\protect\\  Barus Building, 340 Brook Street, Providence, RI 02912, USA}

\author{Nicol\'as Yunes}
\email{nyunes@illinois.edu}
\affiliation{Illinois Center for Advanced Studies of the Universe \& Department of Physics \protect \\ University of Illinois at Urbana-Champaign, Urbana, Illinois 61801, USA}%

\begin{abstract}
We investigate limitations of causality arguments from flat-spacetime amplitudes, based on the eikonal limit of gravitational scattering, to place constraints on modified gravity. We show that causality constraints are only valid in the weak-gravity regime even for transplanckian scattering, and that such constraints are much less stringent than astrophysical ones, obtained for example from gravitational waves emitted in black hole coalescence. Special attention is given to the weakness of causality constraints on dynamical Chern-Simons gravity, but our results apply to other modified gravity theories as well.  In the context of that theory, we also discuss how to obtain a time-delay formula from black hole, neutron stars, and shockwave solutions. For scattering with compact objects, we explicitly show that time delays are greatly suppressed by the ratio of the object's mass to the impact parameter, so time advances only occur greatly outside the cut off of the theory. For the shockwave solution, we find that the time delay is always positive within the regime of validity of the solution. We also comment on the impact of graviton nonlinearities for time-delay calculations in the nonlinear, strong gravity regime. We conclude that amplitude-based causality constraints on modified gravity are typically not stringent relative to other experimental and observational bounds.  
\end{abstract}

\maketitle


\section{Introduction}\label{sec:intro}

While a fully-fledged, physical quantum-gravity theory is yet to be developed, a conservative perspective on the principles of quantum field theory indicates that General Relativity (GR) can be used to compute gravitational scattering amplitudes, provided the energy scales involved in the process are much smaller than the Planck scale. Despite its ultraviolet non-renormalizability issues, from an effective field theory perspective, GR is a quantum field theory for gravitons that yields universal, low-energy predictions \cite{Donoghue:1994dn, Burgess:2003jk, Bjerrum-Bohr:2013bxa, Donoghue:2015hwa, Donoghue:2017pgk}. On top of that, there is a certain regime in which GR can also be used to tackle transplanckian gravitational scattering \cite{tHooft:1987vrq}. This includes, for instance, the physical scattering of macroscopic bodies mediated by gravity (such as interstellar objects scattered by the Sun), where the center of mass energies are at least the sum of the masses of the bodies, each exceeding the Planck mass by over 36 orders of magnitude.  

For concreteness, consider the $2\to 2$ scattering of massive scalar particles via graviton exchanges. Using the Mandelstam variable $s$, the square of the sum of the momenta of the ingoing or outgoing particles, transplanckian gravitational scattering is defined by $\sqrt{s} \gg M_{\rm Pl}$ or $Gs\gg \hbar$. For fundamental quanta, this regime corresponds to the ``ultra-relativistic limit,'' and in such a high-energy limit, the so-called \textit{eikonal approximation} is semi-classical. This point is evident after considering the opposite, subplanckian regime $Gs\ll \hbar$. In this case, the Schwarzschild radius scale $G\sqrt{s}$ is smaller than the Compton wavelength scale $\ell_s \sim \hbar/\sqrt{s}$ and so a quantum description is necessary. On the other hand, for $Gs\gg \hbar$, we have $G\sqrt{s} \gg \ell_s $ and a classical description is possible. 

As we shall see in the next sections, we can obtain semiclassical observables from the eikonal limit of transplanckian scattering, such as the Shapiro time delay and the scattering angle. By Shapiro time delay, or simply time-delay for short, we refer to the difference in how long it takes for a wave packet to propagate between two spacetime events with and without scattering by a massive source (as measured by asymptotic observers). The classroom example for wave-packet scattering is that which light from distant stars experiences as it grazes the surface of the Sun on its way to Earth, which led to the first experimental confirmation of Einstein's theory of GR by Eddington~\cite{Dyson:1920cwa}. A modern example of Shapiro time delay is that suffered by electromagnetic signals emitted by the Cassini spacecraft when it was near Jupiter, as the signals grazed the surface of the Sun on their way to Earth~\cite{Bertotti:2003rm}.
In GR, this time delay is always positive (i.e.~it's a time \textit{delay}), as one intuitively expects based on causality. 

In beyond-Einstein gravity, however, there can be corrections to the Shapiro time delay, and if these corrections are negative enough to overwhelm the GR prediction, there may be a causality problem \cite{Eisenbud:1948paa,Wigner:1955zz} (for discussion of the formulation of the causality problem when gravity is dynamical, see e.g.~\cite{penrose1980schwarzschild, Gao:2000ga}). In \cite{Camanho:2014apa}, the impact of higher-derivative corrections to GR on the gravitational S-matrix was studied. In particular, the authors calculated how the higher-derivative modifications to the S-matrix yield corrections to the Shapiro time delay in transplanckian scattering calculations within the eikonal limit. From these beyond-Einstein corrections to the Shapiro time delay, the authors found constraints on the coefficient of operators yielding modifications to the graviton three-vertex in the perturbative regime. These constraints were required to avoid causality issues in the Shapiro time-delay calculation. This work led to a plethora of similar studies, which employ causality arguments on the S-matrix of different scattering processes to place causality constraints on many low-energy effective field theory extensions of GR \cite{Camanho:2016opx, Goon:2016une, Hinterbichler:2017qyt,Bellazzini:2015cra,deRham:2019ctd,deRham:2020zyh,Tokuda:2020mlf,AccettulliHuber:2020oou,Chen:2021bvg,Caron-Huot:2021rmr,Caron-Huot:2022ugt,Hong:2023zgm,Aoki:2023khq,Caron-Huot:2024tsk,deRham:2022hpx,Serra:2022pzl,Cremonini:2023epg,Kehagias:2024yyp,Cassem:2024djm,Xu:2024iao,Cremonini:2024lxn}.

In this paper, we revisit the validity of these causality arguments for beyond-Einstein gravity theories. More specifically, we consider theories in which GR is modified by the addition of higher-derivative operators, suppressed by powers of some energy scale $M_\lambda \leq M_{\rm Pl}$. The theory is not valid at energies close to or larger than $M_\lambda$, since then, even higher-derivative operators would have had to be included. In the opposite regime, for energies much lower than $M_\lambda$, we expect that all observables will be merely corrected by the extra operators, where, in particular, these beyond-Einstein corrections never dominate the GR predictions. However, in transplanckian scattering, energies larger than $M_{\rm Pl}$ and hence $M_{\lambda}$ are considered, and this begs the question of whether the effective-field-theory picture is valid in such calculations. We find that the answer to this question is affirmative (i.e.~the effective-field-theory picture is valid in transplanckian scattering) \textit{provided} we take the eikonal limit properly and respect its range of applicability.

As we discuss in Sec.~\ref{sec:causality_constraints}, we find that the eikonal limit of gravitational scattering is restricted to the perturbative, weak-field regime of gravity. This is because the definition of gravitons and the gravitational S-matrix in quantum field theory demand a weak-gravity expansion. For instance, in two-body scattering by graviton exchanges in quantum field theory, the eikonal limit is only valid for impact parameters much larger than the Schwarzschild radius scale $G\sqrt{s}$, \emph{even} in the transplanckian regime. Large values of impact parameter correspond to small scattering angles and coincide with the weak-gravity region, where there are no ambiguities in the definition of graviton states and nonlinearities are suppressed by the Planck scale. 

Only when $M_{\lambda} \ll M_{\rm Pl}$ can we hope that causality arguments could inform us about the validity of high-derivative corrections in the action. Failure to note the regime of validity ($M_{\lambda} \ll M_{\rm Pl}$) where causality arguments are valid has resulted in a misuse of causality constraints to incorrectly ``rule out'' modified gravity theories that are effective field theory extensions of GR. For scattering processes in which the value of $M_\lambda$ can only be probed in the strong-gravity regime, gravitational scattering ceases to be useful, and one has to take into account the full nonlinear theory to compute time delays. The latter approach has been successful, especially for metric theories, and has the advantage that the back-reaction of the scattered bodies can be considered in the solution of the theories. For instance, in the transplanckian scattering of massive bodies by gravitons, in the limit of one mass much larger than the other, the metric solution is dominated by the heavier body, and the scattering of the lighter one can be approximated in a probe limit. 

Essentially, causality constraints based on eikonal scattering can give a lower bound on $M_{\lambda}$, at most. Other lower bounds can be obtained from the lack of observational effects, for example in the gravitational waves emitted in the coalescence of black holes~\cite{Yunes2025}. Upper bounds can only be obtained by observations, directly or indirectly, that positively identify a modification from GR, which has not yet occurred in strong-field gravitational observations. Due to the scope of the eikonal approximation, the theoretical lower bound can never be indefinitely large. If one incorrectly sets the theoretical lower bound arbitrarily higher than any lower bound obtained from (the lack of) observations, then one would conclude that the theory is not valid and has to be modified. The question is then, how large can we assume the theoretical lower bound to be? 

The size of the theoretical lower bound depends on the process considered. Let us illustrate the situation in dynamical Chern-Simons (dCS) gravity \cite{Bellucci:1988ff, Bellucci:1990fa, Gates:1991qn, Lue:1998mq, Choi:1999zy, Jackiw:2003pm,Alexander:2009tp}. In such a theory, the higher-derivative operator is proportional to the gravitational Pontryagin density $R\Tilde{R}$ times a dynamical (pseudo)scalar field $\phi$. In \cite{Serra:2022pzl}, the authors studied the eikonal scattering of a $\phi$ quanta out of a heavy scalar, which is meant to represent any source of curvature (e.g.~an astrophysical body), via gravitational interaction. The extra dCS operator induces a new scalar-graviton-graviton cubic vertex that allows the incoming dCS scalar to be converted into an outgoing graviton. This new vertex then induces a negative correction to the GR prediction for the Shapiro time delay that scales with the inverse of the cut-off scale $M_\lambda$. If this correction is too large, for example by choosing $M_\lambda$ to also be very small, then the dCS correction can overwhelm the GR part of the Shapiro time delay, leading to a time advance that is acausal and unphysical. 
By requiring the absence of a negative time delay, the authors of \cite{Serra:2022pzl} then found the bound $M_{\lambda} \gtrsim b^{-1}$, where $b$ is the impact parameter of the scattering. However, for consistency, the impact parameter $b$ cannot be smaller than the Schwarzschild scale, and thus, the lower bound $b^{-1}$ cannot be arbitrarily large. This means that dCS theory cannot be ruled out, and in fact, the current astrophysical constraints are more stringent than this theoretical bound when applied to realistic black holes, as we show in Sec. \ref{sec:validity_causality}.

The remainder of this paper provides a discussion of the validity of eikonal approximation for gravitational scattering in modified gravity. To set up notation and some definitions, in the next section we review the eikonal approximation and how semiclassical observables are computed from the S-matrix. In Sec.~\ref{sec:causality_constraints} we stress some subtleties when taking the eikonal approximation for the gravitational S-matrix, and comment on a limitation associated with the perturbative approach in gravity. We show this explicitly by computing curvature corrections to the graviton three-vertex in perturbative GR. These can be straightforwardly generalized to any modified gravity theories. In Sec.~\ref{sec:validity_causality}, we study causality constraints in dCS gravity and explain why these cannot be more stringent than current astrophysical constraints. We identify the leading diagrams that have to be considered as the impact parameter gets close to the strong gravity regime. Sections \ref{sec:time_delays} and \ref{sec:shockwave_backgrounds} are devoted to calculating time delays in metric theories using geometry, by means of the Shapiro time delay and graviton propagation in shockwave backgrounds, respectively. Section~\ref{sec:conclusion} summarizes our results and presents our conclusions. We work with the mostly-plus signature convention for the metric and set $\hbar =1$ whenever not stated otherwise to make contact with the high-energy community.

\section{Eikonal approximation, phase shift, and time delay}
\label{sec:eikonal}

In this section, we review the basics of scattering amplitude calculations in the eikonal limit, and how this can be used to calculate the Shapiro time delay, following mostly~\cite{Giddings:2011xs, DiVecchia:2023frv} and references therein. Most of this section is a review of well-known high-energy literature, which we summarize here for completeness, to establish notation, and to inform non-high-energy communities that may not be familiar with these theory developments.  

\begin{figure}[t]
    \centering
    \includegraphics[width=1\linewidth]{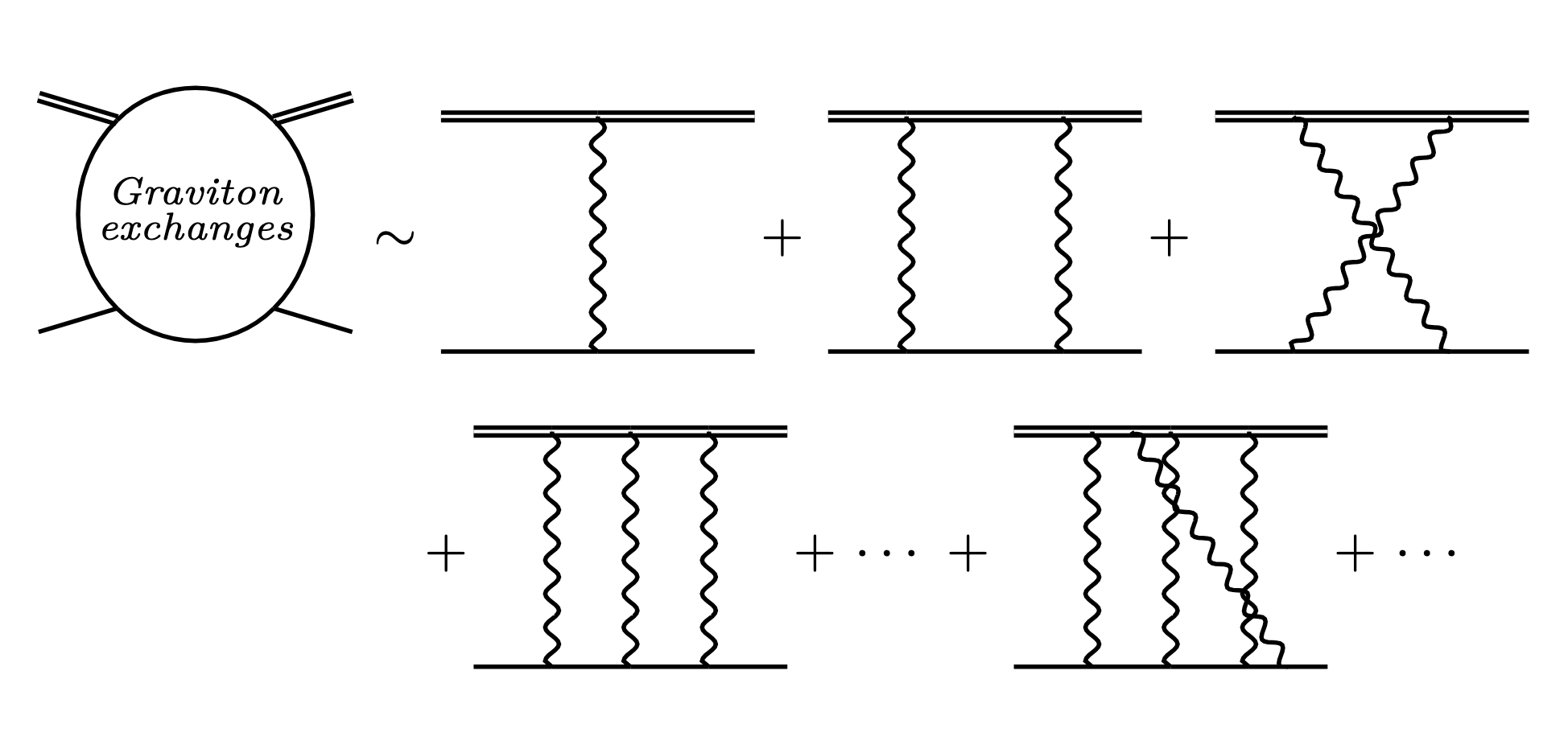}
    \caption{The eikonal amplitude is obtained by summing multi-loop, graviton-tree diagrams. The wiggly lines represent gravitons, while the double and single lines represent quanta minimally coupled to gravity. The dots represent not only higher-loop-order diagrams but also permutations of these. Only in the Born limit, can the amplitude be approximated by the tree-level, single-graviton exchange diagram.}
    \label{fig:eikonal_diagrams}
\end{figure}

Consider the $2\to 2$ scattering of massive scalar particles through graviton exchanges. In the subplanckian case, we consider an expansion in $Gs \ll 1$, corresponding to the Born approximation of the scattering problem, where recall that $s$ is the Mandelstam variable corresponding to the square of the sum of the incoming or outgoing 4-momenta (see e.g. \cite{Giddings:2009gj,Giddings:2011xs}). For the gravitational interaction, the Born amplitude is dominated by a single-graviton exchange diagram. In the transplanckian case, the expansion can be reorganized in powers of the ratio between the (square of the) momentum transferred and the (square of the) center of mass energy, i.e. $t/s \ll 1$ \cite{Giddings:2010pp}. This corresponds to the eikonal approximation \cite{Levy:1969cr}, and it is equivalent to an expansion in $G\sqrt{s}/b \ll 1$ where $b$ is the impact parameter of the scattering event. In the transplanckian case, the eikonal amplitude is dominated by graviton ladder loop diagrams, corresponding to multi-graviton exchanges (see Fig.~\ref{fig:eikonal_diagrams}) \cite{Kabat:1992tb,Amati:1993tb}. 

In the eikonal regime, a resummation of the ladder diagrams gives rise to the following amplitude \cite{Amati:1993tb,Adamo:2021rfq, DiVecchia:2023frv}
\begin{align}\label{amplitude_phase_relation}
    i\mathcal{M}(q_\perp) &= \frac{2\sqrt{\left[s-(m_1+m_2)^2\right]\left[s-(m_1 -m_2)^2\right]}}{\hbar^2} \nonumber\\
    &\times\int d^{D-2} b_\perp e^{-i q_\perp \cdot b_\perp/\hbar}\left(e^{2i\delta(b_\perp)/\hbar}-1\right), 
\end{align}
where $m_1$ and $m_2$ are the masses of the scalar particles, $b_\perp$ is the part of the impact parameter that is orthogonal to the incoming momenta $p_1^\mu$ and $p_2^\mu$, and $q_\perp$ is the transferred momentum conjugate to $b_\perp$. To leading order in $Gs/b^{D-4}$, the eikonal phase $\delta(b_\perp)$ is proportional to the Fourier transform of the tree-level Born amplitude, 
\begin{align}
    2 \delta(b_\perp) &\approx \frac{1}{2\sqrt{\left[s-(m_1+m_2)^2\right]\left[s-(m_1 -m_2)^2\right]}} \nonumber\\
    &\times\int \frac{d^{D-2}q_\perp}{(2\pi)^{D-2}}e^{iq_\perp \cdot b_\perp/\hbar} i\mathcal{M}_0(t= -q_\perp^2),
    \label{eq:Born-limit}
\end{align}
where the Born amplitude is
\begin{align}
    i\mathcal{M}_0 &= \frac{2i \kappa^2}{-t}\left[\frac{1}{2}(s-m_1^2-m_2^2)^2 
    \right.
    \nonumber \\
    &\left.-\frac{2m_1^2m_2^2}{D-2}+\frac{t}{2}(s-m_1^2-m^2_2)\right],
\end{align}
with $\kappa = \sqrt{8\pi G}$. Inserting this expression for ${\cal{M}}_0$ into $\delta(b_\perp)$, one finds
\begin{align}
    2\delta(b_\perp) &\approx \frac{2G}{\hbar} \frac{\frac{1}{2}(s-m_1^2-m_2^2)^2-\frac{2m_1^2m_2^2}{D-2}}{\sqrt{\left[s-(m_1+m_2)^2\right]\left[s-(m_1 -m_2)^2\right]}}\nonumber\\
    &\times \frac{\Gamma\left(\frac{D-4}{2}\right)}{(\pi b^2)^{\frac{D-4}{2}}}.
\end{align}

The constant $D$ in the above equations refers to the dimensionality of spacetime, which is here kept general to allow for dimensional regularization. More specifically, to get the $b$ dependence in four dimensions, we use dimensional regularization and write the spacetime dimension as $D= 4- 2\epsilon$, such that
\begin{equation}
     \frac{\Gamma\left(\frac{D-4}{2}\right)}{(\pi b^2)^{\frac{D-4}{2}}}  = 2\left(\frac{1}{D-4} - \ln b + \cdots\right),
\end{equation}
where the dots stand of terms that do not depend on $b$ and terms that vanish when $\epsilon \to 0$. These terms and the divergence do not appear in any physical observable and we shall drop them in the following. In the ultra-relativistic limit (or massless) case, we have
\begin{equation}\label{eik_phase_grav_scattering}
    2\delta(b_\perp) \approx \frac{Gs}{\hbar}\frac{\Gamma\left(\frac{D-4}{2}\right)}{(\pi b^2)^{\frac{D-4}{2}}} = 2\frac{Gs}{\hbar}\ln\frac{b_0}{b},
\end{equation}
where $b_0$ is an infrared regulator.

The eikonal phase $\delta(b_\perp)$ is related to the phase shift of the S-matrix \cite{Amati:1993tb, Kabat:1992tb}, and the latter is unitary, which allows us to make general statements about higher-order corrections to the eikonal phase as follows. The exponential analytic dependence of the amplitude in Eq.~\eqref{amplitude_phase_relation} on $\delta(b_\perp)$ can be obtained from the large angular momentum limit of the partial-wave decomposition of the S-matrix for arbitrary processes \cite{Bellazzini:2022wzv}. This exponential dependence guarantees the unitarization of the S-matrix, and thus, corrections to the Born limit (i.e.~in our case, corrections to Eq.~\eqref{eq:Born-limit} that are higher order in $G s/b^{D-4}$) must also enter through the same exponential structure \cite{DiVecchia:2023frv}.

\begin{figure}[t]
    \centering
    \includegraphics[width=1\linewidth]{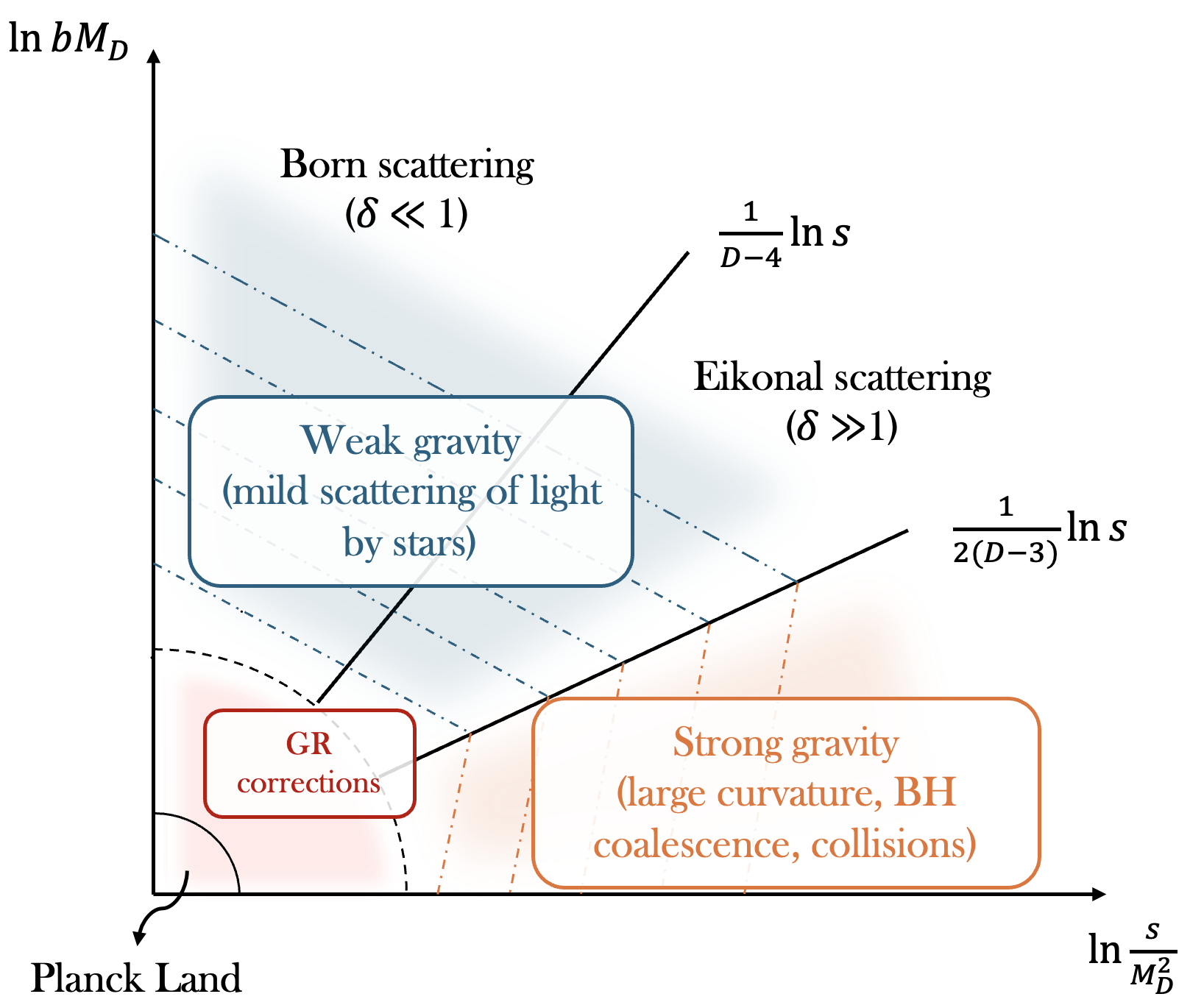}
    \caption{Schematic diagram of the relevant regimes of gravitational scattering at transplanckian energies (not to scale). $M_D$ denotes the Planck energy scale in $D$ dimensions. Both Born and eikonal scattering regions overlap with the weak gravity regime (blue region), and the eikonal case is not valid when gravity is strong (orange region).}
    \label{fig:grav_scat_diagram}
\end{figure}

When one considers the regime where $Gs/b^{D-4} >1$, the eikonal phase $\delta(b_\perp)$ is not simply given by the Born amplitude, but graviton ladder loop diagrams start to dominate \cite{Giddings:2010pp}. In this case, $\delta \gg 1$ and the full integrand in Eq.~\eqref{amplitude_phase_relation} has to be taken into account. The exact result was first obtained in \cite{tHooft:1987vrq}. For our purposes in the next section, it is worth mentioning that for $\delta\gg 1$ the integral in Eq.~\eqref{amplitude_phase_relation} is dominated by a saddle point, and can be evaluated in the stationary-phase approximation. Schematically, the stationary point condition can be used to estimate the total exchange momentum $Q$ as
\begin{equation}
    \frac{\partial }{\partial b}\left(-Q b + 2\delta(b)\right) = 0 \implies Q \sim 2 \frac{\partial \delta}{\partial b}\,,
\end{equation}
so using Eq.~\eqref{eik_phase_grav_scattering}, we find $b\sim [Gs/Q]^{\frac{1}{D-3}}$ or $Q\sim Gs/b^{D-3}$. In the eikonal limit with $s/t\gg1$ or $\sqrt{s}/Q\gg 1$ (since $t \sim Q^2$), we find that $b\gg [G\sqrt{s}]^{\frac{1}{D-3}}$. This last inequality corresponds to impact parameters much larger than the typical Schwarzschild radius of the system, because $G\sqrt{s} \sim r_s$. Thus, ``short-distance'' corrections to the amplitude and phase shift are not significant if $b/r_s \gg 1 $, which overlaps with the weak-gravity regime. Figure~\ref{fig:grav_scat_diagram} shows a diagram of the relevant regimes of gravitational scattering.

In certain regimes, the phase shift $\delta(b_\perp)$ can also be used to compute other semiclassical observables such as the time delay and scattering angle (see \cite{Eisenbud:1948paa,Wigner:1955zz} and references therein). This is accomplished by studying the overlap of wave packets made of superpositions of partial waves \cite{Bellazzini:2022wzv}. In the regime of large $\delta$, the time delay $\tau$ and deflection angle $\theta$ are given by \cite{Bellazzini:2022wzv} (see also \cite{DiVecchia:2023frv})
\begin{equation}\label{time_delay_and_scattering_angle}
    \tau = 2 \hbar \frac{\partial \text{Re}\delta}{\partial \sqrt{s}}, \quad \sin \frac{\theta}{2} = -\frac{\hbar}{2p}\frac{2 \partial\text{Re} \delta}{\partial |b_\perp|},
\end{equation}
where $Q=2p \sin \theta/2$, where recall that $Q$ is the total transferred momentum and $p$ is the magnitude of the momentum of the particles in the center-of-mass frame.

The time-delay formula in Eq.~\eqref{time_delay_and_scattering_angle} can be obtained already for simple central scattering. Consider a wave packet constructed from a superposition of $s$-waves sent to interact with a scattering region of radius $a$. For $r>a$, we have
\begin{equation}
    \psi(r,t) = \frac{1}{r}\int dk\left( A(k) e^{-ik(r+t)} - S(k)A(k)e^{ik(r-t)}\right)dk,
\end{equation}
where $S(k) = e^{i2\delta(k)}$ is the S-matrix. The first term represents the incoming wave packet, that initially (for $t\to -\infty$) is incoming to the scattered region at the speed of light, as evaluated via the stationary phase method
\begin{equation}
    \frac{\partial}{\partial k}\left[k(r_{\rm in}+t)\right] = 0 \implies r_{\rm in}(t) = -t.
\end{equation}
The second term represents the outgoing wave packet, so performing a similar calculation as above, we have (for $t\to \infty$),
 \begin{equation}
\frac{\partial}{\partial k}\left[2 \delta(k) + k(r_{\rm out}-t)\right] = 0 \implies      r_{\rm out}(t) = t - 2\delta'(k).
 \end{equation}
This calculation shows that there is a time delay, exactly proportional to the derivative of the natural logarithm of the S-matrix, relative to the free outgoing propagation, as given in Eq.~\eqref{time_delay_and_scattering_angle}.

\section{Causality constraints from gravitational scattering}\label{sec:causality_constraints}

The approach for computing the time delay $\tau$ by means of scattering amplitudes, which we summarized in the previous section, has been used to place constraints on the coefficients of higher-order operators in extensions of GR. In this paper, we will stress that such an approach has an inherent limitation associated to an expansion in the weak-gravity regime. But before we can do so, we must first review how causality constraints are obtained in this scattering amplitudes approach, which we do in this section, following mostly~\cite{Camanho:2014apa}. 

For the gravitational scattering discussed in the previous section, we can show that the resulting time delay is exactly the Shapiro time delay of GR. From Eqs.~\eqref{eik_phase_grav_scattering} and \eqref{time_delay_and_scattering_angle}, one easily finds that $\theta = G \sqrt{s}/b^{D-3}$, and the eikonal regime implies a small-angle scattering. Using the time-delay formula, we then find
\begin{equation}
    \tau = 4 G \sqrt{s} \ln \frac{b_0}{b},
\end{equation}
which coincides with the leading-order contribution (in a weak-field expansion) to the Shapiro time delay computed by a null geodesic traveling past a Schwarzschild black hole of mass $M$, if we set $2\sqrt{s} \to M$ (see discussion in Sec. \ref{subsec:time_delay_GR}). In the ultra-relativistic limit, we can also find the time delay from the geometry by means of shockwave metrics \cite{tHooft:1987vrq,Dray:1984ha,Aichelburg:1970dh}, and the resulting time delay matches the one computed from the phase shift. 

The seminal paper \cite{Camanho:2014apa} started the recent usage of time delays computed from amplitudes in modified gravity theories to constrain the coefficient of higher-order operators in the action. The idea is that extensions to GR will introduce new vertices into the S-matrix calculation. These new vertices can lead to beyond-GR corrections to the time delay that can be negative and are proportional to the coupling constants. If the coupling constants are chosen to be large enough, these corrections can dominate over the GR contribution to the time delay, and thus, give rise to overall negative time delays (i.e.~time advances). Certainly, this ought to not be allowed on physical grounds, which then allows one to place constraints on modified gravity from purely theoretical causality arguments. In recent years, an explosion of work has been carried out in this direction\footnote{Some of them focus on the subplanckian regime and hence do not assume the eikonal approximation, while others include some non-perturbative GR effects (see e.g. \cite{Cremonini:2024lxn, Caron-Huot:2022ugt}).}~\cite{Camanho:2016opx,Hinterbichler:2017qyt,Bellazzini:2015cra,deRham:2019ctd,Tokuda:2020mlf,AccettulliHuber:2020oou,Chen:2021bvg,Caron-Huot:2021rmr,Caron-Huot:2022ugt,Hong:2023zgm,Aoki:2023khq,Caron-Huot:2024tsk,deRham:2022hpx,Serra:2022pzl,Cremonini:2023epg,Kehagias:2024yyp,Cassem:2024djm,Xu:2024iao,Cremonini:2024lxn}. 

Several generic conditions exist for the validity of the resulting time-delay calculation and the applicability of causality constraints. Three such conditions are that the time delay must be resolvable, smaller than the uncertainty in the time localization of the wave packets used to derive it, and within the applicability range of the effective field theory description \cite{Goon:2016une,deRham:2020zyh}. We here stress that these constraints are also valid only in a regime of weak gravity, which is necessary to define graviton states and its perturbation theory in quantum field theory. In fact, in regions of strong gravity, the self-interacting theory for gravitons obtained from a covariant gravitational theory is strongly coupled. This fact has been discussed pointwisely in the literature\footnote{See \cite{Cremonini:2024lxn} wherein strong gravity effects on time delays in the near horizon regime of a Reissner-Nordstr\"om black hole were discussed.} \cite{Camanho:2014apa, deRham:2020zyh}, but is typically overlooked. Failure to account for strong gravity effects has recently led to incorrect conclusions, as we shall explain in Sec.~\ref{sec:validity_causality}. But before doing so, let us explore how strong-field effects may change our quantum field theory calculations. 

The very definition of a graviton assumes a separation of scales in which the wavelength of the metric perturbations must be smaller than the characteristic curvature radius of the background. In the strong gravity regime, such curvature radius is small and operators that would be irrelevant in flat spacetime become relevant in the infrared. As an illustrative example, consider the metric in Einstein's general relativity plus a matter action,
\begin{equation}
    S = \frac{1}{2\kappa^2} \int d^4 x \sqrt{-g} R(g) + S_{\rm m}.
\end{equation}
The matter action supports a non-trivial background metric $\Bar{g}_{\mu\nu}$. Now consider the traverse-traceless (TT) tensor perturbations of this metric such that $g_{\mu\nu} = \Bar{g}_{\mu\nu} + h_{\mu\nu}$ where $\overline{\nabla}^\mu h_{\mu\nu} = 0 = \Bar{g}^{\mu\nu} h_{\mu\nu}$. The action for the metric perturbation is, up to quadratic order in $h_{\mu\nu}$, 
\begin{equation}
    S_{\rm TT}^{(2)} = \frac{1}{4\kappa^2}\int d^4 x \sqrt{-\Bar{g}}\left[-\frac{1}{2}h^{\alpha\beta}\left(\overline{\nabla}^2 h_{\alpha\beta} + 2 \overline{R}_{\mu\beta\alpha\nu}h^{\mu\nu}\right)\right].
\end{equation}
The leading-order cubic interactions can be obtained by replacing $\Bar{g}_{\mu\nu} \to \Bar{g}_{\mu\nu} + h_{\mu\nu}$ in the Riemann tensor $\Bar{R}_{\mu\beta\alpha\nu}$ of the second term of the above equation, and re-expanding in the metric perturbation to obtain (see e.g. the discussion in \cite{Fanizza:2021ngq})
\begin{align}\label{cubic_vertices}
    S_{\rm TT}^{(3)} &= \frac{1}{4\kappa^2}\left(-\frac{1}{2}\right)\int d^4 x\sqrt{-\Bar{g}}\left(\overline{R}_{\mu\lambda\alpha\nu}h^\lambda_{\;\;\beta} - \overline{R}_{\beta\lambda \alpha\nu}h^{\lambda}_{\;\;\mu} + \right. \nonumber\\
    & + \left. \overline{\nabla}_\alpha \overline{\nabla}_\beta h_{\nu\mu} - \overline{\nabla}_\nu \overline{\nabla}_\beta h_{\alpha\mu} + \overline{\nabla}_\nu \overline{\nabla}_\mu h_{\alpha\beta}\right)h^{\mu\nu}h^{\alpha\beta}.
\end{align}
We see that there are not only derivative terms but also cubic terms proportional to the background curvature tensor in the third-order perturbative expansion of the Einstein-Hilbert action. In regions where the curvature is small, we can safely neglect the terms proportional to the background curvature tensor. Similarly, since $h_{\mu\nu}$ varies significantly more than $\Bar{g}$ over distances smaller than the radius of curvature, we can set $\overline{\nabla}h \to \partial h$ in the weak-gravity region.  For instance, in the Schwarzschild spacetime, we have the Kretschmann scalar
\begin{equation}
   K=  \overline{R}_{\mu\nu\rho\sigma}
    \overline{R}^{\mu\nu\rho\sigma} = \frac{48 G^2 M^2}{r^6},
\end{equation}
and the curvature radius is of ${\cal{O}}(1/\sqrt{K}) \sim {\cal{O}}(r^3/r_s^2)$, where recall that $r_s$ is the Schwarzschild radius. For astrophysical black holes, at distances $r\gg r_s$, the curvature radius is huge and the cubic operators in Eq.~\eqref{cubic_vertices} can be neglected. In this region, the time delay and scattering angle can be computed from geodesics of the linearized Schwarzschild metric and the results match the perturbative amplitude-based calculation that is built from flat spacetime principles, as previously mentioned.

When gravity is strong, however, the above approximations fail and the perturbative theory for $h_{\mu\nu}$ breaks down. In the example above, in regions close to the black hole radius $r_s$, the linearization approximation breaks and the full geometry has to be taken into account. In this region, the time delay effect is very strong, and the scattering angle is large. In the amplitude picture, the eikonal approximation is no longer valid. This begs the following questions: what corrections should be taken into account in order to reproduce the geometric results? What are the corresponding diagrams? 

\begin{figure}
    \centering
    \includegraphics[width=0.5\linewidth]{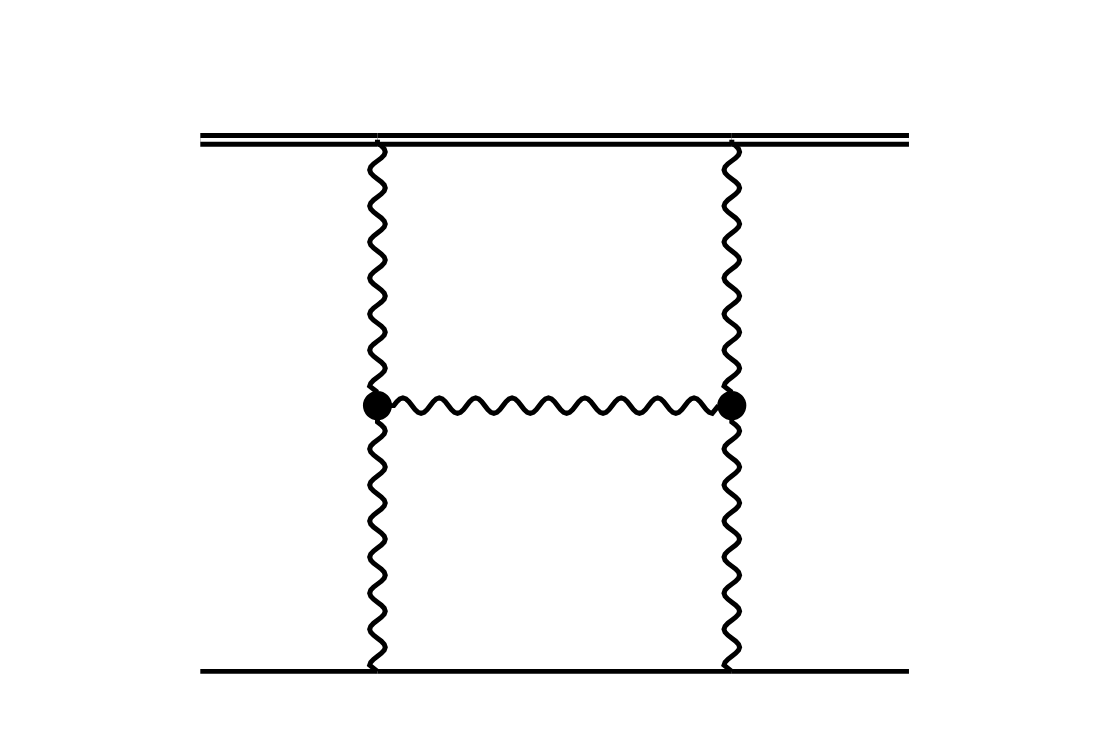}
    \caption{The leading diagram containing graviton self-interacting cubic vertices. The eikonal amplitude gets corrected by diagrams containing such vertices when $b\gtrsim r_s$.}
    \label{fig:gravitational_tree}
\end{figure}

To go beyond linearized gravity, we need to start considering graviton three- and four-vertices. Indeed, for the Schwarzschild case, the nonlinear terms can be resummed to get the full solution from amplitudes \cite{Duff:1973zz,Mougiakakos:2024nku}. For the scattering of two massive bodies, the leading-loop corrections as $b \to G\sqrt{s}$ come from graviton trees between the scalar external lines \cite{Amati:1993tb}. As an example, the order of magnitude of the contribution of the diagram in Fig.~\ref{fig:gravitational_tree} to $\delta(b_\perp)$ can be estimated as follows: each graviton vertex attached to an external lines contributes a factor $\kappa \sqrt{s}$ coming from the vertex ($\kappa s$) times the internal propagator ($1/\sqrt{s}$) contributions (see e.g. the estimates in \cite{Giddings:2009gj, DiVecchia:2023frv}). The vertices connecting internal lines do not have the extra $\sqrt{s}$ factor, but come with the loop and transferred momentum factors. These will give rise to the $b_\perp$-dependent factors in $\delta(b_\perp)$ that can be estimated by dimensional analysis. For the diagram in Fig.~\ref{fig:gravitational_tree}, we have a contribution of ${\cal{O}}[\kappa^6 (\sqrt{s})^4 b^{-2(D-3)}] = {\cal{O}}[Gs(r^2_s/b^2)^{D-3}]$, which differs from the order of the tree-level diagram in the first line of Fig.~\ref{fig:eikonal_diagrams} by a factor of ${\cal{O}}[(r^2_s/b^2)^{D-3}]$.

Summarizing, the eikonal approximation breaks down at $b\sim r_s$, and this can be seen from different approaches: (i) by computing the total transferred momentum from the saddle-point evaluation of the eikonal phase (as discussed in Sec.~\ref{sec:intro}); (ii) by the presence of extra loop diagrams with graviton trees formed from graviton three-vertices; and (iii) from a geometrical perspective, where higher-order curvature corrections cannot be neglected and the scattering angle and time delay increase as $b \to r_s$. When $b \to r_s$, we enter the strong gravity regime, and the typical causality constraints from flat spacetime amplitudes must be reassessed.

The limitation described in the previous paragraph is also discussed in \cite{Camanho:2014apa}, although in different words. In that work, the authors considered a $D>4$ modification to GR by adding the Gauss-Bonnet term to the Einstein-Hilbert Lagrangian. The new terms have a coefficient $\lambda$ with dimension of length squared. Then, the leading-order correction to the GR prediction for $\delta(b_\perp)$ for the scattering of gravitons\footnote{We can apply the discussion of the diagrams of the $2\to 2$ scattering of the scalars to the scattering of gravitons after taking the scalar masses to zero and focusing on individual graviton helicities. This is the case for any theory in which the kinetic terms for these helicities are diagonal and canonically normalized, such as GR. In dCS gravity, we need to work with circular polarizations, $h_+\pm ih_\times$, for that to be the case.} is suppressed by a factor that scales as $\lambda/b^2$ relative to the GR contribution, which is $\mathcal{O}\left({Gs}/{b^{D-4}}\right)$ \cite{Camanho:2009vw,Buchel:2009sk}. Moreover, this correction can overwhelm the GR prediction for the time delay, leading to the possibility of a time advance, when $b^2 \lesssim \lambda$. We then see that the exact constraint on $\lambda$ always depends on how small a $b$ we can trust in our calculations. Although similar arguments have been discussed in \cite{deRham:2019ctd}, we are unaware of estimates on how small $b$ can be before causality arguments cease to be self-consistent. More importantly, many subsequent studies do not even mention this intrinsic limitation. In the following section, we will find a bound on the coefficients of high-order operators for the eikonal approximation to be self-consistent in the weak-gravity limit. 

\section{Validity of causality constraints for beyond Einstein theories}\label{sec:validity_causality}

We have seen in previous sections that causality constraints rely on the accuracy of the approximations used to compute the scattering amplitudes. In this section, we explore more precisely when such constraints are valid and how they apply to dCS gravity. 

\subsection{Regime of validity of weak-field analysis of modified gravity}\label{sec:eikonal_scope_modified_gr}

We start by estimating how small the impact parameter can be before the eikonal approximation breaks down in a generic beyond-Einstein effective field theory. Consider adding a higher-order curvature operator to the Einstein-Hilbert Lagrangian, such that a new cubic graviton vertex is introduced. Suppose that it gives rise to a correction to the eikonal phase of the form
\begin{equation}\label{generically_corrected_phase}
    2\delta(b_\perp) \sim \frac{Gs}{\hbar}\frac{\Gamma\left(\frac{D-4}{2}\right)}{(\pi b^2)^{\frac{D-4}{2}}}\left(1- \frac{\lambda_n}{b^n}\right),
\end{equation}
where $\lambda_n$ is a coefficient with length dimension $n$. If the correction to the GR result within the bracket dominates, then there may be a time advance, which occurs when $\lambda_n > b^n$. 

Can this inequality ever hold? The answer to this question depends strongly on how the non-GR corrections to the gravitational action are treated. When one considers effective field theory modifications to GR, one writes all possible operators that are not forbidden by symmetries, organized through some kind of expansion, such as a derivative expansion or a curvature expansion. One then truncates this series to a given order, which then establishes a cut-off for the effective theory, outside which the latter is not an accurate representation of Nature. For this reason, all non-GR modifications introduced by effective field theories must be small deformations of the GR prediction, which in our language translates to the condition $\lambda_n \ll b^n$ for a scattering experiment. Therefore, it follows that for effective field theory modifications to GR, such as dCS gravity~\cite{Bellucci:1988ff, Bellucci:1990fa, Gates:1991qn, Lue:1998mq, Choi:1999zy, Jackiw:2003pm,Alexander:2009tp}, a time advance can only occur within the regime of validity of the effective field theory deep in the eikonal regime.

When one considers modified gravity theories that are \textit{not} to be understood as effective field theories, i.e.~one imagines the theories as ``exact,'' then there are no restrictions on the size of the non-GR modifications. This, in turn, implies that there are no restrictions\footnote{In this approach the ``exact'' theory still has a distance cutoff, although not set by $\lambda_n^{1/n}$. We comment more about this point in Sec. \ref{sec:conclusion}.}, in principle, on the size of the coupling constant of the theory, which allows the possibility that $\lambda_n > b^n$. As we will show below, however, the approximations used in the scattering amplitudes program do require certain conditions on the coupling to hold. For the eikonal limit to hold, and for us to get the correct time advance from the amplitude, we need $s/t \gg 1$, which imposes a condition on the total transferred momentum $Q$. The latter is obtained from the stationary phase condition
\begin{equation}
    \frac{\partial}{\partial b}\left(-Qb - \beta \frac{Gs}{b^{D-4} \frac{\lambda_n}{b^n}}\right) = 0 \implies b^{D-3+n} \sim \frac{\sqrt{s}}{Q} G\sqrt{s} \lambda_n.
\end{equation}
The eikonal limit\footnote{We remind the reader that in Sec.~\ref{sec:eikonal} we defined the eikonal limit through $s/t \gg 1$, which is equivalent to $\sqrt{s}/Q \gg 1$ because the transferred momentum $Q = \sqrt{t}$, in terms of the Mandelstam variables.} $\sqrt{s}/Q \gg 1$ 
then requires that 
\begin{equation}
    b\gg (\lambda_n G\sqrt{s})^{\frac{1}{D-3+n}}\sim (\lambda_n r_s^{D-3})^{\frac{1}{D-3+n}}.
\end{equation}
Self-consistency with the time-advance assumption yields
\begin{equation}\label{consistent_bound}
    \lambda_n > b^n \gg (\lambda_n r_s^{D-3})^{\frac{n}{D-3+n}} \implies \lambda_n \gg r_s^n,
\end{equation}
or $\lambda \gg (G\sqrt{s})^{\frac{n}{D-3}}$. Since we can consider processes $r_s$ larger than the Planck scale, Eq.~\eqref{consistent_bound} is a more stringent bound than $\lambda_n> 1/M_{\rm Pl}^n$, which is a straightforward condition for the use of perturbative GR as a quantum, but non-renormalizable, field theory. 

\begin{figure}
    \centering
    \includegraphics[width=1\linewidth]{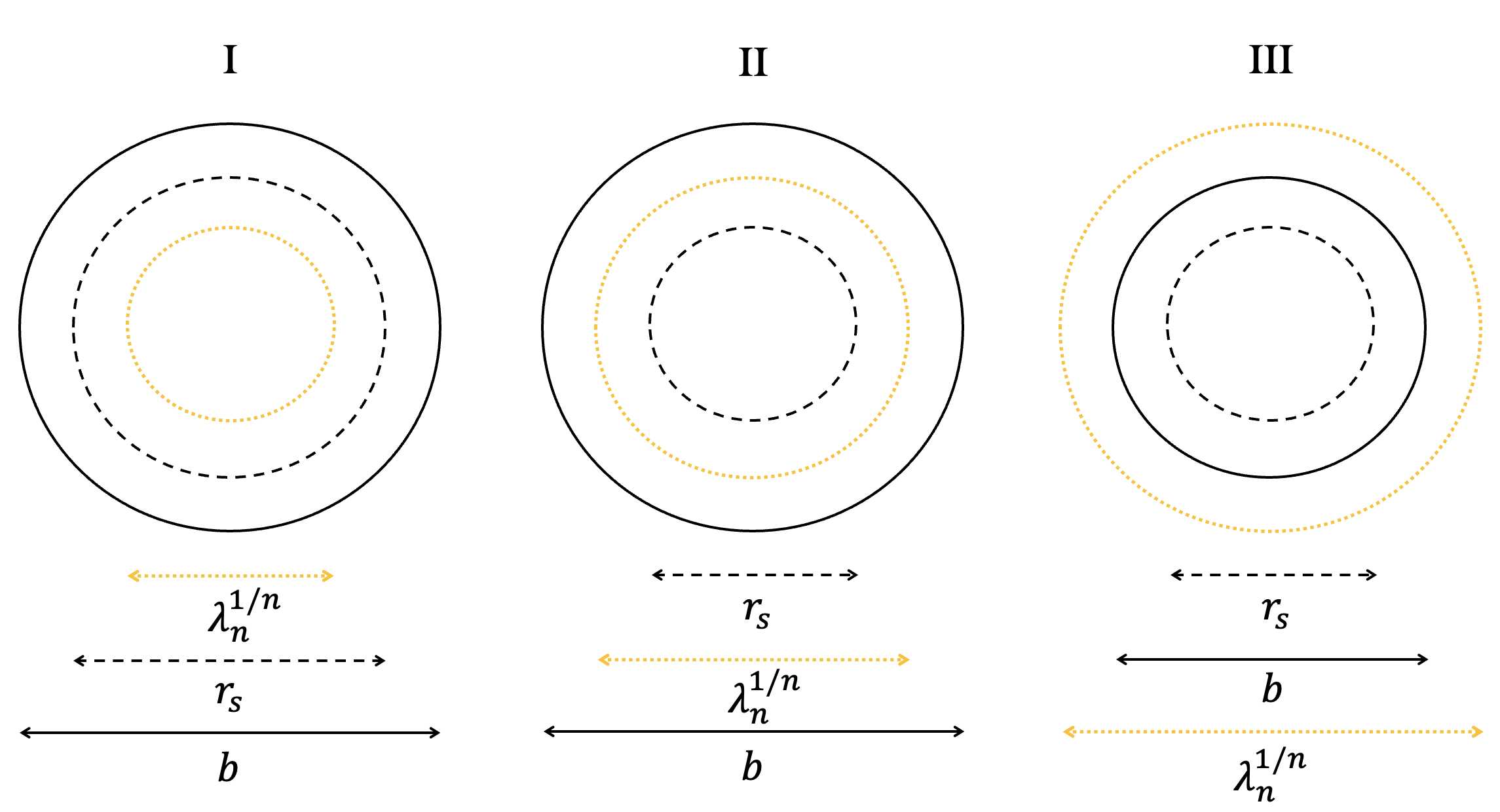}
    \caption{Possible hierarchy of scales. In cases I and II, the correction to the phase shift does not introduce causality issues, since $b>\lambda_n^{1/n}$. For case III, we can have a time advance, but since $b\gg r_s$ for consistency, we should also have $\lambda_n^{1/n}\gg r_s$. This is \textit{only} possible when the corrections to the Einstein-Hilbert action are understood as defining an exact theory. If the corrected action is to be understood as an effective field theory, then $\lambda_n^{1/n} \ll r_s$ for the Einstein-Hilbert action to dominate over the higher-derivative operators, which then disallows case III completely. This figure is not to scale.}
    \label{fig:hierarchy_of_scales}
\end{figure}

In other words, the length scale $(\lambda_n)^{1/n}$ associated with the GR modification has to be much larger than the characteristic Schwarzschild radius of the process considered in order for the time advance be consistently given by the $\tau$ definition in Eq.~\eqref{time_delay_and_scattering_angle}. This condition relates to the assumption of weak gravity that the scattering amplitude calculation is built upon: we can use scattering amplitudes in quantum field theory to calculate modifications to the time delay when the length scale associated with the GR modification is probed in the weak-gravity regime. If the time delay turns out to be negative in a certain case (given a choice of process and external states, for instance), then $\lambda_n$ has to be small, closer to the strong gravity regime, to avoid the causality problem (see Fig.~\ref{fig:hierarchy_of_scales})

Suppose that we can always find a Gedanken process for which a time advance can be consistently computed for arbitrarily values of $r_s$. If $(G\sqrt{s})^{\frac{1}{D-3}}$ is of the order of the Planck scale, and we still have a time advance, then what the causality constraint is telling us is that $(\lambda_n)^{1/n}$ has to be smaller than the Planck length. A particular case of this point was already discussed in \cite{Camanho:2014apa}, in the context of $D>4$ Gauss-Bonnet gravity. In this case, the gravity modification is not useful at all, since we do not know how to handle quantum gravity at Planck scales anyhow. The only way to have such a small value for $\lambda_n$ and be consistent with causality is then to have extra interactions present below some scale that is larger than the Planck length. This is precisely what happens in effective theories descending from string compactifications, where an infinite tower of massive, higher-spin states ``Regge-izes'' the amplitude (see \cite{DiVecchia:2023frv} and references therein).  

\subsection{Causality constraints in dynamical Chern-Simons gravity}\label{sec:causality_const_dCS}

In the rest of this section, we focus on dCS gravity and argue that recent causality constraints on the dCS coefficient are not as stringent as astrophysical constraints, precisely because of the weak-gravity assumption that causality constraints rely on. Let us then consider the  dCS gravity action \cite{Bellucci:1988ff, Bellucci:1990fa, Gates:1991qn, Lue:1998mq, Choi:1999zy, Jackiw:2003pm,Alexander:2009tp},
\begin{align}\label{dCS_action}
    S &= \frac{1}{2\kappa^2}\int d^4x \sqrt{-g}R +\frac{\alpha}{4\kappa} \int d^4 x \sqrt{-g}~ \phi \Tilde{R}R \nonumber\\
    &+\int d^4 x \sqrt{-g} \left[-\frac{1}{2}\left(\nabla^\mu \phi\right)\left(\nabla_\mu \phi\right)-V(\varphi)\right],
\end{align}
where recall that $\kappa^2 = 8\pi G = M^{-2}_{\rm Pl}$, $\alpha = \ell_{\rm CS}^2$ has mass dimension minus two (or length dimension two, in geometric units), and 
\begin{equation}
    \Tilde{R}R = \frac{1}{2} \epsilon^{\mu\nu\alpha\beta}R_{\mu\nu}^{\;\;\;\;\rho\sigma} R_{\rho\sigma \alpha\beta}
\end{equation}
is the Pontryagin invariant, where $\epsilon^{\mu \nu \alpha \beta}$ is the Levi-Civita tensor. In this theory, a scalar field $\phi$ with potential $V(\phi)$ couples non-minimally to curvature through the Pontryagin density. In most works, including this paper, we assume that the scalar field $\phi$ is massless but dynamical. 

\begin{figure}
    \centering
    \includegraphics[width=0.35\linewidth]{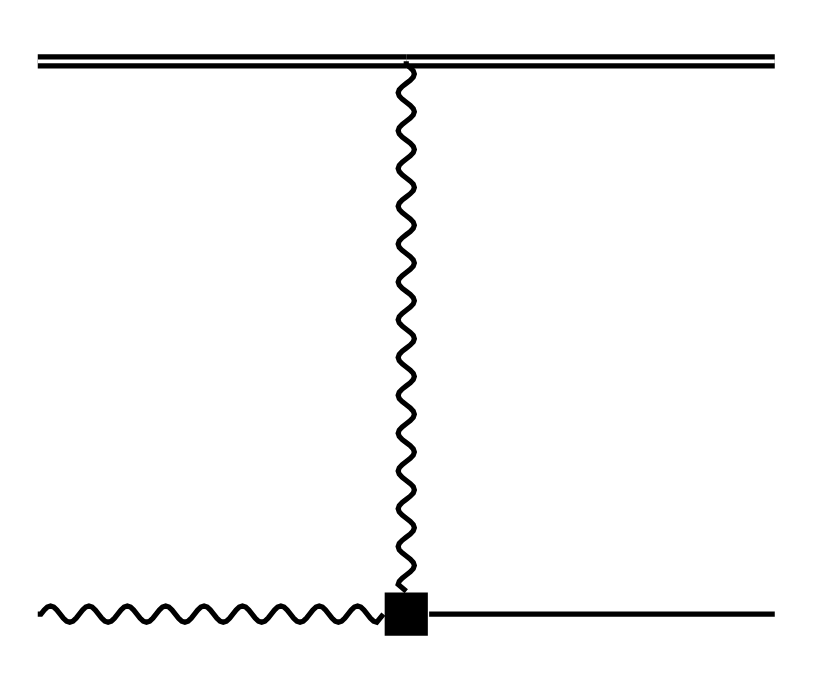}
    \caption{The Born limit of the $\sigma + h \to \sigma +\phi$ process in perturbative dCS gravity is given by this diagram, showing the scattering of a graviton $h$ and a massive scalar field (that could for instance represent a star), and then exchanging a dCS scalar. After canonically normalizing the metric perturbation, the dCS vertex is proportional to $\alpha\kappa$. In the eikonal limit, the full amplitude in the impact parameter space is obtained after exponentiating the diagram.}
    \label{fig:dCS_process}
\end{figure}

Let us now consider scatterings in this effective field theory. To make contact with the previous sections, this effective theory corrects the Einstein-Hilbert action through quadratic curvature terms that are controlled by the length scale $\alpha$, which plays the role of $\lambda_n$, with $n=2$ in this case.
Compared to GR, the dCS coupling introduces a new vertex involving the dCS scalar $\phi$ and two gravitons. This new vertex allows for the process $\sigma + h \to \sigma +\phi$ via graviton exchange, where $\sigma$ is a heavy scalar field that represents the matter sector (e.g.~$\sigma$ could represent a star), see Fig.~\ref{fig:dCS_process}. This process was studied in \cite{Serra:2022pzl} in the eikonal limit and the authors aimed to find causality constraints on the dCS cutoff scale.

With gravitons in the external lines, we need to specify their helicity to fix the amplitude. For $\sigma + h \to \sigma +\phi$, in the eikonal limit, this amplitude is \cite{Serra:2022pzl}
\begin{equation}
    i\mathcal{M} \simeq -\frac{2i\alpha}{M^{2}_{\rm Pl}}\frac{(q_1+iq_2)^2}{q^2}(2m_\sigma\omega)^2,
\end{equation}
for the $++$ polarization, corresponding to $+2$ graviton helicity, where $q_i$ are the components of the transverse exchange momentum. Here $m_\sigma$ is the mass of the heavy field and $\omega$ is the graviton energy in the center-of-mass frame. With a $--$ polarization in the external state we would get $\mathcal{M}^*$ for the scattering amplitude. This result implies that, if we send a graviton to probe the heavy field $\sigma$, there is a non-vanishing probability of this graviton turning into a dCS scalar quanta by exchanging a graviton. 

To check for time advances after the scattering, \cite{Serra:2022pzl} considered all the polarizations of the quanta, $(h^{++}, h^{--}, \phi)$. This resulted in the eikonal phase matrix
\begin{equation}
    2\delta_{\rm GR + dCS} \simeq 4 Gm_\sigma  \omega \begin{pmatrix}
        B & 0 & A\\
        0 & B & A^* \\
        A^* & A & B\\
    \end{pmatrix},
\end{equation}
where
\begin{equation}
    B = -\ln b - \frac{1}{2\epsilon} - \frac{\gamma_E}{2}, \quad A = -\frac{i\alpha}{b^2},
\end{equation}
$\gamma_E$ is the Euler constant, $b$ is the impact parameter, and  dimensional regularization was used with $\epsilon = 2-D/2$. Diagonalizing $\delta_{\rm GR+ dCS}$, we find the eigenvalues
\begin{align}
    \delta_0 &= 4G m_\sigma \omega \left(-\ln b - \frac{1}{2\epsilon} - \frac{\gamma_E}{2}\right),\\ \label{phase_shift_pm}
    \delta_{\pm} &= 4G m_\sigma \omega \left(-\ln b - \frac{1}{2\epsilon} - \frac{\gamma_E}{2}\pm \sqrt{2} \frac{\alpha}{b^2}\right).
\end{align}

The authors in \cite{Serra:2022pzl} then proceeded to find time delays from these expressions. Using Eq.~\eqref{time_delay_and_scattering_angle}, they found, for the $\delta_{\pm}$ modes,
\begin{equation}\label{time_delay_in_dCS_gravity}
    \Delta t_{\pm} = 2 G m_\sigma \left(\ln\frac{b_0}{b} \pm \sqrt{2}\frac{\alpha}{b^2}\right), 
\end{equation}
where $b_0$ is a large reference impact parameter and we neglected $\mathcal{O}(\alpha/b_0^2)$ terms. Reference~\cite{Serra:2022pzl} then claimed that, for $b^2\lesssim \alpha$, one superposition of states would get a time advance when scattered by the massive heavy field $\sigma$ (i.e.~$\Delta t_- < 0$), while the other would receive a time delay (i.e.~$\Delta t_+ > 0$). The question then is the following: can the impact parameter be smaller than the dCS scale and still be in the regime of validity of an effective field theory?

Three length scales are present in this problem: $r_s, \, l_{\rm Pl}$, and $\ell_{\rm CS}$. For $l_{\rm Pl}< r_s$ (which is expected to always be true for real, astrophysical black holes), there are three relevant hierarchies of scales we can think of: 
\begin{itemize}
    \item[(i)] $b>r_s>\ell_{\rm CS}$. The impact parameter is larger than the size of the object, which in turn is larger than the dCS coupling length.
    \item[(ii)] $b>l_{CS}>r_s$. The impact parameter is still larger than the size of the object, but the dCS coupling length is this time also larger than the object.
    \item[(iii)] $\ell_{\rm CS}>b> r_s$. The dCS coupling length is the largest scale in the problem. 
\end{itemize}
Clearly, in cases (i) and (ii), there will never be any causality problems, because $\alpha/b^2$ (or equivalently $\ell_{\rm CS}/b$) never dominates in the expression for the time delay. Causality issues may only appear when the impact parameter is smaller than the dCS length scale. However, for the calculations above to be valid, we should have $b\gg r_s$ (i.e.~the eikonal condition), such that we are always far from the strong gravity regime. Moreover, below Eq.~\eqref{consistent_bound} we have explained that, for the eikonal approximation to work, we must also have that $\lambda_n \gg r_s^n$, which for the dCS case reduces $\ell_{\rm CS}\gg r_s$. Therefore, the calculation in \cite{Serra:2022pzl} is valid and yields a time advance \textit{only} when $r_s\ll b< \ell_{\rm CS}$. The question then becomes how small can $\ell_{\rm CS}$ be (while remaining still much later than $r_s$) such that the time advance is consistent with weak gravity. 

Let us reason by contradiction. Imagine for a second that $\ell_{\rm CS}$ is close to (but larger than) $r_s$, and then ask what this implies for our calculations. If we assume $\ell_{\rm CS} \gtrsim r_s$, then it follows from $\ell_{\rm CS}>b$ and $b > r_s$ that $b \sim r_s$, which, in turn, implies we are in the strong gravity regime, and the time advance cannot be trusted. This is because, as $b$ gets close to $r_s$, loop diagrams from the graviton self-interactions start to contribute to the full amplitude, similar to the pure gravity case discussed in Sec.~\ref{sec:causality_constraints}. 

\begin{figure}
    \centering
    \includegraphics[width=0.75\linewidth]{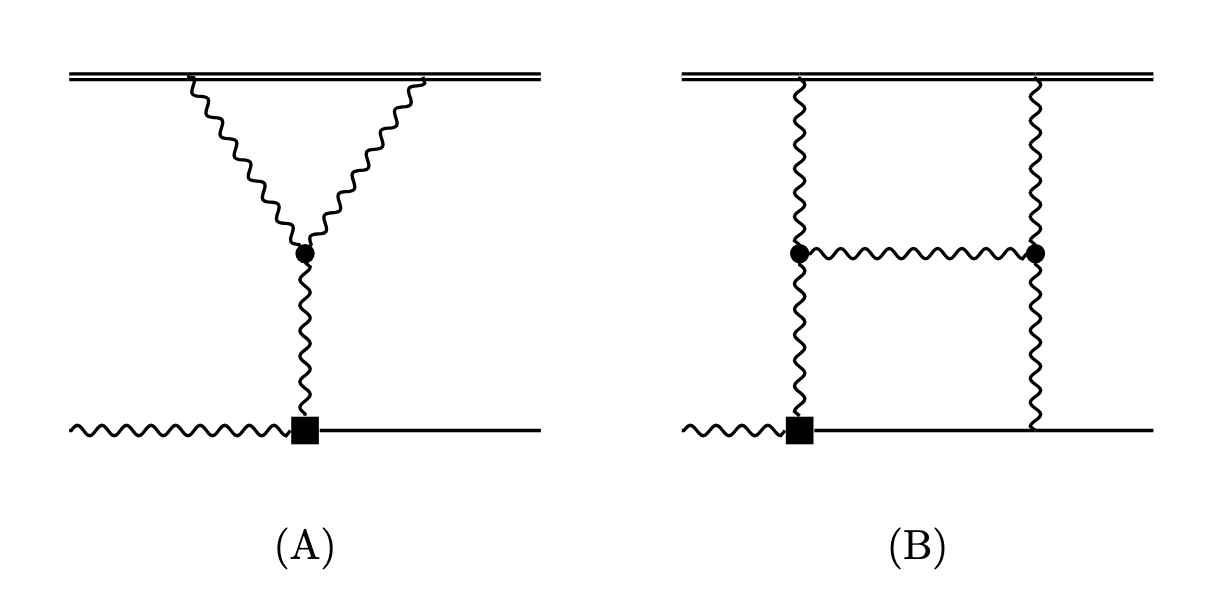}
    \caption{Loop corrections due to graviton nonlinearities. Diagrams like these can be neglected only in the weak-gravity regime.}
    \label{fig:loop_contributions_dCS_process}
\end{figure}

Let us then investigate how large these loop corrections to the eikonal scattering amplitudes can be. 
In the GR case, loop corrections to the eikonal amplitude of graviton-mediated scalar scattering have been studied in \cite{Akhoury:2013yua}. For dCS gravity, the leading diagrams  are shown in Fig.~\ref{fig:loop_contributions_dCS_process}. The diagram (A) in Fig.~\ref{fig:loop_contributions_dCS_process} has a dCS vertex, proportional to $\kappa \alpha$ and, by a similar power-counting argument as in the GR case, this diagram gives a contribution of 
\begin{equation}  
\mathcal{O}\left(\kappa^3(\sqrt{s})^2(\kappa \alpha \sqrt{s}) b^{-3}\right) \sim \mathcal{O}\left(Gs \frac{G\sqrt
s}{b}\frac{\alpha}{b^2}\right)
\end{equation}
to the eikonal phase. We see that these contributions are suppressed by $r_s\alpha/b^3 = r_s \ell_{\rm{CS}}^2/b^3$ relative to the leading-order GR contribution. Meanwhile, the diagram (B) in Fig.~\ref{fig:loop_contributions_dCS_process} yields contributions of
\begin{equation}  
\mathcal{O}\left(\kappa^5(\sqrt{s})^3(\kappa \alpha \sqrt{s}) b^{-4}\right) \sim \mathcal{O}\left(Gs \left(\frac{G\sqrt
s}{b}\right)^2\frac{\alpha}{b^2}\right),
\end{equation}
and these are suppressed by $r_s^2 \alpha/b^4= r_s^2 \ell_{\rm{CS}}^2/b^4$ relative to the GR contribution.

Therefore, to neglect loop corrections from graviton self-interactions, such as the ones in Fig.~\ref{fig:loop_contributions_dCS_process}, we need to take $b\gg (r_s \alpha)^{1/3}=(r_s \ell_{\rm{CS}}^2)^{1/3}$. In the geometric picture, this is precisely the regime in which we can neglect vertices supported by curvature, i.e.~the weak-gravity regime, since
\begin{equation}
    \frac{\alpha \phi}{2}R\Tilde{R} \to \alpha \phi h \partial^2 h \bar{R} \sim \frac{\alpha}{r^2}\left(\frac{G M}{r}\right)\frac{p^2}{M_{\rm Pl}}\phi h_c h_c
\end{equation}
where we expanded one of the background Riemann tensors $\bar{R}_{\mu\nu\rho\sigma}$ in the $R\Tilde{R}$ factor of the dCS coupling to second-order in $h_{\mu\nu},$ and $h_c$ denotes (schematically) the canonically normalized graviton field. Another $\mathcal{O}(h^2)$-term comes from expanding both curvature tensors in the dCS coupling to linear order in $h$: 
\begin{align}
    \frac{\alpha \phi}{2}R\Tilde{R} &\to 2\alpha \phi \epsilon_{\mu\nu\rho\sigma}\bar{R}^{\mu\lambda \alpha\beta}\bar{R}_{\alpha\gamma}^{\;\;\;\;\rho\sigma} h_{\lambda}^{\;\;\nu}h^{\gamma}_{\;\;\beta}\nonumber\\
    & \sim \frac{\alpha}{r^2}\left(\frac{G^2 M^2}{r^2}\right)\frac{l_{\rm Pl}}{r^2} \phi h_c h_c,
\end{align}
The condition to neglect such vertices is $r^3\gg \ell_{\rm CS}^2 r_s$ which is precisely the same condition we found above so that loop corrections could be neglected. 

In fact, the condition for loop corrections to be negligible and for vertices supported by curvature to be negligible is exactly the same as the regime in which the eikonal approximation is self-consistent. Equation \eqref{phase_shift_pm} has the same form as Eq.~\eqref{generically_corrected_phase} with $n=2$ and $\lambda_2= \alpha$. Thus, we must have
\begin{equation}\label{eik_validity_dCS}
    b \gg (\alpha r_s)^{1/3},
\end{equation}
so that the eikonal approximation can be trusted in the weak-gravity regime. Moreover, when the term proportional to $\alpha/b^{2}$ dominates Eq.~\eqref{time_delay_in_dCS_gravity}, we should have $\sqrt{\alpha} \gg r_s$, or $\ell_{\rm CS} \gg r_s$, for the calculated time advance to be compatible with the eikonal limit. 

Now, what does all of this mean and imply with respect to imposing causality constraints on dCS gravity, as in \cite{Serra:2022pzl}? In that work, the scalar $\sigma$ plays the role of a black hole, and the authors combine the time advance together with astrophysical considerations to find a bound on $\ell_{\rm CS}$, which is reportedly more stringent than the current astrophysical bound \cite{Silva:2020acr}. The condition in Eq.~\eqref{consistent_bound}, however, implies that the causality argument cannot be used to ``rule out'' dCS gravity. Suppose $\ell_{\rm CS} \sim 10^{N} r_s$, where $N>1$ is the minimal possible value for which we can neglect the curvature effects, while still having $b<\ell_{\rm CS}$ within the eikonal regime. If $N$ is too close to unity, then $\ell_{\rm CS}$ is close to $r_s$ and $r_s\ll b<\ell_{\rm CS}$ cannot be satisfied. In particular, with this parameterization of $\ell_{\rm CS}$, we have
\begin{equation}
    b^3 \gg r_s l^{2}_{CS} = 10^{-N} \ell^3_{\rm CS} \implies \frac{b}{\ell_{\rm CS}} \gg 10^{-N/3}.
\end{equation}
However, for the dCS-coupling contribution to dominate the time delay, we need $\ell_{\rm CS}>b$. Notice that this violently violates the EFT condition that the dCS effect is a perturbation correction to the GR time delay. Therefore, if we enforce the EFT condition, then no constraint is possible.  

Let us now, for a second, ignore the EFT condition, and imagine that dCS gravity is an exact theory of gravity. Could we then use the causality arguments to place a constraint? To have a time advance and still be within the regime of validity of the eikonal approximation, we must satisfy
\begin{equation}\label{inequality_N}
    1 \gg \left(\frac{b}{\ell_{\rm CS}}\right)^3 \gg 10^{-N}\,,
\end{equation}
which is very difficult to satisfy for small values of $N$. For instance, if $N=3$ (that is, a dCS length scale a thousand times bigger than the Schwarzschild radius), we must have $1> b/\ell_{\rm CS}\gg 0.1$, which is only marginally correct. In fact, if we want the correction factor $r_s \ell_{\rm {CS}}^2/b^3$ to be smaller than 1\%, then we need $(b/\ell_{\rm CS})^3> 10^{-N+2}$. For $N=4$, this gives $(b/\ell_{\rm CS})^3> 0.01$, with a lower bound which is borderline much smaller than 1 and much larger than $10^{-4}$. For larger values of $N$, the inequalities in Eq.~\eqref{inequality_N} are much more easily satisfied. For instance, if $\ell_{\rm CS}$ is a million times larger than $r_{s}$ ($N=6$), we would find $1> b/\ell_{\rm CS}\gg 0.01$ and with $b/\ell_{\rm CS} =0.9$ we would have $r_s \ell^2_{\rm CS}/b^3 \sim 10^{-6}$, such that the corrections could be completely neglected. 

In summary, although we can have time advances when $b<\ell_{\rm CS}$, we still need $b\gg r_s$ to trust the eikonal approximation, and these two inequalities are only compatible when $\ell_{\rm CS} \gg r_s$, which violates the effective field theory requirements. Let us ignore the latter and still ask when the eikonal approximation is valid. If $\ell_{\rm CS} \sim 8$ km , we would need black holes of sizes at most $r_s \sim 0.8$ meters, in order for the time advances to be resolvable within the weakly-coupled eikonal approximation, with corrections of ${\cal{O}}(r_s \ell_{\rm CS}^2/b^3)$ controlled to percent level. Said another way, for astrophysically-realistic black holes, with sizes $r_s \approx 10 $ km, we would need $\ell_{\rm CS} \sim 10^5$ km for the eikonal approximation to the time advance to hold at percent level for reasonable impact parameters (e.g.~for $b \sim 2 \times 10^{4} \; {\rm{km}}$). This value for $\ell_{\rm CS}$ is already excluded by astrophysics constraints~\cite{Silva:2020acr}. Therefore, the flat space amplitude considered does not offer a more stringent bound on $\ell_{\rm CS}$ than the one found in \cite{Silva:2020acr}.

\section{Time delays from geometry in dCS gravity}\label{sec:time_delays}

In the previous sections, we have discussed the limitations of the calculations of time delays based on scattering amplitudes in the eikonal regime. In this section, we remind the reader that, in both weak and strong gravity regimes, time delay can be computed directly from the metric in geometric theories of gravity. Geodesics studies have been the default way to calculate the Shapiro time delay in GR since \cite{Shapiro:1964uw}, which became one of the standard tests of GR in the second half of the twentieth century \cite{Will:2014kxa}. 

Given an asymptotically flat solution $g_{\mu\nu}$ to the field equations of GR or of a modified gravity theory, one can study geodesics in this background and calculate the propagation time between events as measured by asymptotically far-away observers. In this approach, self-consistency is never an issue, since the notion of ``time delay" is well-defined, and we are always working with exact solutions (valid in the strong and weak fields) of GR or of the modified gravity theory. In fact, the limitation of the scattering amplitudes approach can be traced back to the fact that, in the strong gravity regime and for gravitating bodies (like stars or black holes), an expansion around a flat spacetime metric is not self-consistent, since Minkowski spacetime does not approximate well the true solution in this regime. In the rest of this section, we will review how the Shapiro time delay is computed in the standard geometric approach, and then calculate this delay in dCS gravity, as our example modified gravity theory, for light scattered by a black hole and by a regular star.   

\subsection{Shapiro time delay for GR black holes}\label{subsec:time_delay_GR}

To see how the result of the last section intersects with the standard Shapiro time delay, we start by revisiting the computation for a non-rotating, Schwarzschild black hole in GR, following mostly~\cite{Misner:1973prb}. The line element of a Schwarzschild black hole can be written as
\begin{equation}
    ds^2 = -\left(1- \frac{2GM}{r}\right)dt^2 +\frac{dr^2}{\left(1- \frac{2GM}{r}\right)} + r^2 d\Omega^2\,, 
    \label{eq:line-element-Schw}
\end{equation}
in Schwarzschild coordinates $(t,r,\theta,\varphi)$, where $d\Omega^2 = d\theta^2 + \sin^2\theta d\varphi^2$ is the metric on the 2-sphere and $M$ is the black hole mass.

\begin{figure}
    \centering
    \includegraphics[width=0.9\linewidth]{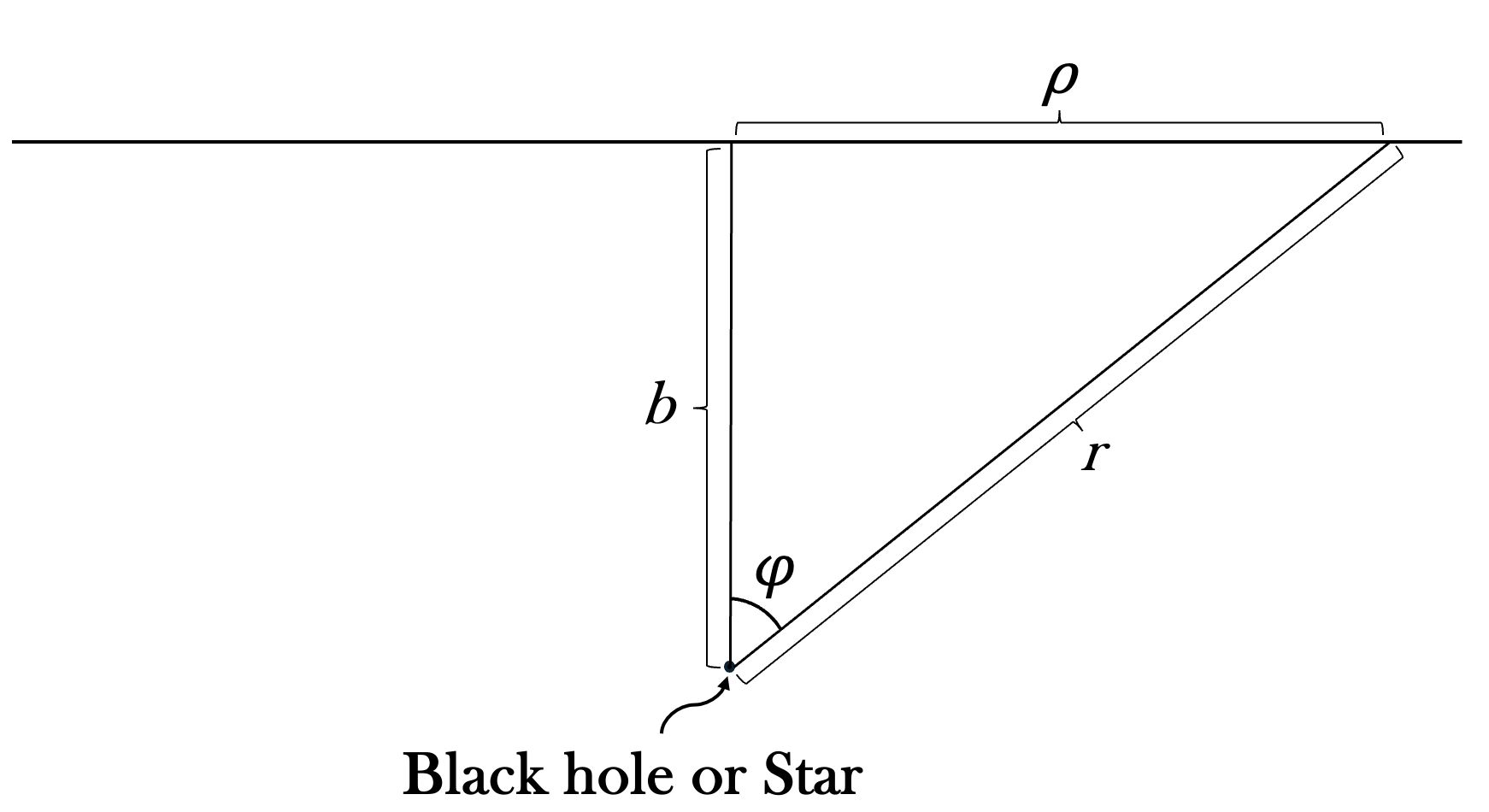}
    \caption{Schwarzschild coordinates of the null geodesic in the $\theta= \pi/2$ plane. There is a spherical object in the origin, potentially a black hole or star (not to scale).}
    \label{fig:Shapiro_time_delay}
\end{figure}

Consider now a light ray propagating in this black hole background, from $(r_1,\varphi_1)$ to $(r_2,\varphi_2)$ along the $\theta = \pi/2$ plane, with impact parameter $b$ from the origin $r=0$ (see Fig.~\ref{fig:Shapiro_time_delay}). In GR and in many modified gravity theories, light rays follow null geodesics of the background. Let then $\rho^2= r^2 - b^2$ be the affine parameter of a null geodesic, such that
\begin{equation}
    dr = \frac{\rho}{\sqrt{\rho^2 + b^2}}d\rho, \quad \tan \varphi = \frac{\rho}{b} \implies d\varphi = \frac{b}{\rho^2+b^2}d\rho.
\end{equation}
Physically, $\rho$ measures the spatial distance along the path from $(r_1,\varphi_1)$ to $(r_2,\varphi_2)$ as seen by asymptotic observers.

With this in hand, we are now ready to calculate the time delay experience by the light ray as it travels from $\rho_1$ to $\rho_2$. The line element for a null ray has zero invariant interval, $ds^2 = 0$, so then, using Eq.~\eqref{eq:line-element-Schw} and the coordinate $\rho$, we find
\begin{equation}
    dt = \frac{\left(1-\frac{2GM}{r(\rho)}\frac{b^2}{r^2(\rho)}\right)^{1/2}}{\left(1-\frac{2GM}{r(\rho)}\right)}d\rho\,.
\end{equation}
Integrating the above expression, we find
\begin{align}
    \delta_b t &= \int_{\rho_1}^{\rho_2} \frac{\left(1-\frac{2GM}{r(\rho)}\frac{b^2}{r^2(\rho)}\right)^{1/2}}{\left(1-\frac{2GM}{r(\rho)}\right)}d\rho\,.
    \label{eq:exact-Shapiro}
\end{align}
Equation~\eqref{eq:exact-Shapiro} is the exact expression for the travel time of a null ray passing outside a non-rotating (spherically-symmetric) gravitating source (be it a star or a black hole).  
Now, if there was no gravitating source, a null ray would simply take an amount of time $(\delta t)_{\rm flat} = \Delta \rho = \rho_2 - \rho_1$, which can be obtained from the above integral by setting $M = 0$. The travel time $\delta_b t$ then contains a time delay relative to $(\delta t)_{\rm flat}$ because $r(\rho) > 2 GM $ implies $\delta_b t > (\delta t)_{\rm flat}$. 

The above result, however, is not the standard expression for the Shapiro time delay that we are used to from elementary textbooks. To obtain the later, one uses that, along the null path, one usually has that $r(\rho)\gg b$, so that $2GM/r \ll 2GM/b \ll 1$, and so
\begin{equation}
    dt = \left[1+\frac{2GM}{r(\rho)} + {\cal{O}}\left(\frac{G^2M^2}{b^2}\right)\right] d\rho\,.
    \label{eq:weak-field-expansion}
\end{equation}
Integrating, one then finds
\begin{align}\label{time_delay_in_GR}
    \delta_{b} t &\simeq \int_{\rho_1}^{\rho_2} \left[1+ \frac{2GM}{\sqrt{\rho^2+b^2}} + {\cal{O}}\left(\frac{G^2M^2}{b^2}\right)\right]d\rho 
    \nonumber\\
    &= \Delta \rho \left[ 1 + \frac{GM}{\Delta \rho} \ln \left(\frac{\sqrt{\rho^2+b^2} +\rho}{\sqrt{\rho^2+b^2}-\rho}\right)\right|_{\rho_1}^{\rho_2}
    \nonumber \\
    &\left. + {\cal{O}}\left(\frac{G^2M^2}{b^2}\right) \right]\nonumber\\
    &= \Delta \rho \left. \left[1 + \frac{2GM}{\Delta \rho} \ln \left(\frac{2\rho}{b}\right) \right|_{\rho_1}^{\rho_2}
    + {\cal{O}}\left(\frac{G^2M^2}{b^2}, \frac{b^2}{\rho^2}\right)\right],
\end{align}
where we assumed that $\rho_{1,2}\gg b$ in the last equality. The second term on the second line of the above equation is a delay relative to the travel time of the null geodesic as measured by asymptotic observers, $\Delta \rho$. This is the standard Shapiro time delay, which we can also write in terms of the radial coordinate of the emitter and receiver events $r_{e} \equiv r(\rho_1)$ and $r_o\equiv r(\rho_2)$, respectively (e.g., see Eq. (10.101) in \cite{poisson2014gravity}) We emphasize that the weak-field expansion carried out in Eq.~\eqref{eq:weak-field-expansion} is only for convenience; the exact result is in Eq.~\eqref{eq:exact-Shapiro}, which can be trivially expanded to any order in $GM/b$ that one desires.

In order to make contact with the results of the previous section, which were obtained with scattering amplitude methods, we will consider the difference between time delays for two different impact parameters $b_0$ and $b$, with $b_0>b$. The largest delay is associated with the smallest impact parameter, so we define
\begin{equation}
    \Delta t \equiv \delta_{b}t - \delta_{b_0}t \approx 2GM \ln\left(\frac{b_0}{b}\right)\,,
    \label{eq:time-delay-final-GR}
\end{equation}
where we expanded in $\rho \gg b$. We will see in the next section that this expression has the same analytic form as the GR contribution to the time delay across a shockwave. Note also that this formula is an approximation valid in the limit $2GM/b \ll 1$ obtained from our exact result, and thus, it can be expanded to any order in $GM/b$ that one desires. 

Before discussing the dCS gravity case, let us investigate the leading contribution from the spin of a rotating (Kerr) black hole to the time delay in GR. In Boyer-Lindquist coordinates, the line element of a Kerr black hole is
\begin{align}\label{Kerr_metric_BL}
    ds^2&= -\left(1-\frac{2GMr}{\Sigma}\right)dt^2 - \frac{4GM a r \sin^2 \theta}{\Sigma}dt d\varphi + \frac{\Sigma}{\Delta}dr^2 \nonumber\\
    &+ \Sigma d\theta^2 + \left(r^2 +a^2 +\frac{2GM a^2 r \sin^2\theta}{\Sigma}\right)\sin^2\theta d\varphi^2,
\end{align}
where
\begin{equation}
    \Delta = r^2 -2GM r + a^2, \quad \Sigma = r^2 + a^2 \cos^2\theta,
\end{equation}
and $a=J/M$ is the ratio between the black hole angular momentum and mass (as measured by observers far away from the black hole). 
Following the same steps as in the Schwarzschild case, a null geodesic will satisfy
\begin{align}
     -\left(1-\frac{2GM}{r(\rho)}\right)dt^2 -\chi(r(\rho))dt d\rho +\Theta(r(\rho))d\rho^2 = 0,
\end{align}
where
\begin{align}
    \chi(r) &= \frac{4GM a b}{r^3},\\
    \Theta(r) &= \left[\frac{\rho^2}{\Delta}+\left(r^2+ a^2 +\frac{2GM a^2}{r}\right)\frac{b^2}{r^4}\right].
\end{align}
This implies that, along the null geodesic,
\begin{equation}\label{tdelay_differential}
    dt = \frac{-\frac{\chi(r(\rho))}{2} + \sqrt{\frac{\chi^2(r(\rho))}{4}+\Theta(r(\rho))\left(1-\frac{2GM}{r}\right)}}{1-\frac{2GM}{r}} d\rho.
\end{equation}
As $r(\rho)\geq b$ along the null path, if $a/b \ll 1$, we can evaluate the expressions above to leading order in $a/r<a/b$ and $2GM/b$ to obtain
\begin{equation}
    \Theta = \frac{1-\frac{2GM}{r}\frac{b^2}{r^2}}{1-\frac{2GM}{r}} +{\cal{O}}\left(\frac{a^2}{b^2}\right) = \frac{1}{1-\frac{2GM}{r}} +{\cal{O}}\left(\frac{a^2}{b^2}, \frac{G^2M^2}{b^2}\right).
    \label{eq:small-spin}
\end{equation}
Therefore, including only the leading-order spin contributions, we find
\begin{equation}
    dt =\frac{1- \frac{2GM ab}{r^3(\rho)}}{1-\frac{2GM}{r(\rho)}}d\rho \simeq \left(1+ \frac{2GM}{r(\rho)}- \frac{2GM ab}{r^3(\rho)}\right)d\rho
\end{equation}
and the time delay is then
\begin{align}
    (\delta_b t)_{\rm Kerr} &\simeq (\delta_b t)_{\rm Schw} - \left.\frac{a}{b}\frac{2GM}{\sqrt{\rho^2 + b^2}}\rho\right|_{\rho_1}^{\rho_2} + {\cal{O}}\left(\frac{a^2}{b^2}, \frac{G^2M^2}{b^2}\right),
    \nonumber \\
    &=\Delta \rho \left[ 1 + \frac{2GM}{\Delta \rho} \ln \left.\left(\frac{\sqrt{\rho^2+b^2} +\rho}{\sqrt{\rho^2+b^2}-\rho}\right)\right|_{\rho_1}^{\rho_2} 
    \right. 
    \nonumber \\
    &\left. - \left.\frac{a}{b}\frac{2GM}{\sqrt{\rho^2 + b^2}} \frac{\rho}{\Delta \rho} \right|_{\rho_1}^{\rho_2} + {\cal{O}}\left(\frac{a^2}{b^2}, \frac{G^2M^2}{b^2}\right)\right]\,,
    \label{eq:slow-rot-Kerr-time-delay}
\end{align}
where $(\delta_b t)_{\rm Schw}$ is the travel time of a null ray that passes a Schwarzschild black hole in Eq.~\eqref{time_delay_in_GR}. The analog of \eqref{eq:time-delay-final-GR} after assuming that $\rho\gg b$ is
\begin{align}
    (\Delta t)_{\rm Kerr} &= (\delta_b t)_{\rm Kerr}- (\delta_{b_0} t)_{\rm Kerr} \nonumber\\
    & \approx 2GM \ln\left(\frac{b_0}{b}\right) - 2 GM \; \frac{a}{b} \left(1 - \frac{b}{b_0}\right).
    \label{eq:slow-rot-Kerr-Shapiro}
\end{align}
Reproducing such a result from a scattering amplitude calculation would require having an incoming state with non-vanishing angular momentum.

Let us discuss the above results in a bit more detail. The first thing we notice is that the spin correction to the travel time is a time advance, and not a time delay, i.e.~the term proportional to $a$ in Eqs.~\eqref{eq:slow-rot-Kerr-time-delay} and~\eqref{eq:slow-rot-Kerr-Shapiro} is strictly negative because $b/b_0 < 1$ by convention. Of course, this time advance contribution does not violate causality because the Kerr spin parameter cannot take arbitrarily large values. One could demand that the term proportional to $a$ be smaller than the second term in Eq.~\eqref{eq:slow-rot-Kerr-time-delay} or in Eq.~\eqref{eq:slow-rot-Kerr-Shapiro} to find a ``causality constraint'' on the spin of black holes. This, of course, would yield a constraint that (i) would be outside the regime of validity of the Kerr solution (i.e.~when $a>M$, the Kerr black hole contains a naked singularity), (ii) would be outside the regime of the slow-rotation approximation used to derive this constraint. The term proportional to $a$ in Eqs.~\eqref{eq:slow-rot-Kerr-time-delay} and~\eqref{eq:slow-rot-Kerr-Shapiro} was derived \textit{assuming} that it would lead to a small correction to the Schwarzschild black hole case (because of the small spin expansion in Eq.~\eqref{eq:small-spin}), and thus, no ``causality constraint'' on spin is actually possible.  

\subsection{Shapiro time delay for dCS gravity black holes}

Let us now consider black holes in dCS gravity. The leading correction to the Kerr metric due to the dCS coupling were calculated in \cite{Yunes:2009hc}. The resulting metric is given by 
\begin{align}
    ds^2 &= \frac{5}{4}\zeta a \left(\frac{GM}{r}\right)^4\left[1+\frac{12}{7}\frac{GM}{r}+\frac{27}{10}\frac{(GM)^2}{r^2}\right]\sin^2\theta dt d\varphi \nonumber\\
    &+ ds^2_{\rm Kerr},
\end{align}
where $ds^2_{\rm Keerr}$ is the Kerr metric of Eq.~\eqref{Kerr_metric_BL} and, in our conventions,
\begin{equation}
    \zeta = \frac{2\alpha^2}{(GM)^4}.
\end{equation}
If $\zeta \ll 1$, we are guaranteed to be within the regime of validity (i.e.~within the cut-off) of dCS gravity as an effective field theory for black-hole calculations.

Let us now consider null geodesics in this dCS-modified rotating black-hole background. Compared to the Kerr case, after we evaluate the metric above on the null geodesic, there will be no changes in $\Theta(r(\rho))$ in Eq.~\eqref{tdelay_differential}, whereas $\chi(r)$ will change as 
\begin{equation}
    \chi(r) \to \chi(r) + \frac{5}{4}\zeta \left(\frac{GM}{r}\right)^4\frac{a b}{r^2}\left[1+\frac{12}{7}\frac{GM}{r}+\frac{27}{10}\frac{(GM)^2}{r^2}\right].
\end{equation}
Then, the leading-order-in-$\zeta$ contribution to the time delay in GR of Eq.~\eqref{eq:slow-rot-Kerr-time-delay}, $(\delta_b t)_{\rm Kerr} \to (\delta_b t)_{\rm Kerr} + \delta_b^{\rm dCS} t$, is
\allowdisplaybreaks[4]
\begin{align}
    \delta_b^{\rm dCS}t &= \int_{\rho_1}^{\rho_2}\left[-\frac{5}{8}\zeta\left(\frac{GM}{r(\rho)}\right)^4\frac{a b}{r^2(\rho)} 
    + {\cal{O}}\left( \frac{G^5 M^5}{r^5} \right)\right]d\rho \nonumber\\
    &= -\frac{5}{64}\zeta \left(\frac{GM}{b}\right)^4 a\left[\frac{5b^3\rho+3b\rho^3}{(b^2+\rho^2)^2}+3 \arctan\left(\frac{\rho}{b}\right)\right]_{\rho_1}^{\rho_2}
    \nonumber \\
    &
    + {\cal{O}}\left( \frac{G^5 M^5}{b^5} \right)
    \nonumber \\
    & \approx \frac{a \zeta}{8} \left(\frac{GM}{b}\right)^4 \left. \left(\frac{b}{\rho}\right)^5 \right|_{\rho_1}^{\rho_2}\,.
\end{align}
Observe that the dCS correction is extremely suppressed by both $\zeta$ and $(GM/b)^4$. Since both $\zeta \ll 1$ (to remain within the regime of validity of dCS gravity as an effective field theory) and $GM/b \ll 1$ (to remain within the far-field expansion of the time delay calculation), this correction can never dominate over the GR contribution in Eq.~\eqref{time_delay_in_GR}. 

Is this result consistent with the scattering amplitude calculation? Yes. The regime $\zeta <1$ corresponds to the limit $\ell_{\rm CS}< r_s$, for which the amplitude approach also does not yield a time advance. We could have carried out the above calculation without expanding in $GM/b \ll 1$, for example by considering light rays that graze the black hole horizon. Had we done so, however, the suppression of this term with respect to $\zeta$ would still be present, even after including nonlinearities that can take factors of $GM/b\sim 1$ into account.

The dCS contribution to the total difference between time delays for impact parameters $b$ and $b_0>b$ is
\begin{align}
     (\Delta t)_{\rm dCS} & \equiv \delta_b^{\rm dCS}t- \delta_{b_0}^{\rm dCS}t \approx  
     - \frac{\zeta}{8} G M \left(\frac{b_0}{G M}\right) 
     \left(\frac{a}{GM}\right)
     \nonumber \\
     &\times \left(\frac{GM}{\rho_1}\right)^5 
     \left(1 - \frac{b}{b_0}\right)
     \left[1  -\left(\frac{\rho_1}{\rho_2}\right)^5 \right].
     \label{eq:dCS-time-delay}
\end{align}
As $\rho_1$ and $\rho_2$ originated from the integral along the null worldline parametrized by $\rho$, we have that $\rho_2>\rho_1$. Hence, the dCS correction to the time delay is always negative. This, however, does not mean that Eq.~\eqref{eq:dCS-time-delay} leads to a time advance instead of a time delay, because this correction must be added on top of the Kerr time delay of Eq.~\eqref{eq:slow-rot-Kerr-Shapiro}. Only when $(\Delta t)_{\rm dCS}$ in Eq.~\eqref{eq:dCS-time-delay} becomes larger than $(\Delta t)_{\rm Kerr}$ in Eq.~\eqref{eq:slow-rot-Kerr-Shapiro} does one obtain a time advance. Note, however, that the dCS correction is \textit{suppressed} relative to the Kerr correction by multiple terms: $\zeta \ll 1$ (so that the theory is within regime of validity), $a/(G M) \ll 1$ (so that the black hole solution be valid in the small spin approximation), $(G M/\rho_1)^5 \ll 1$ (so that the weak-field expansion of gravity be valid). Only when these assumptions, which the above calculations rely on, are broken can the dCS correction lead to a time advance.

The above results \textit{cannot} and should not be compared directly with Eq.~\eqref{time_delay_in_dCS_gravity}. This is because the calculation in this section computes a \textit{different} quantity (one that is actually observable) than that computed with the scattering amplitude calculation. In this subsection, we were concerned with the time delay of a photon that travels through the spacetime of a black hole in dCS gravity. In the scattering amplitude calculation, one is concerned with the time delay between a wave in the scalar field that is sourced by gravitational perturbations relative to a wave in the gravitational perturbations. Since the scalar field in dCS gravity does not couple directly to matter, there is no way to directly observe the scalar wave, and thus, to measure this time difference. Nonetheless, regardless of the situation considered, we see that the only way dCS gravity can lead to a time advance is if one considers the theory outside of its regime of validity (ie.~large $\zeta$) and outside the weak gravity regime. 

\subsection{Shapiro time delay from stars in dCS gravity}

Time delay observations have been carried out accurately with our Sun as the main gravitating object. Let us then consider what the dCS correction to the time delay is for null rays scattered by our Sun. Reference~\cite{Ali-Haimoud:2011zme} has shown that the black hole metric in dCS gravity cannot describe the exterior metric of stars. For $\zeta\ll1$, and in the slow-rotating ($a/GM <1$) and weak-gravity ($GM/r<1$) regimes (both of which are necessary to describe our Sun), the following metric for the exterior of the star was found in \cite{Ali-Haimoud:2011zme}:
\begin{widetext}
\begin{align}
    ds^2_{\rm star} &= -\left(1-\frac{2GM}{r}\right)dt^2 + \frac{dr^2}{1-\frac{2GM}{r}} + r^2\left(d\theta^2 + \sin^2 \theta d\varphi^2\right) -2r^2 \omega(r,\theta)\sin^2\theta dt d\varphi,\\
    \omega(r,\theta) &= \frac{2GM}{r}\frac{a}{r}\frac{1}{r} + 10\pi \frac{a}{(GM)^2}\frac{\alpha^2}{ (GM)^4}\left[(C_1 -1)\left(\frac{GM}{r}\right)^6 +C_1\frac{64}{15}\left(\frac{GM}{r}\right)^9\right],
\end{align}
\end{widetext}
where $C_1 \sim 1$, up to terms of order of the compactness of the star, which is very small for our Sun, and the angular momentum of the sun is $J = M a$, as measured by observers far from the Sun. 

For the computation of time delay, evaluating $dt$ along the null geodesic yields Eq.~\eqref{tdelay_differential} with the same $\Theta(r(\rho))$ as GR in the slow-rotating and weak-gravity limits. However, $\chi(r(\rho))$ is now 
\begin{align}
    \chi(r(\rho)) &= 4\frac{GM}{r}\frac{ab}{r^2} + \frac{256\pi}{3}\frac{ab}{(GM)^2}\frac{\zeta}{2} \left[\left(C_1\frac{GM}{r(\rho)}\right)^9\right.\nonumber\\
    &+\left.(C_1-1)\frac{15}{64}\left(\frac{GM}{r}\right)^6\right],
\end{align}
Then, the dCS coupling enters in the following contribution to the time delay in GR Eq.~\eqref{time_delay_in_GR}, to leading order in $a$:
\allowdisplaybreaks[4]
\begin{align}
    \delta_b^{\rm dCS}t &= \int_{\rho_1}^{\rho_2}\left\{-2 \frac{GM}{r}\frac{ab}{r^2}-\frac{128\pi}{3}\frac{ab}{(GM)^2}\frac{\zeta}{2}\left[C_1\left(\frac{GM}{r}\right)^9   \right.\right.\nonumber\\
    &+\left.\left.(C_1-1)\frac{15}{64}\left(\frac{GM}{r}\right)^6\right] +\mathcal{O}\left(\frac{a^2 b^2}{r^4}\frac{G^2M^2}{r^2}\right)
    \right\}d\rho\nonumber\\
    &\sim a \frac{GM}{b}\left[\left(\frac{b}{\rho}\right)^2 - \frac{3}{4}\left(\frac{b}{\rho}\right)^4\right.\nonumber\\
    &+\left.\pi (C_1-1)\zeta \left(\frac{b}{\rho}\right)^2\left(\frac{GM}{b}\right)^3+\mathcal{O}\left(\frac{a^2}{b^2}, \frac{G^4 M^4}{b^4}\right)\right]_{\rho_1}^{
    \rho_2
    }
\end{align}
which is even more suppressed compared to the black hole case. The dCS correction is so small that it is completely unobservable with current technology. As $C_1 = 1+\mathcal{O}(GM/R)$, this correction is proportional to $\zeta (G M/b)^4(G M/R)$, and $b > R$, where $R$ is the radius of the star. For solar system tests, we then see that $(G M/b)^4(G M/R) \ll C^5$, where $C = GM /R_\odot \sim 2 \times 10^{-6}$ is the gravitational compactness of the Sun. The dCS correction is then much smaller than $3 \times 10^{-29}$, and further suppressed by $\zeta$. One could try to make $\zeta$ ridiculously gigantic, so that its magnitude would overwhelm the $C^5$ suppression, and thus, become dominant over the GR time delay, but this would force us to greatly exit the effective-field-theory regime of dCS gravity.   

\section{Causality constraint from geodesic motion in shockwave backgrounds in dCS gravity}\label{sec:shockwave_backgrounds}

We have so far studied time delays in the context of scattering amplitudes and null geodesics, but there is a third way to carry out the same calculation: graviton propagation on a shockwave background. Reference~\cite{Camanho:2014apa} discussed how to obtain constraints on higher-order corrections to gravity by studying graviton propagation in shockwave background. This idea is based on the results of \cite{Dray:1984ha} on matching the discontinuity of fields through the shock and gravitational time delays computed from the metric (see also \cite{Adamo:2021rfq,Adamo:2022rmp}). In this section, we study whether we can place similar constraints on dCS gravity. Similar to the results of the previous sections, self-consistency with the solutions of the theory will imply that the new dCS corrections to the time delay in GR are always small and do not yield time advances. To set the notation, we start with a short section on the pure GR case before studying the dCS gravity case. 

\subsection{Time delays from shock waves in GR}

A shockwave is a solution to the field equations sourced by a single massless particle. Employing light-cone coordinates, the metric is given by
\begin{equation}\label{shockwave_metric}
    ds^2 = -2 du dv + H(u,x^i)du^2 + \delta_{ij}dx^i dx^j,
\end{equation}
where $\sqrt{2}du = dt + dx$ and $\sqrt{2}dv= dt-dx$ are outgoing and ingoing null coordinates, with $t$ and $\{x,x^i\}$ Cartesian coordinates used by inertial observers, with $x^i = \{y,z\}$. The only non-vanishing components of the Christoffel symbols for this metric in these coordinates are
\begin{equation}
    \Gamma^{i}_{uu} = -\frac{1}{2}\partial^i H, \quad \Gamma^{v}_{uu} = \frac{1}{2}\partial_u H, \quad \Gamma^v_{ui} = \frac{1}{2}\partial_i H.
\end{equation}
All the non-vanishing components of the Riemann tensor are related to 
\begin{equation}
    R_{uju}^{\;\;\;\;\;\;k} = -\frac{1}{2}\partial_j \partial^k H,
\end{equation}
while for the Ricci tensor and scalar, we have
\begin{equation}
    R_{uu} = -\frac{1}{2} \partial_i \partial^i H, \quad R = 0.
\end{equation}
Other curvature invariants that also vanish are 
\begin{equation}
    R_{\mu\mu}R^{\mu\nu} = 0,\quad R^{\mu\nu\rho\sigma}R_{\mu\nu\rho\sigma} =0, \quad \Tilde{R}^{\mu\nu\rho\sigma}R_{\mu\nu\rho\sigma}=0,
\end{equation}
where
\begin{equation}
    \Tilde{R}_{ui}^{\;\;\;\;vj}= \frac{1}{2}\epsilon^{vjuk}R_{uiuk} = -\frac{1}{4}\epsilon^{uvjk}\partial_i \partial_k H.
\end{equation}

The energy-momentum tensor for a massless source propagating in the $-x$ direction (that is, moving along $u=0$) is 
\begin{equation}
    T_{uu} = p \delta(u) \delta^{D-2}(x^i),
\end{equation}
and all other components are zero, where $p$ is the particle's energy (and magnitude of the momentum). Einstein's equations reduce to \cite{Aichelburg:1970dh}
\begin{equation}
    \partial_{i}\partial^i H = - 16\pi G p \delta(u) \delta^{D-2}(x^i).
\end{equation}
For $D=4$ we have 
\begin{equation}\label{H_in_4D}
    H(u, x^i) = -8 Gp \delta(u)\ln r, 
\end{equation}
where $r^2 = x_ix^i$ and for $D>4$, we have
\begin{equation}
    H(u,x^i) = \frac{16\pi G}{\Omega_{D-3}(D-4)}\frac{p}{r^{D-4}}\delta(u),
\end{equation}
which fixes the solution.

\begin{figure}
    \centering
    \includegraphics[width=0.9\linewidth]{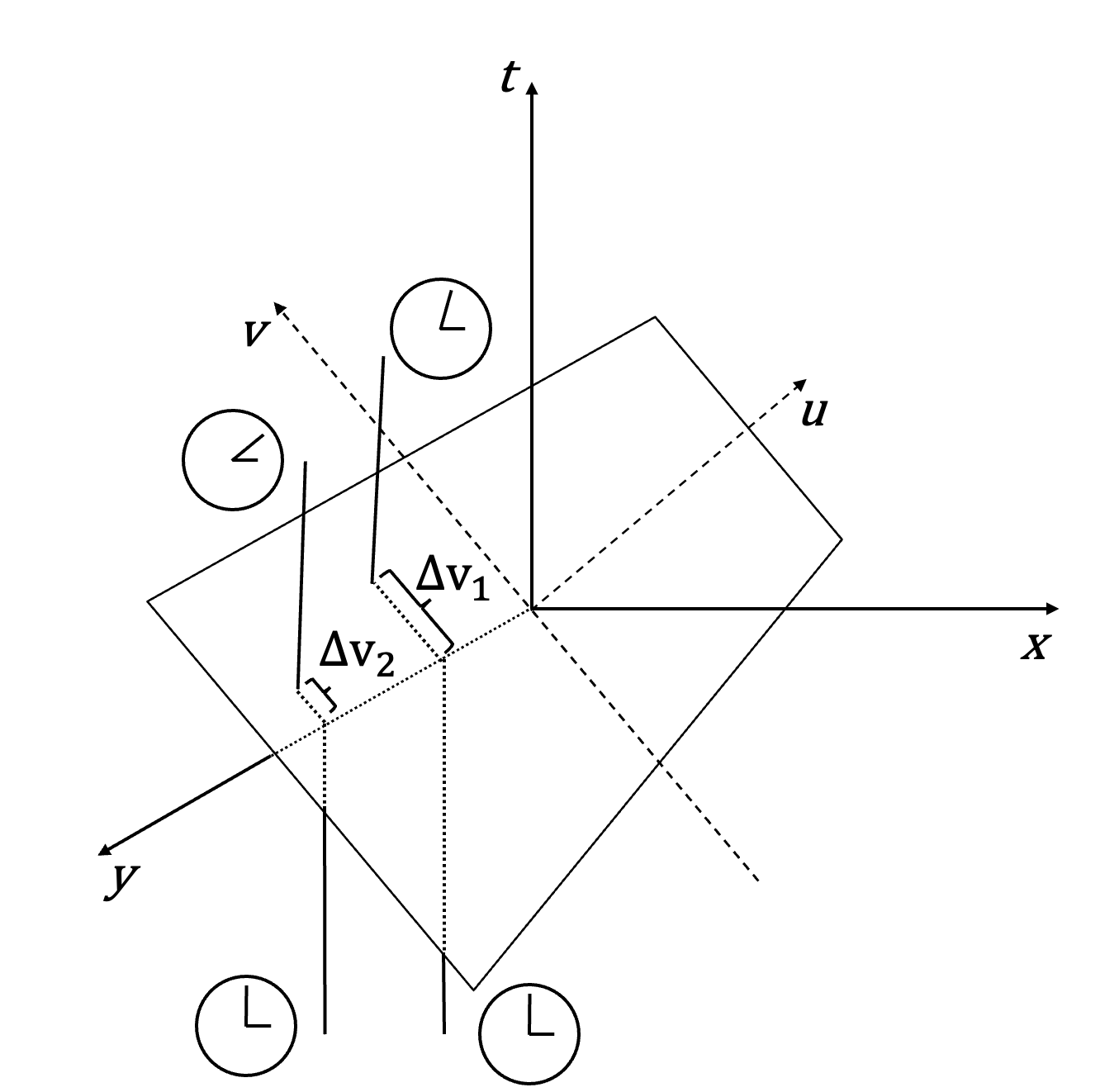}
    \caption{Spacetime diagram in the inertial, Cartesian coordinates $t$ and $x$  illustrating the effect of the shockwave produced by the massless source on nearby clocks. The source is located along the $v$-axis ($u=0$) and, according to the GR solution \eqref{H_in_4D}, it generates a shock in the plane orthogonal to the $u$-axis, along which the gravitational wave propagates (see discussion in the main text). Two initially synchronized clocks located at $x=0=z$ but separated in the $y$-direction become out of sync after crossing the shock. The relative time-delay between them is set by the distance from the source: the closest clock acquires a delay $\Delta v_1$ larger than the other's delay, $\Delta v_2$. The clocks also get an impulse towards the source \cite{Dray:1984ha, Aichelburg:1970dh}.}
    \label{fig:shockwave}
\end{figure}

Let us now consider the following Gedanken experiment. Consider a shockwave spacetime as a background, with a gravitational wave $h_{ij}$ propagating on a geodesic of the background in the $u$-direction, at an impact parameter $b$ in the transverse direction to the shock~\cite{Camanho:2014apa}. The solution to the field equations for $h_{ij}$ will contain a discontinuity along the $v$-direction, when the wave crosses the shockwave, which is usually interpreted as a time delay produced by the massless particle after its passage. See Fig. \ref{fig:shockwave}.

This result can be seen from the equation of motion for the transverse-traceless modes of the metric. In the transverse-traceless gauge, and in the shockwave background, the non-trivial,  $h_{ij}$-dependent Einstein tensor components are
\begin{subequations}
\begin{align}
    G^{(1)}_{uu} &=\frac{1}{2}\left(1+\frac{H}{2}\right)\overline{R}_{uu}h_{xx} - R_{iuju}h^{ij} -\frac{1}{4}\overline{\nabla}_\sigma \overline{\nabla}^\sigma h_{xx},\\  
    G^{(1)}_{uv} &= -\frac{1}{2}\overline{R}_{uu}h_{xx}+\frac{1}{4}\overline{\nabla}_\sigma \overline{\nabla}^\sigma h_{xx},\\
    G^{(1)}_{ui} &=\frac{1}{2\sqrt{2}}\overline{R}_{uu}h_{ix} +\frac{1}{\sqrt{2}}\overline{R}_{uiuj}h_x^{\;\;j}-\frac{1}{2\sqrt{2}}\overline{\nabla}_\sigma \overline{\nabla}^\sigma h_{xi} ,\\
    G^{(1)}_{vv} &= -\frac{1}{4}\overline{\nabla}_\sigma \overline{\nabla}^\sigma h_{xx} ,\\
    G^{(1)}_{vi} &= \frac{1}{2\sqrt{2}}\overline{\nabla}_\sigma \overline{\nabla}^\sigma h_{xi},\\
    G_{ij}^{(1)} &= -\frac{1}{2}\overline{\nabla}_\sigma \overline{\nabla}^\sigma h_{ij}\,,
\end{align}
\end{subequations}
to linear order in $h_{ij}$. Recall that the ${i,j}$ indices only run through directions transverse to $x$. Focusing on the $ij$ equation, we have
\begin{equation}
    2\partial_u \partial_v h_{ij} + H(u, x^i) \partial^2_v h_{ij} - \partial_k \partial^k h_{ij} = 0.
\end{equation}
Close to the shock, the second term is much larger than the third, so let us neglect spatial derivatives of $h_{ij}$. In this case, we then have
\begin{equation}
    2\partial_u h_{ij} + H(u, x^i) \partial_v h_{ij} = 0.
\end{equation}
Writing $H(u, x^i) = f(x^i) \delta(u)$, we have, after integrating around $u=0$,
\begin{equation}
    h_{ij}(u = 0^+) - h_{ij}(u=0^-) = -\frac{f(r)}{2}\partial_v h_{ij}(u=0).
\end{equation}
As shown in \cite{tHooft:1987vrq,Camanho:2014apa}, the discontinuity is proportional to the time delay $\Delta v$:
\begin{equation}
    \Delta v = \frac{f(b)}{2} = -4 G p \ln b,
\end{equation}
where $b$ is the impact parameter and in four dimensions $f(b)$ can be read from Eq.~\eqref{H_in_4D}. 

The time delay obtained above is reminiscent of that found for the scattering of a null ray from the background geometry of a gravitating source in Eq.~\eqref{eq:time-delay-final-GR}. Of course, the result above depends on the energy of the massless and moving source. The physical set up here, however, is quite different from that in the previous section. Nonetheless, it is remarkable that in both cases one obtains similar expressions for the time delay. 

\subsection{Time delays from shock waves in dCS gravity}

Let us now repeat the previous calculation but in dCS gravity. The field equations in dCS gravity that arise from variation of the action in  Eq.~\eqref{dCS_action} are
\begin{align}\label{dCSeom}
    G_{\mu\nu} + 2\tilde{\alpha} \kappa^2C_{\mu\nu} = 8\pi G T_{\mu \nu}^{(\phi)},
\end{align}
where $\tilde{\alpha} = \alpha/\kappa$ and the C-tensor $C_{\mu\nu}$ is given by
\begin{align}
    C^{\mu\nu} &= \nabla_\gamma \phi \epsilon^{\gamma \delta \alpha (\mu}\nabla_\alpha R^{\nu)}_{\;\;\delta}+ \nabla_\gamma \nabla_\delta \phi \Tilde{R}^{\delta (\mu\nu)\gamma}\nonumber\\
    &=\frac{1}{2}\nabla_\gamma \phi \left(\epsilon^{\gamma \delta \alpha \mu}\nabla_\alpha R^{\nu}_{\;\;\delta}+\epsilon^{\gamma \delta \alpha \nu}\nabla_\alpha R^{\mu}_{\;\;\delta}\right)+ \nonumber\\
    &+\frac{1}{4}\nabla_\gamma \nabla_\delta \phi \left(\epsilon^{\alpha\beta\gamma \mu}R^{\nu\delta}_{\;\;\;\;\alpha\beta}+\epsilon^{\alpha\beta\gamma \nu}R^{\mu\delta}_{\;\;\;\;\alpha\beta}\right).
\end{align}
The equation of motion for the scalar field,
\begin{equation}
    g^{\mu\nu}\nabla_\mu \nabla_\nu \phi -V'= \frac{\tilde{\alpha}}{4}\Tilde{R}R\,,
\end{equation}
can be obtained from variation of the action with respect to the scalar field, or by taking the covariant divergence of Eq.~\eqref{dCSeom}.

Let us now linearize Eq.~\eqref{dCSeom} over an arbitrary background metric. To linearize the C-tensor, we first note that
\begin{equation}
    \sqrt{-g} = \sqrt{-\bar{g}}\left(1-\frac{h}{2}\right) + \mathcal{O}(h^2),
\end{equation}
such that
\begin{equation}
    \epsilon^{\gamma \delta \alpha \mu} = \bar{\epsilon}^{\gamma \delta \alpha \mu}\left(1+\frac{h}{2}\right) +\mathcal{O}(h^2)\,,
\end{equation}
and
\begin{equation}
    \nabla_\alpha R^{\nu}_{\;\;\delta} = \overline{\nabla}_\alpha \overline{R}^\nu_{\;\;\delta} + \overline{\nabla}_\alpha\delta R^{\nu}_{\;\;\delta} + \delta\Gamma^{\nu}_{\alpha\lambda} \overline{R}^{\lambda}_{\;\;\delta}- \delta\Gamma^{\lambda}_{\alpha\delta}\overline{R}^{\nu}_{\;\;\lambda}. 
\end{equation}
With this in hand, we then have
\begin{widetext}
\begin{align}\label{pertubedC}
    \delta C^{\mu\nu} &= \overline{\nabla}_\gamma \phi \bar{\epsilon}^{\gamma \delta \alpha (\mu}\left(\frac{h}{2} \overline{\nabla}_\alpha \overline{R}^{\nu)}_{\;\;\delta}+\overline{\nabla}_\alpha\delta R^{\nu)}_{\;\;\delta} + \delta\Gamma^{\nu)}_{\alpha\lambda} \overline{R}^{\lambda}_{\;\;\delta}- \overline{R}^{\nu)}_{\;\;\lambda}\delta\Gamma^{\lambda}_{\alpha\delta}\right)-\frac{1}{2}\nabla_\gamma \phi \delta \Gamma^{\gamma}_{\lambda\delta}\epsilon^{\alpha \beta \lambda (\mu}\overline{R}^{\nu)\delta}_{\;\;\;\;\alpha \beta}+\nonumber\\
    &+ \frac{1}{2}\overline{\nabla}_\gamma \overline{\nabla}_\delta \phi \bar{\epsilon}^{\alpha\beta\gamma(\mu}\left(\frac{h}{2}\overline{R}^{\nu)\delta}_{\;\;\;\;\alpha\beta}+ \delta R^{\nu)\delta}_{\;\;\;\;\alpha\beta}\right) +\nabla_\gamma \delta\phi \epsilon^{\gamma \delta \alpha (\mu}\nabla_\alpha R^{\nu)}_{\;\;\delta}+ \frac{1}{2}\nabla_\gamma \nabla_\delta \delta\phi \epsilon^{\alpha\beta\gamma (\mu}R^{\nu)\delta}_{\;\;\;\;\alpha\beta}.
\end{align}
\end{widetext}
The linearized equation of motion for dCS gravity is then
\begin{equation}\label{pert_dCS_eom}
    \delta G_{\mu\nu} + 2\tilde{\alpha}  \kappa^2 \delta C_{\mu\nu} = 8\pi G \delta T_{\mu\nu}^{(\phi)}.
\end{equation}

What background metric should we use to evaluate the linearized dCS field equations? Reference \cite{Grumiller:2007rv} has shown that the shockwave metric in Eq.~\eqref{shockwave_metric} is also an exact solution to dCS gravity, for different pseudo-scalar analytical behavior. We shall thus evaluate Eq.~\eqref{pert_dCS_eom} over such a background solution (for $V' = 0$) with
\begin{equation}
    \bar{g}_{uv} = -1, \quad \bar{g}_{uu}= H(u,x^i), \quad \bar{g}_{ij} = \delta_{ij},
\end{equation}
supported by a wave packet of background field $\phi$ moving along the $v$-direction with energy $|p|$. In this case, the energy-momentum tensor is localized in $\{u,x^i\}$ and for distances larger than $|p|^{-1}$ we have
\begin{equation}
    T_{uu}^{(\phi)} \approx p \delta(u)\delta^{D-2}(x^i), \quad \partial_u \phi \approx \sqrt{|p|}\delta(u)\delta^{D-2}(x^i).
\end{equation}
The background solution will thus contain a shockwave. 

We are here interested in gravitational wave propagation on this shockwave background in the probe limit for the gravitons, i.e.~neglecting perturbations to the scalar field. We focus on the traceless components in the plane transverse to the $u$ and $v$-directions. The equation of motion for $h_{ij}$ are given by
\begin{equation}
    \delta G_{ij} + 2\tilde{\alpha}  \kappa^2 \delta C_{ij} = 0\,.
\end{equation}
From Eq.~\eqref{pertubedC}, we have that
\begin{equation}
    \delta C_{ij} = \frac{1}{2}\partial_\mu \phi \epsilon^{uvk}_{\;\;\;\;\;\;(i}\nabla_{|v|} \nabla^2 h_{j)k} - \nabla_u \nabla_u \phi \epsilon^{uvk}_{\;\;\;\;\;\;(i}\nabla_{|v|} \nabla^u h_{j)k},
\end{equation}
and so
\begin{align}
    \nabla_\sigma \nabla^\sigma h_{ij} &- 2\tilde{\alpha}  \kappa^2\partial_u \phi \epsilon^{uvk}_{\;\;\;\;\;\;(i}\partial_{|v|} \nabla^2 h_{j)k} \nonumber\\
    &-4\tilde{\alpha}  \kappa^2 \partial_u^2 \phi \epsilon^{uvk}_{\;\;\;\;\;\;(i}\partial_{|v|}^2h_{j)k} = 0.
\end{align}
To find the discontinuity in $u$, we evaluate this equation on a trajectory along the $u$-direction that passes the shock with an impact parameter $b$ in the transverse directions. Note that $b> |p|^{-1}$ for the consistency of the optical (eikonal) approximation of the wave-packet description to be valid. In that limit, $\partial_u^2 \phi$ is subleading compared to $\partial_u \phi$, and we find
\begin{equation}\label{graviton_eom}
    -2\partial_u \partial_v h_{ij} - H \partial^2_v h_{ij}+ 4\tilde{\alpha}  \kappa^2 \partial_u \phi \epsilon^{uvk}_{\;\;\;\;\;\;(i}\partial_{|u}\partial^2_{v|}h_{j)k} =0.
\end{equation}

Let us estimate the behavior of $\partial_u \phi$ along $u=0$. We know that $\partial_u \phi$ should provide a localized energy density of ${\cal{O}}\left((\partial_u \phi)^2\right)$. However, ``localization'' is only really defined relative to a reference scale, so we need to be precise about the scales we are considering (because we want to stay within the regime of validity of our assumptions). In our case, the energy should be localized relative to the energy scale of the wave packet $|p|$, and so
\begin{equation}
    T_{\mu\nu} = p \delta(u)\delta^{(D-2)}(x^i) = \frac{p}{b^{D-1}}\delta(u/b)\delta^{(D-2)}\delta(x^i/b),
\end{equation}
and, around $u\sim0$, we should have
\begin{equation}
    \partial_u \phi \sim \sqrt{\frac{|p|}{b^{D-1}}}.
\end{equation}

We are now ready to compute the time delay.
Using the above scaling result in Eq.~\eqref{graviton_eom}, integrating this equation over $v$ and $u \in (-\epsilon, \epsilon)$, evaluating the result at $r = b$, and taking the $\epsilon \to 0$ limit, we obtain
\begin{equation}
    \left[h_{ij}\right]^+_- +\frac{f(b)}{2}\left.\partial_v h_{ij}\right|_{u=0} -2\tilde{\alpha}  \kappa^2 \sqrt{\frac{|p|}{b^{D-1}}} \epsilon^{uvk}_{\;\;\;\;\;\;(i}\left.\partial_v h_{j)k}\right|_{u=0} = 0,
\end{equation}
where we defined $[h]^+_- = h(u=0^+)-h(u=0^-)$. Decomposing the metric perturbation in circular polarizations, $h_{ij} = e^A_{ij} h_{A}$, and using $\epsilon^{uvk}_{\;\;\;\;\;\;i}e^{ij}_B = \epsilon_B e_B^{kj}$ (no sum over $B$), gives us
\begin{equation}
    [h_B]^{+}_- = \left(-\frac{f(b)}{2} + 4\tilde{\alpha}  \kappa^2 \sqrt{\frac{|p|}{b^{D-1}}} \epsilon_B\right)\left.\partial_v h_B\right|_{u=0},
\end{equation}
from which we read the $D=4$ time delay
\begin{equation}\label{time_delay_dCS}
    \Delta v_B = 4G|p|\left[\ln\left(\frac{1}{b}\right) - \frac{8\pi}{\sqrt{|p|b}} \frac{\tilde{\alpha}}{b}\epsilon_B\right].
\end{equation}
This is the dCS-corrected time delay one obtains by considering a gravitational wave crossing a shockwave sourced by the dCS pseudoscalar. Observe that the dCS term is proportional to 
\begin{align}
    \frac{1}{\sqrt{|p| b}} \frac{\tilde{\alpha}}{b} = \frac{1}{\sqrt{M b}} \frac{\tilde{\alpha}}{b} \sim \left(\frac{G M}{b}\right)^{3/2} \frac{{\alpha}}{(G M)^2}, 
\end{align}
where we replaced $p$ with $M$, which represents the mass scale of the background shockwave. We shall show below that this dCS correction is small. 

Clearly, if $\tilde\alpha$ is large enough, one could obtain a time advance for one of the polarizations, but is this result self-consistent? We shall now show that the values of $\alpha$ that lead to acausal behavior are deeply outside the regime of validity of the shock solution if we demand that the wave packet has energy below the effective-field-theory scale $\sim \tilde{\alpha} ^{-1}$. First, note that the localization of the wave packet is only valid when $b\gg|p|^{-1}$, such that
\begin{equation}
    |p|\gg \frac{1}{b} \implies \frac{1}{\sqrt{|p|b^3}}<\frac{1}{b}.
\end{equation}
Moreover, since Eq.~\eqref{dCS_action} is an effective field theory valid below the dCS energy scale $\tilde{\alpha}^{-1}$, we \textit{must} require an upper bound for the energy of the source, $|p|< \tilde{\alpha}^{-1}$. This is a necessary condition for the packet to be resolvable within the effective field theory. Combining this condition and the inequality above, we find that
\begin{equation}
    \frac{1}{\tilde{\alpha}}>|p|\gg \frac{1}{b} \implies \frac{\tilde{\alpha}}{\sqrt{|p|b^3}}< \frac{\tilde{\alpha}}{b}\ll 1.
\end{equation}
The dCS contribution to the time delay can overcome the GR contribution only when $\ln(1/b)$ is very small. However, this can never happen, since by ``time delay'' we really mean delay measured by clocks in the asymptotic region away from the shock. Therefore, we should actually compare the dCS contribution with $\ln(b_0/b)$ where $b_0/b\gg1$. Hence, the term proportional to $\tilde\alpha$ in Eq.~\eqref{time_delay_dCS} will always be a small correction to the GR result in situations for which the calculations above are valid.

\section{Discussion and Conclusion}\label{sec:conclusion}

In this paper, we have revisited the assumptions of causality constraints on higher-order corrections to GR based on eikonal scattering amplitudes. We explicitly discussed a limitation associated with the perturbative expansion in the weak-gravity regime. The assumption about this regime is often overlooked in the literature and has given rise to wrong claims about the strength of constraints in modified gravity, and in particular, in dCS theory. We have revisited those claims and shown that causality arguments cannot be used to rule out modified theories, including dCS gravity, and that the current astrophysical constraints are more stringent than the ones based on flat-space amplitudes. Our arguments are based on first principles and assumptions of the eikonal approximation, and hence they are general enough to be applicable to a large class of modified theories of gravity.

We further supported the above conclusions with additional calculations of time delays experienced by light as it passes near a black hole and a star in dCS gravity, and experienced by gravitational waves scattering off a new shock-wave background also in dCS gravity. In all cases, we find that the GR prediction for the time delay is dCS corrected by terms that are suppressed by both the ratio of the naive cutoff energy scale of the theory to the characteristic mass-energy of the system, as well as by the a large power of the ratio of the mass of the object to the impact parameter. The latter, in particular, suppresses the dCS correction for all real astrophysical systems. Therefore, for the time delay to become a time advance, and thus violate causality, one would need to make the dCS coupling constant so large that one would be deeply outside the regime of validity of the effective field theory, reinforcing the conclusions presented above. 

At the core of the eikonal approximation is the fact that although it can handle transplanckian scattering (or more generically, scattering at energies larger than the cutoff of the theory considered), each individual graviton (or corresponding mediator) has subplanckian energy. This is compatible with energy-momentum conservation when $Gs/b^{D-4}>1$ (where $s$ is the square of the sum of the ingoing or outgoing momenta, $b$ is the impact parameter and $D$ is the number of spacetime dimensions), because there are many gravitons being exchanged; see discussion of momentum fractionation in \cite{Giddings:2010pp}. Hence, if we stick to the eikonal regime, where $s/t \gg 1$ (with $t$ the momentum transferred), values of $\sqrt{s}$ larger than the naive cutoff of the theory can be considered. However, when gravity is dynamical, care has to be taken due to graviton nonlinearities. 

As discussed in Sec. \ref{sec:eikonal_scope_modified_gr}, time advances in beyond Einstein theories can only occur at distance scales smaller than $1/M_\lambda$, where $M_\lambda$ is the energy scale appearing in the leading higher-derivative correction to the Einstein-Hilbert action. Thus, one may think that the study on constraints of effective field theory modifications to gravity from time advance 
considerations is obviously flawed since distances smaller than $1/M_\lambda$ are considered. This is not so because the ladder diagram resummation described in Sec. \ref{sec:eikonal} guarantees that processes at energies larger than $M_\lambda$ can be consistently considered deep in the eikonal regime. For this reason, one might neglect higher order terms and effectively consider the modified theory as ``exact'' as long as the calculations are restricted to the eikonal limit or deep in the range of validity of the theory.

Then, questioning the validity of the causality constraints becomes questioning the scope of the time advance calculation. The eikonal approximation requires the weak gravity regime, and for pure GR this means values of impact parameter $b$ much larger than the Schwarzschild radius $r_s$ associated with the process. However, as we show in this paper, this does not
extend to modified gravity because the weak gravity requirement implies additional conditions on top of $b>r_s$. Because of this, time advance constraints on the cutoff of modified
gravity theories are limited to the regime of validity of the eikonal approximation, which then renders these
constraints weaker than the current astrophysical ones.

Consider dCS gravity as an example [see Eq. \eqref{dCS_action}]. It comes with an energy scale $1/\sqrt{\alpha}$, and since it is a perturbatively non-renormalizable gravitational theory, we expect higher-derivative terms to be suppressed by higher powers of $\alpha/\kappa$. However, due to the new $\phi h h$ cubic vertex, new diagrams start to significantly contribute to gravitational scattering at energy scales $\Lambda_{\rm dCS} \sim (M_{\rm Pl}/\alpha)^{1/3}$, where $\alpha$ is the coupling constant of the theory \cite{Alexander:2021ssr}. For $1/\sqrt{\alpha} < M_{\rm Pl}$, we have $\Lambda_{\rm dCS} >1/\sqrt{\alpha}> \kappa/\alpha$. Therefore, $\kappa/\alpha$ is the cutoff for processes away from the eikonal regime, especially for subplanckian scattering\footnote{We thank Gregory Gabadadze for discussions about the dCS cutoff scale.}. Moreover, as discussed in Sec. \ref{sec:validity_causality}, to neglect graviton nonlinearities and remain within the validity of the eikonal approximation, we need $b \gg (\alpha r_{\rm s})^{1/3}$ or [see Eq. \eqref{eik_validity_dCS}],
\begin{equation}
    \frac{1}{b} \ll \left(\frac{1}{\alpha r_{\rm s}}\right)^{1/3} \sim \left(\frac{M_{\rm Pl}}{\alpha}\right)^{1/3} \left(\frac{M_{\rm Pl}}{\sqrt{s}}\right)^{1/3}.
\end{equation}
In other words, the relevant energy scale $1/b$ is much smaller than the naive cutoff $(M_{\rm Pl}/\alpha)^{1/3}$ in the transplanckian regime. 

For dCS gravity, although $b>r_s$ is not sufficient for neglecting $\mathcal{O}\left(\alpha r_s/b^3\right)$ nonlinearities, the latter might still be subleading\footnote{We thank Francesco Serra for discussions about this point.} compared to $\alpha/b^2$ when $b\gg r_s$. For this to be the case in a consistent way, the nonlinearities have to appear in the eikonal phase as an overall contribution containing a factor of $\alpha/b^2$. Even if that is the case, the time advance requires $b<\ell_{\rm CS}$ and hence we need $r_s \ll \ell_{\rm CS}$ to trust the calculation. This significantly relaxes the extent of causality constraints for dCS gravity, as discussed in Sec. \ref{sec:causality_const_dCS}. 
 
Since we discussed dCS gravity, a few comments on its ultraviolet completion are worth mentioning. The main point of the causality constraints on higher-dimensional Gauss-Bonnet gravity, presented in \cite{Camanho:2014apa}, was not to rule the theory out; rather, the main goal was to show how extra higher-spin states in string theory make higher-dimensional Gauss-Bonnet gravity unitary in the eikonal regime, \emph{for values of the impact parameter arbitrarily close to the string scale}. In some sense, this feature is one of the reasons why string theory is considered a successful perturbative, higher-dimensional quantum gravity theory. Similarly, we expect dCS gravity to be corrected and the eikonal amplitude to receive contributions from other states within a broader theory. In fact, four-dimensional effective gravity descending from heterotic strings has a dCS coupling, with $\phi$ being the zero mode of the Kalb-Rammond field in the internal manifold \cite{Alexander:2009tp}. Thus, even if the causality constraints were more stringent than the astrophysical ones, which is not the case as shown here, the best one could argue is that \emph{pure} dCS gravity is constrained. But being non-renormalizable, dCS gravity is naturally expected to be a sector within a larger theory, with the dCS coupling generated after integrating out high-energy modes. For instance, on top of the string theory example, a dCS coupling is also generated from integrating out chiral fermions \cite{Alexander:2024vav}, which can be easily embedded in extensions of the Standard Model. 

Another dCS gravity feature worth noting is the topological nature of the Pontryagin term. Since $R\Tilde{R}$ can be written as a total derivative, it only affects the equations of motion when $\phi$ is not constant. Moreover, although $R\tilde{R}$ sources $\phi$, it vanishes for backgrounds with spherical, homogeneous or isotropic isometries. Hence, in the absence of a potential to source its dynamics, $\phi$ remains a constant for many relevant metric ansatze (which is the reason why so many GR exact solutions are also valid in dCS gravity, see e.g. \cite{Jackiw:2003pm,Grumiller:2007rv,Alexander:2009tp}). Then, testing dCS gravity on Solar System scales is incredibly difficult, because not only gravity is weak, but also because $R\tilde{R}$ is dominated by matter-spin contributions which are always subleading compared to isotropic ones~\cite{Nakamura:2018yaw}. Along the same lines, Earth-based experiments, such as torsion-balance experiments (see \cite{Adelberger:2003zx} and the discussion in \cite{Cassem:2024djm}), cannot constrain dCS gravity at all since, in these setups, the dCS gravity predictions reduce to the standard GR ones.

\acknowledgments
SA thanks Dan Kabat for discussions. HB thanks Simon Caron-Huot for discussions on eikonal scattering. We also thank Robert Brandenberger, Pablo A. Cano, Simon Caron-Huot, Sera Cremonini, Cyril Creque-Sarbinowski, Gregory Gabadadze, Francesco Serra, Javi Serra, Enrico Trincherini, and Leonardo G. Trombetta for comments on a earlier version of this work. We acknowledge support from the Simons Foundation through Award No. 896696. NY also acknowledges support from the NSF through award PHY-2207650 and NASA through Grant No. 80NSSC22K0806.



\bibliography{References}

\begin{thebibliography}{72}%
\makeatletter
\providecommand \@ifxundefined [1]{%
 \@ifx{#1\undefined}
}%
\providecommand \@ifnum [1]{%
 \ifnum #1\expandafter \@firstoftwo
 \else \expandafter \@secondoftwo
 \fi
}%
\providecommand \@ifx [1]{%
 \ifx #1\expandafter \@firstoftwo
 \else \expandafter \@secondoftwo
 \fi
}%
\providecommand \natexlab [1]{#1}%
\providecommand \enquote  [1]{``#1''}%
\providecommand \bibnamefont  [1]{#1}%
\providecommand \bibfnamefont [1]{#1}%
\providecommand \citenamefont [1]{#1}%
\providecommand \href@noop [0]{\@secondoftwo}%
\providecommand \href [0]{\begingroup \@sanitize@url \@href}%
\providecommand \@href[1]{\@@startlink{#1}\@@href}%
\providecommand \@@href[1]{\endgroup#1\@@endlink}%
\providecommand \@sanitize@url [0]{\catcode `\\12\catcode `\$12\catcode `\&12\catcode `\#12\catcode `\^12\catcode `\_12\catcode `\%12\relax}%
\providecommand \@@startlink[1]{}%
\providecommand \@@endlink[0]{}%
\providecommand \url  [0]{\begingroup\@sanitize@url \@url }%
\providecommand \@url [1]{\endgroup\@href {#1}{\urlprefix }}%
\providecommand \urlprefix  [0]{URL }%
\providecommand \Eprint [0]{\href }%
\providecommand \doibase [0]{https://doi.org/}%
\providecommand \selectlanguage [0]{\@gobble}%
\providecommand \bibinfo  [0]{\@secondoftwo}%
\providecommand \bibfield  [0]{\@secondoftwo}%
\providecommand \translation [1]{[#1]}%
\providecommand \BibitemOpen [0]{}%
\providecommand \bibitemStop [0]{}%
\providecommand \bibitemNoStop [0]{.\EOS\space}%
\providecommand \EOS [0]{\spacefactor3000\relax}%
\providecommand \BibitemShut  [1]{\csname bibitem#1\endcsname}%
\let\auto@bib@innerbib\@empty
\bibitem [{\citenamefont {Donoghue}(1994)}]{Donoghue:1994dn}%
  \BibitemOpen
  \bibfield  {author} {\bibinfo {author} {\bibfnamefont {J.~F.}\ \bibnamefont {Donoghue}},\ }\bibfield  {title} {\bibinfo {title} {{General relativity as an effective field theory: The leading quantum corrections}},\ }\href {https://doi.org/10.1103/PhysRevD.50.3874} {\bibfield  {journal} {\bibinfo  {journal} {Phys. Rev. D}\ }\textbf {\bibinfo {volume} {50}},\ \bibinfo {pages} {3874} (\bibinfo {year} {1994})},\ \Eprint {https://arxiv.org/abs/gr-qc/9405057} {arXiv:gr-qc/9405057} \BibitemShut {NoStop}%
\bibitem [{\citenamefont {Burgess}(2004)}]{Burgess:2003jk}%
  \BibitemOpen
  \bibfield  {author} {\bibinfo {author} {\bibfnamefont {C.~P.}\ \bibnamefont {Burgess}},\ }\bibfield  {title} {\bibinfo {title} {{Quantum gravity in everyday life: General relativity as an effective field theory}},\ }\href {https://doi.org/10.12942/lrr-2004-5} {\bibfield  {journal} {\bibinfo  {journal} {Living Rev. Rel.}\ }\textbf {\bibinfo {volume} {7}},\ \bibinfo {pages} {5} (\bibinfo {year} {2004})},\ \Eprint {https://arxiv.org/abs/gr-qc/0311082} {arXiv:gr-qc/0311082} \BibitemShut {NoStop}%
\bibitem [{\citenamefont {Bjerrum-Bohr}\ \emph {et~al.}(2014)\citenamefont {Bjerrum-Bohr}, \citenamefont {Donoghue},\ and\ \citenamefont {Vanhove}}]{Bjerrum-Bohr:2013bxa}%
  \BibitemOpen
  \bibfield  {author} {\bibinfo {author} {\bibfnamefont {N.~E.~J.}\ \bibnamefont {Bjerrum-Bohr}}, \bibinfo {author} {\bibfnamefont {J.~F.}\ \bibnamefont {Donoghue}},\ and\ \bibinfo {author} {\bibfnamefont {P.}~\bibnamefont {Vanhove}},\ }\bibfield  {title} {\bibinfo {title} {{On-shell Techniques and Universal Results in Quantum Gravity}},\ }\href {https://doi.org/10.1007/JHEP02(2014)111} {\bibfield  {journal} {\bibinfo  {journal} {JHEP}\ }\textbf {\bibinfo {volume} {02}},\ \bibinfo {pages} {111}},\ \Eprint {https://arxiv.org/abs/1309.0804} {arXiv:1309.0804 [hep-th]} \BibitemShut {NoStop}%
\bibitem [{\citenamefont {Donoghue}\ and\ \citenamefont {Holstein}(2015)}]{Donoghue:2015hwa}%
  \BibitemOpen
  \bibfield  {author} {\bibinfo {author} {\bibfnamefont {J.~F.}\ \bibnamefont {Donoghue}}\ and\ \bibinfo {author} {\bibfnamefont {B.~R.}\ \bibnamefont {Holstein}},\ }\bibfield  {title} {\bibinfo {title} {{Low Energy Theorems of Quantum Gravity from Effective Field Theory}},\ }\href {https://doi.org/10.1088/0954-3899/42/10/103102} {\bibfield  {journal} {\bibinfo  {journal} {J. Phys. G}\ }\textbf {\bibinfo {volume} {42}},\ \bibinfo {pages} {103102} (\bibinfo {year} {2015})},\ \Eprint {https://arxiv.org/abs/1506.00946} {arXiv:1506.00946 [gr-qc]} \BibitemShut {NoStop}%
\bibitem [{\citenamefont {Donoghue}\ \emph {et~al.}(2017)\citenamefont {Donoghue}, \citenamefont {Ivanov},\ and\ \citenamefont {Shkerin}}]{Donoghue:2017pgk}%
  \BibitemOpen
  \bibfield  {author} {\bibinfo {author} {\bibfnamefont {J.~F.}\ \bibnamefont {Donoghue}}, \bibinfo {author} {\bibfnamefont {M.~M.}\ \bibnamefont {Ivanov}},\ and\ \bibinfo {author} {\bibfnamefont {A.}~\bibnamefont {Shkerin}},\ }\href@noop {} {\bibinfo {title} {{EPFL Lectures on General Relativity as a Quantum Field Theory}}} (\bibinfo {year} {2017}),\ \Eprint {https://arxiv.org/abs/1702.00319} {arXiv:1702.00319 [hep-th]} \BibitemShut {NoStop}%
\bibitem [{\citenamefont {'t~Hooft}(1987)}]{tHooft:1987vrq}%
  \BibitemOpen
  \bibfield  {author} {\bibinfo {author} {\bibfnamefont {G.}~\bibnamefont {'t~Hooft}},\ }\bibfield  {title} {\bibinfo {title} {{Graviton Dominance in Ultrahigh-Energy Scattering}},\ }\href {https://doi.org/10.1016/0370-2693(87)90159-6} {\bibfield  {journal} {\bibinfo  {journal} {Phys. Lett. B}\ }\textbf {\bibinfo {volume} {198}},\ \bibinfo {pages} {61} (\bibinfo {year} {1987})}\BibitemShut {NoStop}%
\bibitem [{\citenamefont {Dyson}\ \emph {et~al.}(1920)\citenamefont {Dyson}, \citenamefont {Eddington},\ and\ \citenamefont {Davidson}}]{Dyson:1920cwa}%
  \BibitemOpen
  \bibfield  {author} {\bibinfo {author} {\bibfnamefont {F.~W.}\ \bibnamefont {Dyson}}, \bibinfo {author} {\bibfnamefont {A.~S.}\ \bibnamefont {Eddington}},\ and\ \bibinfo {author} {\bibfnamefont {C.}~\bibnamefont {Davidson}},\ }\bibfield  {title} {\bibinfo {title} {{A Determination of the Deflection of Light by the Sun's Gravitational Field, from Observations Made at the Total Eclipse of May 29, 1919}},\ }\href {https://doi.org/10.1098/rsta.1920.0009} {\bibfield  {journal} {\bibinfo  {journal} {Phil. Trans. Roy. Soc. Lond. A}\ }\textbf {\bibinfo {volume} {220}},\ \bibinfo {pages} {291} (\bibinfo {year} {1920})}\BibitemShut {NoStop}%
\bibitem [{\citenamefont {Bertotti}\ \emph {et~al.}(2003)\citenamefont {Bertotti}, \citenamefont {Iess},\ and\ \citenamefont {Tortora}}]{Bertotti:2003rm}%
  \BibitemOpen
  \bibfield  {author} {\bibinfo {author} {\bibfnamefont {B.}~\bibnamefont {Bertotti}}, \bibinfo {author} {\bibfnamefont {L.}~\bibnamefont {Iess}},\ and\ \bibinfo {author} {\bibfnamefont {P.}~\bibnamefont {Tortora}},\ }\bibfield  {title} {\bibinfo {title} {{A test of general relativity using radio links with the Cassini spacecraft}},\ }\href {https://doi.org/10.1038/nature01997} {\bibfield  {journal} {\bibinfo  {journal} {Nature}\ }\textbf {\bibinfo {volume} {425}},\ \bibinfo {pages} {374} (\bibinfo {year} {2003})}\BibitemShut {NoStop}%
\bibitem [{\citenamefont {Eisenbud}(1948)}]{Eisenbud:1948paa}%
  \BibitemOpen
  \bibfield  {author} {\bibinfo {author} {\bibfnamefont {L.}~\bibnamefont {Eisenbud}},\ }\emph {\bibinfo {title} {{The formal properties of nuclear collisions}}},\ \href@noop {} {Ph.D. thesis},\ \bibinfo  {school} {Princeton U.} (\bibinfo {year} {1948})\BibitemShut {NoStop}%
\bibitem [{\citenamefont {Wigner}(1955)}]{Wigner:1955zz}%
  \BibitemOpen
  \bibfield  {author} {\bibinfo {author} {\bibfnamefont {E.~P.}\ \bibnamefont {Wigner}},\ }\bibfield  {title} {\bibinfo {title} {{Lower Limit for the Energy Derivative of the Scattering Phase Shift}},\ }\href {https://doi.org/10.1103/PhysRev.98.145} {\bibfield  {journal} {\bibinfo  {journal} {Phys. Rev.}\ }\textbf {\bibinfo {volume} {98}},\ \bibinfo {pages} {145} (\bibinfo {year} {1955})}\BibitemShut {NoStop}%
\bibitem [{\citenamefont {Penrose}\ and\ \citenamefont {Tipler}(1980)}]{penrose1980schwarzschild}%
  \BibitemOpen
  \bibfield  {author} {\bibinfo {author} {\bibfnamefont {R.}~\bibnamefont {Penrose}}\ and\ \bibinfo {author} {\bibfnamefont {F.}~\bibnamefont {Tipler}},\ }\href@noop {} {\emph {\bibinfo {title} {On Schwarzschild causality----a problem for “Lorentz covariant” general relativity}}}\ (\bibinfo  {publisher} {Academic New York},\ \bibinfo {year} {1980})\BibitemShut {NoStop}%
\bibitem [{\citenamefont {Gao}\ and\ \citenamefont {Wald}(2000)}]{Gao:2000ga}%
  \BibitemOpen
  \bibfield  {author} {\bibinfo {author} {\bibfnamefont {S.}~\bibnamefont {Gao}}\ and\ \bibinfo {author} {\bibfnamefont {R.~M.}\ \bibnamefont {Wald}},\ }\bibfield  {title} {\bibinfo {title} {{Theorems on gravitational time delay and related issues}},\ }\href {https://doi.org/10.1088/0264-9381/17/24/305} {\bibfield  {journal} {\bibinfo  {journal} {Class. Quant. Grav.}\ }\textbf {\bibinfo {volume} {17}},\ \bibinfo {pages} {4999} (\bibinfo {year} {2000})},\ \Eprint {https://arxiv.org/abs/gr-qc/0007021} {arXiv:gr-qc/0007021} \BibitemShut {NoStop}%
\bibitem [{\citenamefont {Camanho}\ \emph {et~al.}(2016)\citenamefont {Camanho}, \citenamefont {Edelstein}, \citenamefont {Maldacena},\ and\ \citenamefont {Zhiboedov}}]{Camanho:2014apa}%
  \BibitemOpen
  \bibfield  {author} {\bibinfo {author} {\bibfnamefont {X.~O.}\ \bibnamefont {Camanho}}, \bibinfo {author} {\bibfnamefont {J.~D.}\ \bibnamefont {Edelstein}}, \bibinfo {author} {\bibfnamefont {J.}~\bibnamefont {Maldacena}},\ and\ \bibinfo {author} {\bibfnamefont {A.}~\bibnamefont {Zhiboedov}},\ }\bibfield  {title} {\bibinfo {title} {{Causality Constraints on Corrections to the Graviton Three-Point Coupling}},\ }\href {https://doi.org/10.1007/JHEP02(2016)020} {\bibfield  {journal} {\bibinfo  {journal} {JHEP}\ }\textbf {\bibinfo {volume} {02}},\ \bibinfo {pages} {020}},\ \Eprint {https://arxiv.org/abs/1407.5597} {arXiv:1407.5597 [hep-th]} \BibitemShut {NoStop}%
\bibitem [{\citenamefont {Camanho}\ \emph {et~al.}(2017)\citenamefont {Camanho}, \citenamefont {Lucena~G\'omez},\ and\ \citenamefont {Rahman}}]{Camanho:2016opx}%
  \BibitemOpen
  \bibfield  {author} {\bibinfo {author} {\bibfnamefont {X.~O.}\ \bibnamefont {Camanho}}, \bibinfo {author} {\bibfnamefont {G.}~\bibnamefont {Lucena~G\'omez}},\ and\ \bibinfo {author} {\bibfnamefont {R.}~\bibnamefont {Rahman}},\ }\bibfield  {title} {\bibinfo {title} {{Causality Constraints on Massive Gravity}},\ }\href {https://doi.org/10.1103/PhysRevD.96.084007} {\bibfield  {journal} {\bibinfo  {journal} {Phys. Rev. D}\ }\textbf {\bibinfo {volume} {96}},\ \bibinfo {pages} {084007} (\bibinfo {year} {2017})},\ \Eprint {https://arxiv.org/abs/1610.02033} {arXiv:1610.02033 [hep-th]} \BibitemShut {NoStop}%
\bibitem [{\citenamefont {Goon}\ and\ \citenamefont {Hinterbichler}(2017)}]{Goon:2016une}%
  \BibitemOpen
  \bibfield  {author} {\bibinfo {author} {\bibfnamefont {G.}~\bibnamefont {Goon}}\ and\ \bibinfo {author} {\bibfnamefont {K.}~\bibnamefont {Hinterbichler}},\ }\bibfield  {title} {\bibinfo {title} {{Superluminality, black holes and EFT}},\ }\href {https://doi.org/10.1007/JHEP02(2017)134} {\bibfield  {journal} {\bibinfo  {journal} {JHEP}\ }\textbf {\bibinfo {volume} {02}},\ \bibinfo {pages} {134}},\ \Eprint {https://arxiv.org/abs/1609.00723} {arXiv:1609.00723 [hep-th]} \BibitemShut {NoStop}%
\bibitem [{\citenamefont {Hinterbichler}\ \emph {et~al.}(2018)\citenamefont {Hinterbichler}, \citenamefont {Joyce},\ and\ \citenamefont {Rosen}}]{Hinterbichler:2017qyt}%
  \BibitemOpen
  \bibfield  {author} {\bibinfo {author} {\bibfnamefont {K.}~\bibnamefont {Hinterbichler}}, \bibinfo {author} {\bibfnamefont {A.}~\bibnamefont {Joyce}},\ and\ \bibinfo {author} {\bibfnamefont {R.~A.}\ \bibnamefont {Rosen}},\ }\bibfield  {title} {\bibinfo {title} {{Massive Spin-2 Scattering and Asymptotic Superluminality}},\ }\href {https://doi.org/10.1007/JHEP03(2018)051} {\bibfield  {journal} {\bibinfo  {journal} {JHEP}\ }\textbf {\bibinfo {volume} {03}},\ \bibinfo {pages} {051}},\ \Eprint {https://arxiv.org/abs/1708.05716} {arXiv:1708.05716 [hep-th]} \BibitemShut {NoStop}%
\bibitem [{\citenamefont {Bellazzini}\ \emph {et~al.}(2016)\citenamefont {Bellazzini}, \citenamefont {Cheung},\ and\ \citenamefont {Remmen}}]{Bellazzini:2015cra}%
  \BibitemOpen
  \bibfield  {author} {\bibinfo {author} {\bibfnamefont {B.}~\bibnamefont {Bellazzini}}, \bibinfo {author} {\bibfnamefont {C.}~\bibnamefont {Cheung}},\ and\ \bibinfo {author} {\bibfnamefont {G.~N.}\ \bibnamefont {Remmen}},\ }\bibfield  {title} {\bibinfo {title} {{Quantum Gravity Constraints from Unitarity and Analyticity}},\ }\href {https://doi.org/10.1103/PhysRevD.93.064076} {\bibfield  {journal} {\bibinfo  {journal} {Phys. Rev. D}\ }\textbf {\bibinfo {volume} {93}},\ \bibinfo {pages} {064076} (\bibinfo {year} {2016})},\ \Eprint {https://arxiv.org/abs/1509.00851} {arXiv:1509.00851 [hep-th]} \BibitemShut {NoStop}%
\bibitem [{\citenamefont {de~Rham}\ and\ \citenamefont {Tolley}(2020{\natexlab{a}})}]{deRham:2019ctd}%
  \BibitemOpen
  \bibfield  {author} {\bibinfo {author} {\bibfnamefont {C.}~\bibnamefont {de~Rham}}\ and\ \bibinfo {author} {\bibfnamefont {A.~J.}\ \bibnamefont {Tolley}},\ }\bibfield  {title} {\bibinfo {title} {{Speed of gravity}},\ }\href {https://doi.org/10.1103/PhysRevD.101.063518} {\bibfield  {journal} {\bibinfo  {journal} {Phys. Rev. D}\ }\textbf {\bibinfo {volume} {101}},\ \bibinfo {pages} {063518} (\bibinfo {year} {2020}{\natexlab{a}})},\ \Eprint {https://arxiv.org/abs/1909.00881} {arXiv:1909.00881 [hep-th]} \BibitemShut {NoStop}%
\bibitem [{\citenamefont {de~Rham}\ and\ \citenamefont {Tolley}(2020{\natexlab{b}})}]{deRham:2020zyh}%
  \BibitemOpen
  \bibfield  {author} {\bibinfo {author} {\bibfnamefont {C.}~\bibnamefont {de~Rham}}\ and\ \bibinfo {author} {\bibfnamefont {A.~J.}\ \bibnamefont {Tolley}},\ }\bibfield  {title} {\bibinfo {title} {{Causality in curved spacetimes: The speed of light and gravity}},\ }\href {https://doi.org/10.1103/PhysRevD.102.084048} {\bibfield  {journal} {\bibinfo  {journal} {Phys. Rev. D}\ }\textbf {\bibinfo {volume} {102}},\ \bibinfo {pages} {084048} (\bibinfo {year} {2020}{\natexlab{b}})},\ \Eprint {https://arxiv.org/abs/2007.01847} {arXiv:2007.01847 [hep-th]} \BibitemShut {NoStop}%
\bibitem [{\citenamefont {Tokuda}\ \emph {et~al.}(2020)\citenamefont {Tokuda}, \citenamefont {Aoki},\ and\ \citenamefont {Hirano}}]{Tokuda:2020mlf}%
  \BibitemOpen
  \bibfield  {author} {\bibinfo {author} {\bibfnamefont {J.}~\bibnamefont {Tokuda}}, \bibinfo {author} {\bibfnamefont {K.}~\bibnamefont {Aoki}},\ and\ \bibinfo {author} {\bibfnamefont {S.}~\bibnamefont {Hirano}},\ }\bibfield  {title} {\bibinfo {title} {{Gravitational positivity bounds}},\ }\href {https://doi.org/10.1007/JHEP11(2020)054} {\bibfield  {journal} {\bibinfo  {journal} {JHEP}\ }\textbf {\bibinfo {volume} {11}},\ \bibinfo {pages} {054}},\ \Eprint {https://arxiv.org/abs/2007.15009} {arXiv:2007.15009 [hep-th]} \BibitemShut {NoStop}%
\bibitem [{\citenamefont {Accettulli~Huber}\ \emph {et~al.}(2020)\citenamefont {Accettulli~Huber}, \citenamefont {Brandhuber}, \citenamefont {De~Angelis},\ and\ \citenamefont {Travaglini}}]{AccettulliHuber:2020oou}%
  \BibitemOpen
  \bibfield  {author} {\bibinfo {author} {\bibfnamefont {M.}~\bibnamefont {Accettulli~Huber}}, \bibinfo {author} {\bibfnamefont {A.}~\bibnamefont {Brandhuber}}, \bibinfo {author} {\bibfnamefont {S.}~\bibnamefont {De~Angelis}},\ and\ \bibinfo {author} {\bibfnamefont {G.}~\bibnamefont {Travaglini}},\ }\bibfield  {title} {\bibinfo {title} {{Eikonal phase matrix, deflection angle and time delay in effective field theories of gravity}},\ }\href {https://doi.org/10.1103/PhysRevD.102.046014} {\bibfield  {journal} {\bibinfo  {journal} {Phys. Rev. D}\ }\textbf {\bibinfo {volume} {102}},\ \bibinfo {pages} {046014} (\bibinfo {year} {2020})},\ \Eprint {https://arxiv.org/abs/2006.02375} {arXiv:2006.02375 [hep-th]} \BibitemShut {NoStop}%
\bibitem [{\citenamefont {Chen}\ \emph {et~al.}(2022)\citenamefont {Chen}, \citenamefont {de~Rham}, \citenamefont {Margalit},\ and\ \citenamefont {Tolley}}]{Chen:2021bvg}%
  \BibitemOpen
  \bibfield  {author} {\bibinfo {author} {\bibfnamefont {C.~Y.~R.}\ \bibnamefont {Chen}}, \bibinfo {author} {\bibfnamefont {C.}~\bibnamefont {de~Rham}}, \bibinfo {author} {\bibfnamefont {A.}~\bibnamefont {Margalit}},\ and\ \bibinfo {author} {\bibfnamefont {A.~J.}\ \bibnamefont {Tolley}},\ }\bibfield  {title} {\bibinfo {title} {{A cautionary case of casual causality}},\ }\href {https://doi.org/10.1007/JHEP03(2022)025} {\bibfield  {journal} {\bibinfo  {journal} {JHEP}\ }\textbf {\bibinfo {volume} {03}},\ \bibinfo {pages} {025}},\ \Eprint {https://arxiv.org/abs/2112.05031} {arXiv:2112.05031 [hep-th]} \BibitemShut {NoStop}%
\bibitem [{\citenamefont {Caron-Huot}\ \emph {et~al.}(2021)\citenamefont {Caron-Huot}, \citenamefont {Mazac}, \citenamefont {Rastelli},\ and\ \citenamefont {Simmons-Duffin}}]{Caron-Huot:2021rmr}%
  \BibitemOpen
  \bibfield  {author} {\bibinfo {author} {\bibfnamefont {S.}~\bibnamefont {Caron-Huot}}, \bibinfo {author} {\bibfnamefont {D.}~\bibnamefont {Mazac}}, \bibinfo {author} {\bibfnamefont {L.}~\bibnamefont {Rastelli}},\ and\ \bibinfo {author} {\bibfnamefont {D.}~\bibnamefont {Simmons-Duffin}},\ }\bibfield  {title} {\bibinfo {title} {{Sharp boundaries for the swampland}},\ }\href {https://doi.org/10.1007/JHEP07(2021)110} {\bibfield  {journal} {\bibinfo  {journal} {JHEP}\ }\textbf {\bibinfo {volume} {07}},\ \bibinfo {pages} {110}},\ \Eprint {https://arxiv.org/abs/2102.08951} {arXiv:2102.08951 [hep-th]} \BibitemShut {NoStop}%
\bibitem [{\citenamefont {Caron-Huot}\ \emph {et~al.}(2023)\citenamefont {Caron-Huot}, \citenamefont {Li}, \citenamefont {Parra-Martinez},\ and\ \citenamefont {Simmons-Duffin}}]{Caron-Huot:2022ugt}%
  \BibitemOpen
  \bibfield  {author} {\bibinfo {author} {\bibfnamefont {S.}~\bibnamefont {Caron-Huot}}, \bibinfo {author} {\bibfnamefont {Y.-Z.}\ \bibnamefont {Li}}, \bibinfo {author} {\bibfnamefont {J.}~\bibnamefont {Parra-Martinez}},\ and\ \bibinfo {author} {\bibfnamefont {D.}~\bibnamefont {Simmons-Duffin}},\ }\bibfield  {title} {\bibinfo {title} {{Causality constraints on corrections to Einstein gravity}},\ }\href {https://doi.org/10.1007/JHEP05(2023)122} {\bibfield  {journal} {\bibinfo  {journal} {JHEP}\ }\textbf {\bibinfo {volume} {05}},\ \bibinfo {pages} {122}},\ \Eprint {https://arxiv.org/abs/2201.06602} {arXiv:2201.06602 [hep-th]} \BibitemShut {NoStop}%
\bibitem [{\citenamefont {Hong}\ \emph {et~al.}(2023)\citenamefont {Hong}, \citenamefont {Wang},\ and\ \citenamefont {Zhou}}]{Hong:2023zgm}%
  \BibitemOpen
  \bibfield  {author} {\bibinfo {author} {\bibfnamefont {D.-Y.}\ \bibnamefont {Hong}}, \bibinfo {author} {\bibfnamefont {Z.-H.}\ \bibnamefont {Wang}},\ and\ \bibinfo {author} {\bibfnamefont {S.-Y.}\ \bibnamefont {Zhou}},\ }\bibfield  {title} {\bibinfo {title} {{Causality bounds on scalar-tensor EFTs}},\ }\href {https://doi.org/10.1007/JHEP10(2023)135} {\bibfield  {journal} {\bibinfo  {journal} {JHEP}\ }\textbf {\bibinfo {volume} {10}},\ \bibinfo {pages} {135}},\ \Eprint {https://arxiv.org/abs/2304.01259} {arXiv:2304.01259 [hep-th]} \BibitemShut {NoStop}%
\bibitem [{\citenamefont {Aoki}\ \emph {et~al.}(2024)\citenamefont {Aoki}, \citenamefont {Noumi}, \citenamefont {Saito}, \citenamefont {Sato}, \citenamefont {Shirai}, \citenamefont {Tokuda},\ and\ \citenamefont {Yamazaki}}]{Aoki:2023khq}%
  \BibitemOpen
  \bibfield  {author} {\bibinfo {author} {\bibfnamefont {K.}~\bibnamefont {Aoki}}, \bibinfo {author} {\bibfnamefont {T.}~\bibnamefont {Noumi}}, \bibinfo {author} {\bibfnamefont {R.}~\bibnamefont {Saito}}, \bibinfo {author} {\bibfnamefont {S.}~\bibnamefont {Sato}}, \bibinfo {author} {\bibfnamefont {S.}~\bibnamefont {Shirai}}, \bibinfo {author} {\bibfnamefont {J.}~\bibnamefont {Tokuda}},\ and\ \bibinfo {author} {\bibfnamefont {M.}~\bibnamefont {Yamazaki}},\ }\bibfield  {title} {\bibinfo {title} {{Gravitational positivity for phenomenologists: Dark gauge boson in the swampland}},\ }\href {https://doi.org/10.1103/PhysRevD.110.016002} {\bibfield  {journal} {\bibinfo  {journal} {Phys. Rev. D}\ }\textbf {\bibinfo {volume} {110}},\ \bibinfo {pages} {016002} (\bibinfo {year} {2024})},\ \Eprint {https://arxiv.org/abs/2305.10058} {arXiv:2305.10058 [hep-ph]} \BibitemShut {NoStop}%
\bibitem [{\citenamefont {Caron-Huot}\ and\ \citenamefont {Tokuda}(2024)}]{Caron-Huot:2024tsk}%
  \BibitemOpen
  \bibfield  {author} {\bibinfo {author} {\bibfnamefont {S.}~\bibnamefont {Caron-Huot}}\ and\ \bibinfo {author} {\bibfnamefont {J.}~\bibnamefont {Tokuda}},\ }\bibfield  {title} {\bibinfo {title} {{String loops and gravitational positivity bounds: imprint of light particles at high energies}},\ }\href {https://doi.org/10.1007/JHEP11(2024)055} {\bibfield  {journal} {\bibinfo  {journal} {JHEP}\ }\textbf {\bibinfo {volume} {11}},\ \bibinfo {pages} {055}},\ \Eprint {https://arxiv.org/abs/2406.07606} {arXiv:2406.07606 [hep-th]} \BibitemShut {NoStop}%
\bibitem [{\citenamefont {de~Rham}\ \emph {et~al.}(2022)\citenamefont {de~Rham}, \citenamefont {Kundu}, \citenamefont {Reece}, \citenamefont {Tolley},\ and\ \citenamefont {Zhou}}]{deRham:2022hpx}%
  \BibitemOpen
  \bibfield  {author} {\bibinfo {author} {\bibfnamefont {C.}~\bibnamefont {de~Rham}}, \bibinfo {author} {\bibfnamefont {S.}~\bibnamefont {Kundu}}, \bibinfo {author} {\bibfnamefont {M.}~\bibnamefont {Reece}}, \bibinfo {author} {\bibfnamefont {A.~J.}\ \bibnamefont {Tolley}},\ and\ \bibinfo {author} {\bibfnamefont {S.-Y.}\ \bibnamefont {Zhou}},\ }\bibfield  {title} {\bibinfo {title} {{Snowmass White Paper: UV Constraints on IR Physics}},\ }in\ \href@noop {} {\emph {\bibinfo {booktitle} {{Snowmass 2021}}}}\ (\bibinfo {year} {2022})\ \Eprint {https://arxiv.org/abs/2203.06805} {arXiv:2203.06805 [hep-th]} \BibitemShut {NoStop}%
\bibitem [{\citenamefont {Serra}\ \emph {et~al.}(2022)\citenamefont {Serra}, \citenamefont {Serra}, \citenamefont {Trincherini},\ and\ \citenamefont {Trombetta}}]{Serra:2022pzl}%
  \BibitemOpen
  \bibfield  {author} {\bibinfo {author} {\bibfnamefont {F.}~\bibnamefont {Serra}}, \bibinfo {author} {\bibfnamefont {J.}~\bibnamefont {Serra}}, \bibinfo {author} {\bibfnamefont {E.}~\bibnamefont {Trincherini}},\ and\ \bibinfo {author} {\bibfnamefont {L.~G.}\ \bibnamefont {Trombetta}},\ }\bibfield  {title} {\bibinfo {title} {{Causality constraints on black holes beyond GR}},\ }\href {https://doi.org/10.1007/JHEP08(2022)157} {\bibfield  {journal} {\bibinfo  {journal} {JHEP}\ }\textbf {\bibinfo {volume} {08}},\ \bibinfo {pages} {157}},\ \Eprint {https://arxiv.org/abs/2205.08551} {arXiv:2205.08551 [hep-th]} \BibitemShut {NoStop}%
\bibitem [{\citenamefont {Cremonini}\ \emph {et~al.}(2024{\natexlab{a}})\citenamefont {Cremonini}, \citenamefont {McPeak},\ and\ \citenamefont {Tang}}]{Cremonini:2023epg}%
  \BibitemOpen
  \bibfield  {author} {\bibinfo {author} {\bibfnamefont {S.}~\bibnamefont {Cremonini}}, \bibinfo {author} {\bibfnamefont {B.}~\bibnamefont {McPeak}},\ and\ \bibinfo {author} {\bibfnamefont {Y.}~\bibnamefont {Tang}},\ }\bibfield  {title} {\bibinfo {title} {{Electric shocks: bounding Einstein-Maxwell theory with time delays on boosted RN backgrounds}},\ }\href {https://doi.org/10.1007/JHEP05(2024)192} {\bibfield  {journal} {\bibinfo  {journal} {JHEP}\ }\textbf {\bibinfo {volume} {05}},\ \bibinfo {pages} {192}},\ \Eprint {https://arxiv.org/abs/2312.17328} {arXiv:2312.17328 [hep-th]} \BibitemShut {NoStop}%
\bibitem [{\citenamefont {Kehagias}\ and\ \citenamefont {Riotto}(2024)}]{Kehagias:2024yyp}%
  \BibitemOpen
  \bibfield  {author} {\bibinfo {author} {\bibfnamefont {A.}~\bibnamefont {Kehagias}}\ and\ \bibinfo {author} {\bibfnamefont {A.}~\bibnamefont {Riotto}},\ }\href@noop {} {\bibinfo {title} {{Can We Detect Deviations from Einstein's Gravity in Black Hole Ringdowns?}}} (\bibinfo {year} {2024}),\ \Eprint {https://arxiv.org/abs/2411.12428} {arXiv:2411.12428 [gr-qc]} \BibitemShut {NoStop}%
\bibitem [{\citenamefont {Cassem}\ and\ \citenamefont {Hertzberg}(2024)}]{Cassem:2024djm}%
  \BibitemOpen
  \bibfield  {author} {\bibinfo {author} {\bibfnamefont {A.}~\bibnamefont {Cassem}}\ and\ \bibinfo {author} {\bibfnamefont {M.~P.}\ \bibnamefont {Hertzberg}},\ }\href@noop {} {\bibinfo {title} {{Unitarity, Causality, and Solar System Bounds, Significantly Limit Using Gravitational Waves to Test General Relativity}}} (\bibinfo {year} {2024}),\ \Eprint {https://arxiv.org/abs/2408.12118} {arXiv:2408.12118 [hep-th]} \BibitemShut {NoStop}%
\bibitem [{\citenamefont {Xu}\ \emph {et~al.}(2025)\citenamefont {Xu}, \citenamefont {Hong}, \citenamefont {Wang},\ and\ \citenamefont {Zhou}}]{Xu:2024iao}%
  \BibitemOpen
  \bibfield  {author} {\bibinfo {author} {\bibfnamefont {H.}~\bibnamefont {Xu}}, \bibinfo {author} {\bibfnamefont {D.-Y.}\ \bibnamefont {Hong}}, \bibinfo {author} {\bibfnamefont {Z.-H.}\ \bibnamefont {Wang}},\ and\ \bibinfo {author} {\bibfnamefont {S.-Y.}\ \bibnamefont {Zhou}},\ }\bibfield  {title} {\bibinfo {title} {{Positivity bounds on parity-violating scalar-tensor EFTs}},\ }\href {https://doi.org/10.1088/1475-7516/2025/01/102} {\bibfield  {journal} {\bibinfo  {journal} {JCAP}\ }\textbf {\bibinfo {volume} {01}},\ \bibinfo {pages} {102}},\ \Eprint {https://arxiv.org/abs/2410.09794} {arXiv:2410.09794 [hep-th]} \BibitemShut {NoStop}%
\bibitem [{\citenamefont {Cremonini}\ \emph {et~al.}(2024{\natexlab{b}})\citenamefont {Cremonini}, \citenamefont {McPeak}, \citenamefont {Moezzi},\ and\ \citenamefont {Rajaguru}}]{Cremonini:2024lxn}%
  \BibitemOpen
  \bibfield  {author} {\bibinfo {author} {\bibfnamefont {S.}~\bibnamefont {Cremonini}}, \bibinfo {author} {\bibfnamefont {B.}~\bibnamefont {McPeak}}, \bibinfo {author} {\bibfnamefont {M.}~\bibnamefont {Moezzi}},\ and\ \bibinfo {author} {\bibfnamefont {M.}~\bibnamefont {Rajaguru}},\ }\href@noop {} {\bibinfo {title} {{Causality bounds from charged shockwaves in 5d}}} (\bibinfo {year} {2024}{\natexlab{b}}),\ \Eprint {https://arxiv.org/abs/2412.06891} {arXiv:2412.06891 [hep-th]} \BibitemShut {NoStop}%
\bibitem [{\citenamefont {Yunes}\ \emph {et~al.}(2025)\citenamefont {Yunes}, \citenamefont {Siemens},\ and\ \citenamefont {Yagi}}]{Yunes2025}%
  \BibitemOpen
  \bibfield  {author} {\bibinfo {author} {\bibfnamefont {N.}~\bibnamefont {Yunes}}, \bibinfo {author} {\bibfnamefont {X.}~\bibnamefont {Siemens}},\ and\ \bibinfo {author} {\bibfnamefont {K.}~\bibnamefont {Yagi}},\ }\bibfield  {title} {\bibinfo {title} {Gravitational-wave tests of general relativity with ground-based detectors and pulsar-timing arrays},\ }\href {https://doi.org/10.1007/s41114-024-00054-9} {\bibfield  {journal} {\bibinfo  {journal} {Living Reviews in Relativity}\ }\textbf {\bibinfo {volume} {28}},\ \bibinfo {pages} {3} (\bibinfo {year} {2025})}\BibitemShut {NoStop}%
\bibitem [{\citenamefont {Bellucci}\ and\ \citenamefont {Gates}(1988)}]{Bellucci:1988ff}%
  \BibitemOpen
  \bibfield  {author} {\bibinfo {author} {\bibfnamefont {S.}~\bibnamefont {Bellucci}}\ and\ \bibinfo {author} {\bibfnamefont {S.~J.}\ \bibnamefont {Gates}, \bibfnamefont {Jr.}},\ }\bibfield  {title} {\bibinfo {title} {{$D=10$, $N=1$ Superspace Supergravity and the Lorentz Chern-simons Form}},\ }\href {https://doi.org/10.1016/0370-2693(88)90647-8} {\bibfield  {journal} {\bibinfo  {journal} {Phys. Lett. B}\ }\textbf {\bibinfo {volume} {208}},\ \bibinfo {pages} {456} (\bibinfo {year} {1988})}\BibitemShut {NoStop}%
\bibitem [{\citenamefont {Bellucci}\ \emph {et~al.}(1990)\citenamefont {Bellucci}, \citenamefont {Depireux},\ and\ \citenamefont {Gates}}]{Bellucci:1990fa}%
  \BibitemOpen
  \bibfield  {author} {\bibinfo {author} {\bibfnamefont {S.}~\bibnamefont {Bellucci}}, \bibinfo {author} {\bibfnamefont {D.~A.}\ \bibnamefont {Depireux}},\ and\ \bibinfo {author} {\bibfnamefont {S.~J.}\ \bibnamefont {Gates}, \bibfnamefont {Jr.}},\ }\bibfield  {title} {\bibinfo {title} {{Consistent and Universal Inclusion of the Lorentz {Chern-Simons} Form in $D=10$, $N=1$ Supergravity Theories}},\ }\href {https://doi.org/10.1016/0370-2693(90)91741-S} {\bibfield  {journal} {\bibinfo  {journal} {Phys. Lett. B}\ }\textbf {\bibinfo {volume} {238}},\ \bibinfo {pages} {315} (\bibinfo {year} {1990})}\BibitemShut {NoStop}%
\bibitem [{\citenamefont {Gates}\ and\ \citenamefont {Nishino}(1992)}]{Gates:1991qn}%
  \BibitemOpen
  \bibfield  {author} {\bibinfo {author} {\bibfnamefont {S.~J.}\ \bibnamefont {Gates}, \bibfnamefont {Jr.}}\ and\ \bibinfo {author} {\bibfnamefont {H.}~\bibnamefont {Nishino}},\ }\bibfield  {title} {\bibinfo {title} {{Remarks on the N=2 supersymmetric Chern-Simons theories}},\ }\href {https://doi.org/10.1016/0370-2693(92)90277-B} {\bibfield  {journal} {\bibinfo  {journal} {Phys. Lett. B}\ }\textbf {\bibinfo {volume} {281}},\ \bibinfo {pages} {72} (\bibinfo {year} {1992})}\BibitemShut {NoStop}%
\bibitem [{\citenamefont {Lue}\ \emph {et~al.}(1999)\citenamefont {Lue}, \citenamefont {Wang},\ and\ \citenamefont {Kamionkowski}}]{Lue:1998mq}%
  \BibitemOpen
  \bibfield  {author} {\bibinfo {author} {\bibfnamefont {A.}~\bibnamefont {Lue}}, \bibinfo {author} {\bibfnamefont {L.-M.}\ \bibnamefont {Wang}},\ and\ \bibinfo {author} {\bibfnamefont {M.}~\bibnamefont {Kamionkowski}},\ }\bibfield  {title} {\bibinfo {title} {{Cosmological signature of new parity violating interactions}},\ }\href {https://doi.org/10.1103/PhysRevLett.83.1506} {\bibfield  {journal} {\bibinfo  {journal} {Phys. Rev. Lett.}\ }\textbf {\bibinfo {volume} {83}},\ \bibinfo {pages} {1506} (\bibinfo {year} {1999})},\ \Eprint {https://arxiv.org/abs/astro-ph/9812088} {arXiv:astro-ph/9812088} \BibitemShut {NoStop}%
\bibitem [{\citenamefont {Choi}\ \emph {et~al.}(2000)\citenamefont {Choi}, \citenamefont {Hwang},\ and\ \citenamefont {Hwang}}]{Choi:1999zy}%
  \BibitemOpen
  \bibfield  {author} {\bibinfo {author} {\bibfnamefont {K.}~\bibnamefont {Choi}}, \bibinfo {author} {\bibfnamefont {J.-c.}\ \bibnamefont {Hwang}},\ and\ \bibinfo {author} {\bibfnamefont {K.~W.}\ \bibnamefont {Hwang}},\ }\bibfield  {title} {\bibinfo {title} {{String theoretic axion coupling and the evolution of cosmic structures}},\ }\href {https://doi.org/10.1103/PhysRevD.61.084026} {\bibfield  {journal} {\bibinfo  {journal} {Phys. Rev. D}\ }\textbf {\bibinfo {volume} {61}},\ \bibinfo {pages} {084026} (\bibinfo {year} {2000})},\ \Eprint {https://arxiv.org/abs/hep-ph/9907244} {arXiv:hep-ph/9907244} \BibitemShut {NoStop}%
\bibitem [{\citenamefont {Jackiw}\ and\ \citenamefont {Pi}(2003)}]{Jackiw:2003pm}%
  \BibitemOpen
  \bibfield  {author} {\bibinfo {author} {\bibfnamefont {R.}~\bibnamefont {Jackiw}}\ and\ \bibinfo {author} {\bibfnamefont {S.~Y.}\ \bibnamefont {Pi}},\ }\bibfield  {title} {\bibinfo {title} {{Chern-Simons modification of general relativity}},\ }\href {https://doi.org/10.1103/PhysRevD.68.104012} {\bibfield  {journal} {\bibinfo  {journal} {Phys. Rev. D}\ }\textbf {\bibinfo {volume} {68}},\ \bibinfo {pages} {104012} (\bibinfo {year} {2003})},\ \Eprint {https://arxiv.org/abs/gr-qc/0308071} {arXiv:gr-qc/0308071} \BibitemShut {NoStop}%
\bibitem [{\citenamefont {Alexander}\ and\ \citenamefont {Yunes}(2009)}]{Alexander:2009tp}%
  \BibitemOpen
  \bibfield  {author} {\bibinfo {author} {\bibfnamefont {S.}~\bibnamefont {Alexander}}\ and\ \bibinfo {author} {\bibfnamefont {N.}~\bibnamefont {Yunes}},\ }\bibfield  {title} {\bibinfo {title} {{Chern-Simons Modified General Relativity}},\ }\href {https://doi.org/10.1016/j.physrep.2009.07.002} {\bibfield  {journal} {\bibinfo  {journal} {Phys. Rept.}\ }\textbf {\bibinfo {volume} {480}},\ \bibinfo {pages} {1} (\bibinfo {year} {2009})},\ \Eprint {https://arxiv.org/abs/0907.2562} {arXiv:0907.2562 [hep-th]} \BibitemShut {NoStop}%
\bibitem [{\citenamefont {Giddings}(2013)}]{Giddings:2011xs}%
  \BibitemOpen
  \bibfield  {author} {\bibinfo {author} {\bibfnamefont {S.~B.}\ \bibnamefont {Giddings}},\ }\bibfield  {title} {\bibinfo {title} {{The gravitational S-matrix: Erice lectures}},\ }\href {https://doi.org/10.1142/9789814522489_0005} {\bibfield  {journal} {\bibinfo  {journal} {Subnucl. Ser.}\ }\textbf {\bibinfo {volume} {48}},\ \bibinfo {pages} {93} (\bibinfo {year} {2013})},\ \Eprint {https://arxiv.org/abs/1105.2036} {arXiv:1105.2036 [hep-th]} \BibitemShut {NoStop}%
\bibitem [{\citenamefont {Di~Vecchia}\ \emph {et~al.}(2024)\citenamefont {Di~Vecchia}, \citenamefont {Heissenberg}, \citenamefont {Russo},\ and\ \citenamefont {Veneziano}}]{DiVecchia:2023frv}%
  \BibitemOpen
  \bibfield  {author} {\bibinfo {author} {\bibfnamefont {P.}~\bibnamefont {Di~Vecchia}}, \bibinfo {author} {\bibfnamefont {C.}~\bibnamefont {Heissenberg}}, \bibinfo {author} {\bibfnamefont {R.}~\bibnamefont {Russo}},\ and\ \bibinfo {author} {\bibfnamefont {G.}~\bibnamefont {Veneziano}},\ }\bibfield  {title} {\bibinfo {title} {{The gravitational eikonal: From particle, string and brane collisions to black-hole encounters}},\ }\href {https://doi.org/10.1016/j.physrep.2024.06.002} {\bibfield  {journal} {\bibinfo  {journal} {Phys. Rept.}\ }\textbf {\bibinfo {volume} {1083}},\ \bibinfo {pages} {1} (\bibinfo {year} {2024})},\ \Eprint {https://arxiv.org/abs/2306.16488} {arXiv:2306.16488 [hep-th]} \BibitemShut {NoStop}%
\bibitem [{\citenamefont {Giddings}\ and\ \citenamefont {Porto}(2010)}]{Giddings:2009gj}%
  \BibitemOpen
  \bibfield  {author} {\bibinfo {author} {\bibfnamefont {S.~B.}\ \bibnamefont {Giddings}}\ and\ \bibinfo {author} {\bibfnamefont {R.~A.}\ \bibnamefont {Porto}},\ }\bibfield  {title} {\bibinfo {title} {{The Gravitational S-matrix}},\ }\href {https://doi.org/10.1103/PhysRevD.81.025002} {\bibfield  {journal} {\bibinfo  {journal} {Phys. Rev. D}\ }\textbf {\bibinfo {volume} {81}},\ \bibinfo {pages} {025002} (\bibinfo {year} {2010})},\ \Eprint {https://arxiv.org/abs/0908.0004} {arXiv:0908.0004 [hep-th]} \BibitemShut {NoStop}%
\bibitem [{\citenamefont {Giddings}\ \emph {et~al.}(2010)\citenamefont {Giddings}, \citenamefont {Schmidt-Sommerfeld},\ and\ \citenamefont {Andersen}}]{Giddings:2010pp}%
  \BibitemOpen
  \bibfield  {author} {\bibinfo {author} {\bibfnamefont {S.~B.}\ \bibnamefont {Giddings}}, \bibinfo {author} {\bibfnamefont {M.}~\bibnamefont {Schmidt-Sommerfeld}},\ and\ \bibinfo {author} {\bibfnamefont {J.~R.}\ \bibnamefont {Andersen}},\ }\bibfield  {title} {\bibinfo {title} {{High energy scattering in gravity and supergravity}},\ }\href {https://doi.org/10.1103/PhysRevD.82.104022} {\bibfield  {journal} {\bibinfo  {journal} {Phys. Rev. D}\ }\textbf {\bibinfo {volume} {82}},\ \bibinfo {pages} {104022} (\bibinfo {year} {2010})},\ \Eprint {https://arxiv.org/abs/1005.5408} {arXiv:1005.5408 [hep-th]} \BibitemShut {NoStop}%
\bibitem [{\citenamefont {Levy}\ and\ \citenamefont {Sucher}(1969)}]{Levy:1969cr}%
  \BibitemOpen
  \bibfield  {author} {\bibinfo {author} {\bibfnamefont {M.}~\bibnamefont {Levy}}\ and\ \bibinfo {author} {\bibfnamefont {J.}~\bibnamefont {Sucher}},\ }\bibfield  {title} {\bibinfo {title} {{Eikonal approximation in quantum field theory}},\ }\href {https://doi.org/10.1103/PhysRev.186.1656} {\bibfield  {journal} {\bibinfo  {journal} {Phys. Rev.}\ }\textbf {\bibinfo {volume} {186}},\ \bibinfo {pages} {1656} (\bibinfo {year} {1969})}\BibitemShut {NoStop}%
\bibitem [{\citenamefont {Kabat}\ and\ \citenamefont {Ortiz}(1992)}]{Kabat:1992tb}%
  \BibitemOpen
  \bibfield  {author} {\bibinfo {author} {\bibfnamefont {D.~N.}\ \bibnamefont {Kabat}}\ and\ \bibinfo {author} {\bibfnamefont {M.}~\bibnamefont {Ortiz}},\ }\bibfield  {title} {\bibinfo {title} {{Eikonal quantum gravity and Planckian scattering}},\ }\href {https://doi.org/10.1016/0550-3213(92)90627-N} {\bibfield  {journal} {\bibinfo  {journal} {Nucl. Phys. B}\ }\textbf {\bibinfo {volume} {388}},\ \bibinfo {pages} {570} (\bibinfo {year} {1992})},\ \Eprint {https://arxiv.org/abs/hep-th/9203082} {arXiv:hep-th/9203082} \BibitemShut {NoStop}%
\bibitem [{\citenamefont {Amati}\ \emph {et~al.}(1993)\citenamefont {Amati}, \citenamefont {Ciafaloni},\ and\ \citenamefont {Veneziano}}]{Amati:1993tb}%
  \BibitemOpen
  \bibfield  {author} {\bibinfo {author} {\bibfnamefont {D.}~\bibnamefont {Amati}}, \bibinfo {author} {\bibfnamefont {M.}~\bibnamefont {Ciafaloni}},\ and\ \bibinfo {author} {\bibfnamefont {G.}~\bibnamefont {Veneziano}},\ }\bibfield  {title} {\bibinfo {title} {{Effective action and all order gravitational eikonal at Planckian energies}},\ }\href {https://doi.org/10.1016/0550-3213(93)90367-X} {\bibfield  {journal} {\bibinfo  {journal} {Nucl. Phys. B}\ }\textbf {\bibinfo {volume} {403}},\ \bibinfo {pages} {707} (\bibinfo {year} {1993})}\BibitemShut {NoStop}%
\bibitem [{\citenamefont {Adamo}\ \emph {et~al.}(2022{\natexlab{a}})\citenamefont {Adamo}, \citenamefont {Cristofoli},\ and\ \citenamefont {Tourkine}}]{Adamo:2021rfq}%
  \BibitemOpen
  \bibfield  {author} {\bibinfo {author} {\bibfnamefont {T.}~\bibnamefont {Adamo}}, \bibinfo {author} {\bibfnamefont {A.}~\bibnamefont {Cristofoli}},\ and\ \bibinfo {author} {\bibfnamefont {P.}~\bibnamefont {Tourkine}},\ }\bibfield  {title} {\bibinfo {title} {{Eikonal amplitudes from curved backgrounds}},\ }\href {https://doi.org/10.21468/SciPostPhys.13.2.032} {\bibfield  {journal} {\bibinfo  {journal} {SciPost Phys.}\ }\textbf {\bibinfo {volume} {13}},\ \bibinfo {pages} {032} (\bibinfo {year} {2022}{\natexlab{a}})},\ \Eprint {https://arxiv.org/abs/2112.09113} {arXiv:2112.09113 [hep-th]} \BibitemShut {NoStop}%
\bibitem [{\citenamefont {Bellazzini}\ \emph {et~al.}(2023)\citenamefont {Bellazzini}, \citenamefont {Isabella},\ and\ \citenamefont {Riva}}]{Bellazzini:2022wzv}%
  \BibitemOpen
  \bibfield  {author} {\bibinfo {author} {\bibfnamefont {B.}~\bibnamefont {Bellazzini}}, \bibinfo {author} {\bibfnamefont {G.}~\bibnamefont {Isabella}},\ and\ \bibinfo {author} {\bibfnamefont {M.~M.}\ \bibnamefont {Riva}},\ }\bibfield  {title} {\bibinfo {title} {{Classical vs quantum eikonal scattering and its causal structure}},\ }\href {https://doi.org/10.1007/JHEP04(2023)023} {\bibfield  {journal} {\bibinfo  {journal} {JHEP}\ }\textbf {\bibinfo {volume} {04}},\ \bibinfo {pages} {023}},\ \Eprint {https://arxiv.org/abs/2211.00085} {arXiv:2211.00085 [hep-th]} \BibitemShut {NoStop}%
\bibitem [{\citenamefont {Dray}\ and\ \citenamefont {'t~Hooft}(1985)}]{Dray:1984ha}%
  \BibitemOpen
  \bibfield  {author} {\bibinfo {author} {\bibfnamefont {T.}~\bibnamefont {Dray}}\ and\ \bibinfo {author} {\bibfnamefont {G.}~\bibnamefont {'t~Hooft}},\ }\bibfield  {title} {\bibinfo {title} {{The Gravitational Shock Wave of a Massless Particle}},\ }\href {https://doi.org/10.1016/0550-3213(85)90525-5} {\bibfield  {journal} {\bibinfo  {journal} {Nucl. Phys. B}\ }\textbf {\bibinfo {volume} {253}},\ \bibinfo {pages} {173} (\bibinfo {year} {1985})}\BibitemShut {NoStop}%
\bibitem [{\citenamefont {Aichelburg}\ and\ \citenamefont {Sexl}(1971)}]{Aichelburg:1970dh}%
  \BibitemOpen
  \bibfield  {author} {\bibinfo {author} {\bibfnamefont {P.~C.}\ \bibnamefont {Aichelburg}}\ and\ \bibinfo {author} {\bibfnamefont {R.~U.}\ \bibnamefont {Sexl}},\ }\bibfield  {title} {\bibinfo {title} {{On the Gravitational field of a massless particle}},\ }\href {https://doi.org/10.1007/BF00758149} {\bibfield  {journal} {\bibinfo  {journal} {Gen. Rel. Grav.}\ }\textbf {\bibinfo {volume} {2}},\ \bibinfo {pages} {303} (\bibinfo {year} {1971})}\BibitemShut {NoStop}%
\bibitem [{\citenamefont {Fanizza}\ \emph {et~al.}(2021)\citenamefont {Fanizza}, \citenamefont {Gasperini}, \citenamefont {Pavone},\ and\ \citenamefont {Tedesco}}]{Fanizza:2021ngq}%
  \BibitemOpen
  \bibfield  {author} {\bibinfo {author} {\bibfnamefont {G.}~\bibnamefont {Fanizza}}, \bibinfo {author} {\bibfnamefont {M.}~\bibnamefont {Gasperini}}, \bibinfo {author} {\bibfnamefont {E.}~\bibnamefont {Pavone}},\ and\ \bibinfo {author} {\bibfnamefont {L.}~\bibnamefont {Tedesco}},\ }\bibfield  {title} {\bibinfo {title} {{Linearized propagation equations for metric fluctuations in a general (non-vacuum) background geometry}},\ }\href {https://doi.org/10.1088/1475-7516/2021/07/021} {\bibfield  {journal} {\bibinfo  {journal} {JCAP}\ }\textbf {\bibinfo {volume} {07}},\ \bibinfo {pages} {021}},\ \Eprint {https://arxiv.org/abs/2105.13041} {arXiv:2105.13041 [gr-qc]} \BibitemShut {NoStop}%
\bibitem [{\citenamefont {Duff}(1973)}]{Duff:1973zz}%
  \BibitemOpen
  \bibfield  {author} {\bibinfo {author} {\bibfnamefont {M.~J.}\ \bibnamefont {Duff}},\ }\bibfield  {title} {\bibinfo {title} {{Quantum Tree Graphs and the Schwarzschild Solution}},\ }\href {https://doi.org/10.1103/PhysRevD.7.2317} {\bibfield  {journal} {\bibinfo  {journal} {Phys. Rev. D}\ }\textbf {\bibinfo {volume} {7}},\ \bibinfo {pages} {2317} (\bibinfo {year} {1973})}\BibitemShut {NoStop}%
\bibitem [{\citenamefont {Mougiakakos}\ and\ \citenamefont {Vanhove}(2024)}]{Mougiakakos:2024nku}%
  \BibitemOpen
  \bibfield  {author} {\bibinfo {author} {\bibfnamefont {S.}~\bibnamefont {Mougiakakos}}\ and\ \bibinfo {author} {\bibfnamefont {P.}~\bibnamefont {Vanhove}},\ }\bibfield  {title} {\bibinfo {title} {{Schwarzschild Metric from Scattering Amplitudes to All Orders in GN}},\ }\href {https://doi.org/10.1103/PhysRevLett.133.111601} {\bibfield  {journal} {\bibinfo  {journal} {Phys. Rev. Lett.}\ }\textbf {\bibinfo {volume} {133}},\ \bibinfo {pages} {111601} (\bibinfo {year} {2024})},\ \Eprint {https://arxiv.org/abs/2405.14421} {arXiv:2405.14421 [hep-th]} \BibitemShut {NoStop}%
\bibitem [{\citenamefont {Camanho}\ and\ \citenamefont {Edelstein}(2010)}]{Camanho:2009vw}%
  \BibitemOpen
  \bibfield  {author} {\bibinfo {author} {\bibfnamefont {X.~O.}\ \bibnamefont {Camanho}}\ and\ \bibinfo {author} {\bibfnamefont {J.~D.}\ \bibnamefont {Edelstein}},\ }\bibfield  {title} {\bibinfo {title} {{Causality constraints in AdS/CFT from conformal collider physics and Gauss-Bonnet gravity}},\ }\href {https://doi.org/10.1007/JHEP04(2010)007} {\bibfield  {journal} {\bibinfo  {journal} {JHEP}\ }\textbf {\bibinfo {volume} {04}},\ \bibinfo {pages} {007}},\ \Eprint {https://arxiv.org/abs/0911.3160} {arXiv:0911.3160 [hep-th]} \BibitemShut {NoStop}%
\bibitem [{\citenamefont {Buchel}\ \emph {et~al.}(2010)\citenamefont {Buchel}, \citenamefont {Escobedo}, \citenamefont {Myers}, \citenamefont {Paulos}, \citenamefont {Sinha},\ and\ \citenamefont {Smolkin}}]{Buchel:2009sk}%
  \BibitemOpen
  \bibfield  {author} {\bibinfo {author} {\bibfnamefont {A.}~\bibnamefont {Buchel}}, \bibinfo {author} {\bibfnamefont {J.}~\bibnamefont {Escobedo}}, \bibinfo {author} {\bibfnamefont {R.~C.}\ \bibnamefont {Myers}}, \bibinfo {author} {\bibfnamefont {M.~F.}\ \bibnamefont {Paulos}}, \bibinfo {author} {\bibfnamefont {A.}~\bibnamefont {Sinha}},\ and\ \bibinfo {author} {\bibfnamefont {M.}~\bibnamefont {Smolkin}},\ }\bibfield  {title} {\bibinfo {title} {{Holographic GB gravity in arbitrary dimensions}},\ }\href {https://doi.org/10.1007/JHEP03(2010)111} {\bibfield  {journal} {\bibinfo  {journal} {JHEP}\ }\textbf {\bibinfo {volume} {03}},\ \bibinfo {pages} {111}},\ \Eprint {https://arxiv.org/abs/0911.4257} {arXiv:0911.4257 [hep-th]} \BibitemShut {NoStop}%
\bibitem [{\citenamefont {Akhoury}\ \emph {et~al.}(2021)\citenamefont {Akhoury}, \citenamefont {Saotome},\ and\ \citenamefont {Sterman}}]{Akhoury:2013yua}%
  \BibitemOpen
  \bibfield  {author} {\bibinfo {author} {\bibfnamefont {R.}~\bibnamefont {Akhoury}}, \bibinfo {author} {\bibfnamefont {R.}~\bibnamefont {Saotome}},\ and\ \bibinfo {author} {\bibfnamefont {G.}~\bibnamefont {Sterman}},\ }\bibfield  {title} {\bibinfo {title} {{High Energy Scattering in Perturbative Quantum Gravity at Next to Leading Power}},\ }\href {https://doi.org/10.1103/PhysRevD.103.064036} {\bibfield  {journal} {\bibinfo  {journal} {Phys. Rev. D}\ }\textbf {\bibinfo {volume} {103}},\ \bibinfo {pages} {064036} (\bibinfo {year} {2021})},\ \Eprint {https://arxiv.org/abs/1308.5204} {arXiv:1308.5204 [hep-th]} \BibitemShut {NoStop}%
\bibitem [{\citenamefont {Silva}\ \emph {et~al.}(2021)\citenamefont {Silva}, \citenamefont {Holgado}, \citenamefont {C\'ardenas-Avenda\~no},\ and\ \citenamefont {Yunes}}]{Silva:2020acr}%
  \BibitemOpen
  \bibfield  {author} {\bibinfo {author} {\bibfnamefont {H.~O.}\ \bibnamefont {Silva}}, \bibinfo {author} {\bibfnamefont {A.~M.}\ \bibnamefont {Holgado}}, \bibinfo {author} {\bibfnamefont {A.}~\bibnamefont {C\'ardenas-Avenda\~no}},\ and\ \bibinfo {author} {\bibfnamefont {N.}~\bibnamefont {Yunes}},\ }\bibfield  {title} {\bibinfo {title} {{Astrophysical and theoretical physics implications from multimessenger neutron star observations}},\ }\href {https://doi.org/10.1103/PhysRevLett.126.181101} {\bibfield  {journal} {\bibinfo  {journal} {Phys. Rev. Lett.}\ }\textbf {\bibinfo {volume} {126}},\ \bibinfo {pages} {181101} (\bibinfo {year} {2021})},\ \Eprint {https://arxiv.org/abs/2004.01253} {arXiv:2004.01253 [gr-qc]} \BibitemShut {NoStop}%
\bibitem [{\citenamefont {Shapiro}(1964)}]{Shapiro:1964uw}%
  \BibitemOpen
  \bibfield  {author} {\bibinfo {author} {\bibfnamefont {I.~I.}\ \bibnamefont {Shapiro}},\ }\bibfield  {title} {\bibinfo {title} {{Fourth Test of General Relativity}},\ }\href {https://doi.org/10.1103/PhysRevLett.13.789} {\bibfield  {journal} {\bibinfo  {journal} {Phys. Rev. Lett.}\ }\textbf {\bibinfo {volume} {13}},\ \bibinfo {pages} {789} (\bibinfo {year} {1964})}\BibitemShut {NoStop}%
\bibitem [{\citenamefont {Will}(2014)}]{Will:2014kxa}%
  \BibitemOpen
  \bibfield  {author} {\bibinfo {author} {\bibfnamefont {C.~M.}\ \bibnamefont {Will}},\ }\bibfield  {title} {\bibinfo {title} {{The Confrontation between General Relativity and Experiment}},\ }\href {https://doi.org/10.12942/lrr-2014-4} {\bibfield  {journal} {\bibinfo  {journal} {Living Rev. Rel.}\ }\textbf {\bibinfo {volume} {17}},\ \bibinfo {pages} {4} (\bibinfo {year} {2014})},\ \Eprint {https://arxiv.org/abs/1403.7377} {arXiv:1403.7377 [gr-qc]} \BibitemShut {NoStop}%
\bibitem [{\citenamefont {Misner}\ \emph {et~al.}(1973)\citenamefont {Misner}, \citenamefont {Thorne},\ and\ \citenamefont {Wheeler}}]{Misner:1973prb}%
  \BibitemOpen
  \bibfield  {author} {\bibinfo {author} {\bibfnamefont {C.~W.}\ \bibnamefont {Misner}}, \bibinfo {author} {\bibfnamefont {K.~S.}\ \bibnamefont {Thorne}},\ and\ \bibinfo {author} {\bibfnamefont {J.~A.}\ \bibnamefont {Wheeler}},\ }\href@noop {} {\emph {\bibinfo {title} {{Gravitation}}}}\ (\bibinfo  {publisher} {W. H. Freeman},\ \bibinfo {address} {San Francisco},\ \bibinfo {year} {1973})\BibitemShut {NoStop}%
\bibitem [{\citenamefont {Poisson}\ and\ \citenamefont {Will}(2014)}]{poisson2014gravity}%
  \BibitemOpen
  \bibfield  {author} {\bibinfo {author} {\bibfnamefont {E.}~\bibnamefont {Poisson}}\ and\ \bibinfo {author} {\bibfnamefont {C.~M.}\ \bibnamefont {Will}},\ }\href@noop {} {\emph {\bibinfo {title} {Gravity: Newtonian, post-newtonian, relativistic}}}\ (\bibinfo  {publisher} {Cambridge University Press},\ \bibinfo {year} {2014})\BibitemShut {NoStop}%
\bibitem [{\citenamefont {Yunes}\ and\ \citenamefont {Pretorius}(2009)}]{Yunes:2009hc}%
  \BibitemOpen
  \bibfield  {author} {\bibinfo {author} {\bibfnamefont {N.}~\bibnamefont {Yunes}}\ and\ \bibinfo {author} {\bibfnamefont {F.}~\bibnamefont {Pretorius}},\ }\bibfield  {title} {\bibinfo {title} {{Dynamical Chern-Simons Modified Gravity. I. Spinning Black Holes in the Slow-Rotation Approximation}},\ }\href {https://doi.org/10.1103/PhysRevD.79.084043} {\bibfield  {journal} {\bibinfo  {journal} {Phys. Rev. D}\ }\textbf {\bibinfo {volume} {79}},\ \bibinfo {pages} {084043} (\bibinfo {year} {2009})},\ \Eprint {https://arxiv.org/abs/0902.4669} {arXiv:0902.4669 [gr-qc]} \BibitemShut {NoStop}%
\bibitem [{\citenamefont {Ali-Haimoud}\ and\ \citenamefont {Chen}(2011)}]{Ali-Haimoud:2011zme}%
  \BibitemOpen
  \bibfield  {author} {\bibinfo {author} {\bibfnamefont {Y.}~\bibnamefont {Ali-Haimoud}}\ and\ \bibinfo {author} {\bibfnamefont {Y.}~\bibnamefont {Chen}},\ }\bibfield  {title} {\bibinfo {title} {{Slowly-rotating stars and black holes in dynamical Chern-Simons gravity}},\ }\href {https://doi.org/10.1103/PhysRevD.84.124033} {\bibfield  {journal} {\bibinfo  {journal} {Phys. Rev. D}\ }\textbf {\bibinfo {volume} {84}},\ \bibinfo {pages} {124033} (\bibinfo {year} {2011})},\ \Eprint {https://arxiv.org/abs/1110.5329} {arXiv:1110.5329 [astro-ph.HE]} \BibitemShut {NoStop}%
\bibitem [{\citenamefont {Adamo}\ \emph {et~al.}(2022{\natexlab{b}})\citenamefont {Adamo}, \citenamefont {Cristofoli},\ and\ \citenamefont {Ilderton}}]{Adamo:2022rmp}%
  \BibitemOpen
  \bibfield  {author} {\bibinfo {author} {\bibfnamefont {T.}~\bibnamefont {Adamo}}, \bibinfo {author} {\bibfnamefont {A.}~\bibnamefont {Cristofoli}},\ and\ \bibinfo {author} {\bibfnamefont {A.}~\bibnamefont {Ilderton}},\ }\bibfield  {title} {\bibinfo {title} {{Classical physics from amplitudes on curved backgrounds}},\ }\href {https://doi.org/10.1007/JHEP08(2022)281} {\bibfield  {journal} {\bibinfo  {journal} {JHEP}\ }\textbf {\bibinfo {volume} {08}},\ \bibinfo {pages} {281}},\ \Eprint {https://arxiv.org/abs/2203.13785} {arXiv:2203.13785 [hep-th]} \BibitemShut {NoStop}%
\bibitem [{\citenamefont {Grumiller}\ and\ \citenamefont {Yunes}(2008)}]{Grumiller:2007rv}%
  \BibitemOpen
  \bibfield  {author} {\bibinfo {author} {\bibfnamefont {D.}~\bibnamefont {Grumiller}}\ and\ \bibinfo {author} {\bibfnamefont {N.}~\bibnamefont {Yunes}},\ }\bibfield  {title} {\bibinfo {title} {{How do Black Holes Spin in Chern-Simons Modified Gravity?}},\ }\href {https://doi.org/10.1103/PhysRevD.77.044015} {\bibfield  {journal} {\bibinfo  {journal} {Phys. Rev. D}\ }\textbf {\bibinfo {volume} {77}},\ \bibinfo {pages} {044015} (\bibinfo {year} {2008})},\ \Eprint {https://arxiv.org/abs/0711.1868} {arXiv:0711.1868 [gr-qc]} \BibitemShut {NoStop}%
\bibitem [{\citenamefont {Alexander}\ \emph {et~al.}(2021)\citenamefont {Alexander}, \citenamefont {Gabadadze}, \citenamefont {Jenks},\ and\ \citenamefont {Yunes}}]{Alexander:2021ssr}%
  \BibitemOpen
  \bibfield  {author} {\bibinfo {author} {\bibfnamefont {S.}~\bibnamefont {Alexander}}, \bibinfo {author} {\bibfnamefont {G.}~\bibnamefont {Gabadadze}}, \bibinfo {author} {\bibfnamefont {L.}~\bibnamefont {Jenks}},\ and\ \bibinfo {author} {\bibfnamefont {N.}~\bibnamefont {Yunes}},\ }\bibfield  {title} {\bibinfo {title} {{Chern-Simons caps for rotating black holes}},\ }\href {https://doi.org/10.1103/PhysRevD.104.064033} {\bibfield  {journal} {\bibinfo  {journal} {Phys. Rev. D}\ }\textbf {\bibinfo {volume} {104}},\ \bibinfo {pages} {064033} (\bibinfo {year} {2021})},\ \Eprint {https://arxiv.org/abs/2104.00019} {arXiv:2104.00019 [hep-th]} \BibitemShut {NoStop}%
\bibitem [{\citenamefont {Alexander}\ \emph {et~al.}(2024)\citenamefont {Alexander}, \citenamefont {Bernardo},\ and\ \citenamefont {Creque-Sarbinowski}}]{Alexander:2024vav}%
  \BibitemOpen
  \bibfield  {author} {\bibinfo {author} {\bibfnamefont {S.}~\bibnamefont {Alexander}}, \bibinfo {author} {\bibfnamefont {H.}~\bibnamefont {Bernardo}},\ and\ \bibinfo {author} {\bibfnamefont {C.}~\bibnamefont {Creque-Sarbinowski}},\ }\bibfield  {title} {\bibinfo {title} {{Nontriviality of dynamical Chern-Simons gravity and the standard model}},\ }\href {https://doi.org/10.1103/PhysRevD.110.025015} {\bibfield  {journal} {\bibinfo  {journal} {Phys. Rev. D}\ }\textbf {\bibinfo {volume} {110}},\ \bibinfo {pages} {025015} (\bibinfo {year} {2024})},\ \Eprint {https://arxiv.org/abs/2403.15657} {arXiv:2403.15657 [hep-th]} \BibitemShut {NoStop}%
\bibitem [{\citenamefont {Nakamura}\ \emph {et~al.}(2019)\citenamefont {Nakamura}, \citenamefont {Kikuchi}, \citenamefont {Yamada}, \citenamefont {Asada},\ and\ \citenamefont {Yunes}}]{Nakamura:2018yaw}%
  \BibitemOpen
  \bibfield  {author} {\bibinfo {author} {\bibfnamefont {Y.}~\bibnamefont {Nakamura}}, \bibinfo {author} {\bibfnamefont {D.}~\bibnamefont {Kikuchi}}, \bibinfo {author} {\bibfnamefont {K.}~\bibnamefont {Yamada}}, \bibinfo {author} {\bibfnamefont {H.}~\bibnamefont {Asada}},\ and\ \bibinfo {author} {\bibfnamefont {N.}~\bibnamefont {Yunes}},\ }\bibfield  {title} {\bibinfo {title} {{Weakly-gravitating objects in dynamical Chern\textendash{}Simons gravity and constraints with gravity probe B}},\ }\href {https://doi.org/10.1088/1361-6382/ab04c5} {\bibfield  {journal} {\bibinfo  {journal} {Class. Quant. Grav.}\ }\textbf {\bibinfo {volume} {36}},\ \bibinfo {pages} {105006} (\bibinfo {year} {2019})},\ \Eprint {https://arxiv.org/abs/1810.13313} {arXiv:1810.13313 [gr-qc]} \BibitemShut {NoStop}%
\bibitem [{\citenamefont {Adelberger}\ \emph {et~al.}(2003)\citenamefont {Adelberger}, \citenamefont {Heckel},\ and\ \citenamefont {Nelson}}]{Adelberger:2003zx}%
  \BibitemOpen
  \bibfield  {author} {\bibinfo {author} {\bibfnamefont {E.~G.}\ \bibnamefont {Adelberger}}, \bibinfo {author} {\bibfnamefont {B.~R.}\ \bibnamefont {Heckel}},\ and\ \bibinfo {author} {\bibfnamefont {A.~E.}\ \bibnamefont {Nelson}},\ }\bibfield  {title} {\bibinfo {title} {{Tests of the gravitational inverse square law}},\ }\href {https://doi.org/10.1146/annurev.nucl.53.041002.110503} {\bibfield  {journal} {\bibinfo  {journal} {Ann. Rev. Nucl. Part. Sci.}\ }\textbf {\bibinfo {volume} {53}},\ \bibinfo {pages} {77} (\bibinfo {year} {2003})},\ \Eprint {https://arxiv.org/abs/hep-ph/0307284} {arXiv:hep-ph/0307284} \BibitemShut {NoStop}%
\end{thebibliography}%

\end{document}